\documentclass[pre,twocolumn,floatfix,superscriptaddress,longbibliography,nofootinbib]{revtex4-1}

\usepackage[utf8]{inputenc}
\usepackage{amsmath,amssymb,amsfonts}
\usepackage{hyperref}
\DeclareMathOperator{\Tr}{Tr}

\usepackage[justification=centerlast,format=plain]{caption}

\usepackage{xcolor}
\newcommand{\rev}[1]{{\color{black}#1}}

\usepackage{graphicx}
\usepackage{subcaption}
\usepackage{ctable}

\begin{document}

\title{High-quality Thermal Gibbs Sampling with Quantum Annealing Hardware}

\author{Jon Nelson}
\affiliation{Center for Non Linear Studies, Los Alamos National Laboratory, Los Alamos, NM, USA}
\affiliation{Advanced Network Science Initiative, Los Alamos National Laboratory, Los Alamos, NM, USA}

\author{Marc Vuffray}
\affiliation{Theoretical Division, Los Alamos National Laboratory, Los Alamos, NM, USA}

\author{Andrey Y. Lokhov}
\affiliation{Theoretical Division, Los Alamos National Laboratory, Los Alamos, NM, USA}

\author{Tameem Albash}
\affiliation{Department of Electrical and Computer Engineering, University of New Mexico, Albuquerque, NM, USA}
\affiliation{Department of Physics and Astronomy and Center for Quantum Information and Control, University of New Mexico, Albuquerque, NM, USA}

\author{Carleton Coffrin}
\affiliation{Advanced Network Science Initiative, Los Alamos National Laboratory, Los Alamos, NM, USA}


\begin{abstract}
Quantum Annealing (QA) was originally intended for accelerating the solution of combinatorial optimization tasks that have natural encodings as Ising models.
However, recent experiments on QA hardware platforms have demonstrated that, in the operating regime corresponding to weak interactions, the QA hardware behaves like a noisy Gibbs sampler at a hardware-specific effective temperature.
This work builds on those insights and identifies a class of small hardware-native Ising models that are robust to noise effects and proposes a procedure for executing these models on QA hardware to maximize Gibbs sampling performance.
Experimental results indicate that the proposed protocol results in high-quality Gibbs samples from a hardware-specific effective temperature. \rev{Furthermore, we show that this effective temperature can be adjusted by modulating the annealing time and energy scale.}
The procedure proposed in this work provides an approach to using QA hardware for Ising model sampling presenting potential new opportunities for applications in machine learning and physics simulation.
\end{abstract}


\maketitle

\section{Introduction}

The computational task of sampling -- that is producing independent configurations of random variables from a given distribution -- is believed to be among the most challenging computational problems.
In particular, many sampling tasks cannot be performed in polynomial time, unless strong and widely accepted conjectures in approximation theory are refuted \cite{sinclair1989approximate,jerrum1986random,jerrum1993polynomial}.
Consequently, state-of-the-art algorithms for general purpose sampling are based on Monte Carlo methods that require significant computing resources to sample from distributions of practical interest. 

Gibbs distributions, i.e., distributions with the form $P(\sigma) \propto \exp(-\beta H(\sigma))$, have close ties to modeling many physical systems as they capture the equilibrium configurations that matter takes at a given inverse temperature $\beta$ \cite{gallavotti2013statistical}.
It has also been observed that Gibbs distributions provide a powerful foundation for building machine learning models and optimization algorithms \cite{Job_2018,Perdomo_Ortiz_2018}.
However, the computational challenge of sampling from these distributions is often prohibitive for practical applications of machine learning and optimization.
It has been observed that a variety of the emerging analog computing devices, including those based on optical \cite{inagaki2016coherent} and quantum \cite{arute2019quantum} technologies, can be used as fast Gibbs samplers because of the natural connection between probabilistic physical systems and state sampling.
The emergence of these novel computing devices has the potential for a dramatic impact on state-of-the-art approaches for physics simulation, machine learning and optimization.

From its inception, Quantum Annealing \cite{PhysRevE.58.5355,Brooke1999,Brooke2001,Farhi2001} (QA) has been designed for solving optimization tasks and not sampling tasks.
However, in practice, various non-ideal properties of QA hardware platforms lead to outputs that are reminiscent of thermal Gibbs distributions \cite{dwave_boltzmann,10.3389/fict.2016.00023,PhysRevApplied.8.064025,PhysRevApplied.11.044083}. 
Recent experiments have demonstrated that, in the operating regime corresponding to weak \rev{input interactions}, QA hardware behaves like a {\em noisy Gibbs sampler} \cite{2012.08827} at a hardware-specific effective temperature, that is, a sampler from a mixture of Gibbs distributions with fluctuating parameters caused by noise. 
Furthermore, it was shown in \cite{2012.08827} that samples from this noisy mixture of distributions is indistinguishable from a single Gibbs distribution with spurious additions to interaction structure of the programmed model.
In applications where sampling from a specific Gibbs distribution is required, this distortion of the model parameters may present an undesirable feature because of this mismatch in model structure.

Building on the insights of \cite{2012.08827}, this work demonstrates that it is possible to leverage QA hardware to perform high-quality Gibbs sampling from some types of input models.
In particular, we identify a class of hardware-native Ising models where the D-Wave 2000Q platform consistently produces samples with a \rev{total variation distance} less than $5\%$ from a desired target Gibbs distribution. \rev{To this end, this work explores the full range of operational input parameters, in contrast with \cite{2012.08827} that focused on quantifying the effect of noise on the structure of the output distribution and hence considered a narrow range of weak input couplings.}
We also show that changing the annealing time enables the QA hardware to sample from a range of inverse temperatures between 1.9 and 5.1, which is an essential feature in a variety of applications.
This work leverages the combination of three insights to achieve high-quality Gibbs sampling on the QA hardware:
(1) it focuses on the hardware-native Ising models that are resilient to the hardware's noise; 
(2) the input models are re-scaled to avoid distortions caused by the transverse field in the computational model; and
(3) both the annealing time and the model's energy scale are leveraged to control the effective temperature of the target distribution.
This results in a prescriptive procedure for conducting high-quality Ising model Gibbs sampling on QA hardware without the need to tune the input model or perform post-processing to correct for distorted output distributions.
If required, both of these procedures can be leveraged to further improve the results presented in this work, as done in \cite{dwave_boltzmann,PhysRevA.94.022308,PhysRevX.7.041052}.
\rev{While this work shows that, under specific conditions, Gibbs sampling is possible using quantum annealing hardware, it does not illustrate how it can be used in specific sampling applications.}

This work begins by reviewing the computational model of quantum annealing and previous works exploring sampling applications in Section \ref{sec:background}.
Building on these previous works, a procedure for conducting high-quality thermal Gibbs sampling is proposed in Section \ref{sec:sampling} and evaluated on quantum annealing hardware. Section \ref{sec:discussion} explores theoretical models to provide insights into the mechanisms that under-pin the proposed sampling approach and Section \ref{sec:conclusion} concludes with a discussion of future directions.

\section{Problem Formulation \&  Related Work}
\label{sec:background}

The foundational physical system of interest to this work is the Ising model, a class of graphical models where the nodes, $V$, represent classical spins (i.e., $\sigma_i \in \{-1,1\} ~\forall i \in V$), and the edges, $E \subseteq N \times N$, represent pairwise spin interactions.
A local field $h_i ~\forall i \in V$ is specified for each spin and an interaction strength $J_{ij} ~\forall i,j \in E$ is specified for each edge. 
The  energy of a given spin configuration is given by the Hamiltonian:

\begin{align}
    H_{\textrm{Ising}}(\sigma) &= - \sum_{i,j \in E} J_{ij} \sigma_i \sigma_j - \sum_{i \in V} h_i \sigma_i \ . \label{eq:ising_eng}
\end{align}

This Ising model is widely used to encode challenging computational problems arising in the study of magnetic materials, machine learning, and optimization \cite{hopfield1982neural,panjwani1995markov,lokhov2018optimal,Kochenberger2014}.
The particular computational task of interest to this work is to produce i.i.d. samples $\sigma$ from the Gibbs distribution associated with the Ising model at the inverse temperature\footnote{Throughout this work, $\alpha$ is used to denote the energy scale of a Gibbs distribution and $\beta$ is used to indicate the effective inverse temperature of the quantum annealer.} $\alpha$, that is,

\begin{align}
    \mu(\sigma) \propto \exp \left( -\alpha H_{\textrm{Ising}}(\sigma) \right) \ . \label{eq:ising_prob}
\end{align}

Throughout this work, we assume that the parameters $J,h$ are in the range of $-1$ to $1$, to provide a consistent scaling for $\alpha$.
Given the computational challenge of Ising model sampling at finite temperature, our objective is to leverage QA hardware to conduct this task.

\subsection{A Brief Review of Quantum Annealing}
The central idea of quantum annealing is to use the transverse field Ising model combined with an annealing process to find the low-energy configurations of a classical Ising model.
The elementary unit of this model is a qubit $i \in V$ described by the standard vector of Pauli matrices $\{\widehat{\sigma}^x, \widehat{\sigma}^y, \widehat{\sigma}^z \}$ along the three spatial directions $\{x,y,z\}$.
The hardware platform provides a programmable Ising model Hamiltonian on the $z$-axis \cite{PhysRevE.58.5355},

\begin{equation}
\widehat{H}_{\text{Ising}} = - \sum_{ij \in E} J_{ij} \widehat{\sigma}^{z}_i \widehat{\sigma}^{z}_j - \sum_{i \in V}  h_{i} \widehat{\sigma}^{z}_i,
\end{equation}

which encodes the Ising energy function \eqref{eq:ising_eng} with a one-to-one mapping of qubits to Ising spins. Note, that the eigenvalues of the Ising Hamiltonian operator are in bijection with the $2^N$ possible assignments of the classical Ising model from Eq.~\eqref{eq:ising_eng}.
The quantum annealing protocol strives to find the low-energy assignments to a user-specified $\widehat{H}_{\text{Ising}}$ model by conducting an analog interpolation process of the following transverse field Ising model Hamiltonian:

\begin{equation}
    \widehat{H}(s) = - A(s) \sum_{i \in V} \widehat{\sigma}^{x}_i + B(s) \widehat{H}_{\text{Ising}}. \label{eq:tising}
\end{equation}

The interpolation process starts with $s = 0$ and ends with $s = 1$. The two interpolation functions $A(s)$ and $B(s)$ are designed such that $A(0) \gg B(0)$ and $A(1) \ll B(1)$, that is, starting with a Hamiltonian dominated by $- \sum_{i \in V}\widehat{\sigma}^{x}_i$ and slowly transitioning to a Hamiltonian dominated by $\widehat{H}_{\text{Ising}}$.
In the closed system setting and when this transition process is sufficiently slow, \rev{the quantum annealing is considered adiabatic quantum computing. It is known that under these conditions, quantum annealing} will find the ground states of $\widehat{H}_{\mathrm{Ising}}$, and therefore minimizing configurations of ${H}_{\mathrm{Ising}}$, with high probability \cite{Einstein:adiabatic,Born:28,Ehrenfest:adiabatic,Kato:50,Jansen:07}.
To that end, the length of the annealing process $t$ (in microseconds) is a user controllable parameter.
The outcome of the quantum annealing process is specified by the binary string $\sigma$, where each element $\sigma_i$ takes a value $+1$ or $-1$ and corresponds to the observation of the spin projection of qubit $i$ in the computational basis denoted by $z$.

\subsection{Quantum Annealing and Sampling}
Given the energy minimizing nature of quantum annealing, it is not immediately clear how it may be applied to the sampling task posed in \eqref{eq:ising_prob}.
It is reasonable to postulate that an adiabatic system would behave similarly to Gibbs sampling at the zero temperature limit (i.e., $\alpha = \infty$), that is drawing i.i.d. samples from the ground states of $H_{\text{Ising}}$.
Interestingly, this is not usually the case as the transverse field component of the Hamiltonian, $\widehat{\sigma}^{x}_i$, induces biases in the annealing process to prefer some ground states over others \cite{Matsuda_2009,Boixo2013,Zhang2017,PhysRevA.100.030303}.
Nevertheless, protocols have been proposed to help increase the fairness of ground state sampling using quantum annealing \cite{2007.08487,2107.06468}.  
In contrast to previous works, this work is concerned with sampling from thermal distributions (i.e., $\alpha < \infty$), which requires accurate sampling at all of the energy levels of $H_{\text{Ising}}$, presenting a formidable challenge in model accuracy.

Real-world QA hardware is an open quantum system that is subject to a wide variety of non-ideal properties, including thermal excitations and relaxations that impact the results of the computation \cite{PhysRevA.91.062320,Boixo2016,Smirnov_2018}.
In this regard, the relevant adiabatic theorem is the open system adiabatic theorem \cite{Abo2007,Avr2012,Ven2016}: if the dynamics is governed by a master equation of Lindblad form \cite{Lin1976} and if the annealing is done sufficiently slowly, the system will converge with high probability to the steady state of the dynamics. This result is promising because, for certain master equations, the thermal Gibbs state is the steady-state solution \cite{Dav1974,Alb2012}.

In addition to the fundamental impacts of open quantum systems, the D-Wave hardware documentation highlights five other sources of deviations from ideal system operations called {\it integrated control errors} (ICE) \cite{dwave_docs}, which include: background susceptibility; flux noise; Digital-to-Analog Conversion quantization; Input/Ouput system effects; and variable scale across qubits.
Consequently, it has long been observed that output distributions of the QA hardware produced by D-Wave Systems are reminiscent of a Gibbs distribution of the input Hamiltonian $H_{\text{Ising}}$ \cite{dwave_boltzmann,PhysRevA.94.022308,10.3389/fict.2016.00023,PhysRevApplied.8.064025,PhysRevApplied.11.044083} with a hardware-specific effective temperature of $\beta \approx 10$ \cite{9465651,2012.08827}.
The prevailing interpretation of the hardware's output distribution is the {\em Freeze Out} model, which proposes that the output reflects a quantum Gibbs distribution occurring at an input-dependent point towards the end of the annealing process where some small amount of $\widehat{\sigma}^{x}_i$ remains \cite{PhysRevA.92.052323}.
This model has been notably successful for training quantum machine learning models \cite{PhysRevX.8.021050,Winci_2020} and generating statistics for quantum Gibbs distributions \cite{Harris162,King2018,Kingeabe2824}.
However,  these quantum Gibbs distributions are inaccurate when targeting a desired classical Gibbs distribution for sampling applications \cite{PhysRevA.94.022308,Perdomo_Ortiz_2018,PhysRevResearch.2.023020,Li_2020}.
The recent insight from \cite{2012.08827} is that when this hardware is operated at a low-energy scale (i.e., $|J|,|h| \leq 0.050$) it behaves as a thermalized classical Gibbs sampler from $H_{\text{Ising}}$ but suffers from a notable amount of distortion from instantaneous noise in the local field parameters, $h$, on the order of $0.036$ \cite{9465651,9319535}, \rev{behaving as a so-called noisy Gibbs sampler}.

Inspired by the classical Gibbs sampling insights of \cite{2012.08827}, this work demonstrates that there exists a class of Ising Hamiltonians where D-Wave's 2000Q hardware produces high-quality Gibbs samples, with minimal distortions from noise or transverse fields.
In particular, we consider the class of $H_{\text{Ising}}$ with $J,h \in \{-1,0,1\}$, $|V| \leq 16$ that are representable on the D-Wave 2000Q hardware graph.
The primary insight of this work is to operate the QA hardware at an energy scale, $\alpha_{in}$, that is large enough to be noise-resilient but small enough to avoid the degeneracy breaking properties of the transverse field.
It is not guaranteed that such a `sweet-spot' exists but this work demonstrates that the range of $0.2 \leq \alpha_{in} \leq 0.4$ on the platform considered achieves the desired properties.
\rev{Section \ref{sec:discussion}} of the paper provides a qualitative study postulating why a sweet-spot occurs at this particular energy scale.
Additionally, this work shows that in the proposed energy scale the annealing time has a consistent effect of tuning the effective temperature (i.e., $\alpha$) of the Gibbs samples generated by the hardware.
Combined these observations present new opportunities for leveraging QA hardware for conducting the Ising model sampling task described by \eqref{eq:ising_prob}.

\section{Exploring Gibbs Sampling with Quantum Annealing Hardware}
\label{sec:sampling}

This work is concerned with the following four parameters:
$H_{\mathrm{Ising}}$, the Ising model that one would like to sample from (restricted to $J,h \in \{-1,0,1\}$);
$\alpha_{in} \in [0,1]$, a scaling factor that will be used when programming the QA hardware;
$t$, the annealing time of the QA protocol; and
$\alpha_{out}\in [0,\alpha_{\mathrm{max}}]$, a scaling factor of the distribution output by the QA hardware, \rev{where $\alpha_{max}$ is an estimated upper bound and is further described in section \ref{alphaout}}.
Given $H_{\mathrm{Ising}}$, the QA hardware is programmed with the re-scaled model $H^{in}_{\mathrm{Ising}} = \alpha_{in} H_{\mathrm{Ising}}$ and executed with an annealing time $t$. \rev{Note that rescaling is equivalent to replacing $J$ with $\alpha_{in} \cdot J$ and likewise with $h$.}
The empirical distribution output by the hardware $\nu$ is then compared to Gibbs distributions of $H_{\mathrm{Ising}}$ at different effective temperatures, i.e.,  $\mu(\alpha, \sigma) \propto \exp \left( -\alpha H_{\mathrm{Ising}}(\sigma) \right)$.
We define $\alpha_{out}$ as,
\begin{align}
\alpha_{out} = \mathrm{argmin}_{\alpha} \;\; \mathrm{TV}(\nu, \mu(\alpha)),
\label{eq:alpha_out}
\end{align}
that is, the effective temperature of the closest Gibbs distribution of $H_{\mathrm{Ising}}$ to the results output by the hardware, using total variation distance (TV) as the measure of closeness.
Details of the total variation metric are provided in Appendix \ref{app:total-variation}.
The remainder of this section explores how changes in the $\alpha_{in}$ and $t$ parameters impact the output distribution of the D-Wave 2000Q Quantum Annealer located at Los Alamos National Laboratory, known as \texttt{DW\_2000Q\_LANL}.
The unexpected finding is that there exists range of $\alpha_{in}$ values where the QA hardware is a high-quality Gibbs sampler.

\subsection{Experiment Design}

\subsubsection{Ising Model Selection}

Given the challenges of sampling from embedded Ising models \cite{PhysRevResearch.2.023020}, this work focuses exclusively on $H_{\mathrm{ising}}$ Hamiltonians that have native representation in the QA hardware.
In particular, the D-Wave 2000Q system implements a ${\cal C}_{16}$ chimera graph \cite{6802426}, which consists of a $16 \times 16$ grid of unit cells each containing 8 qubits (4 horizontal and 4 vertical).
In each unit cell, every horizontal qubit is connected to every vertical qubit through couplers and various qubits in adjacent unit cells are also connected through couplers.
The coupling strength $J_{ij}$ between qubits $i$ and $j$ can be programmed to values in the continuous range $\left[-1, 1\right]$, and the local fields $h_i$ can be programmed to values in the range $\left[-2, 2\right]$.
\rev{As part of the execution process these unitless quantities are transformed into the hardware's energy scales of $0.0$ to $6.36$ GHz for $A(s)/h$ and $0.07$ to $14.56$ GHz for $B(s)/h$, where $h$ is Planck's constant.}

In selecting the Ising models for testing Gibbs sampling with QA, we strive to design models that exhibit a variety of sampling difficulties.
Previous works have highlighted that QA will prefer some degenerate ground states over others as a result of residual effects from the annealing process \cite{Matsuda_2009,Boixo2013,Zhang2017,PhysRevA.100.030303}.
As ground states are the most heavily weighted states in the low-temperature Gibbs distribution, these asymmetries introduce significant sampling errors.
In this work, we elicit this effect by constructing a set of 13 models with ground-state degeneracies ranging from 1 to 38 to capture both easy and challenging instances to sample from with a quantum annealer.

In particular, this work considers seven {\em GSD} (ground-state degeneracy) cases with degeneracy values of 2, 4, 6, 8, 10, 24, and 38, without local fields, and six additional {\em GSD-F} models with 1, 2, 3, 4, 5, and 6 ground state degeneracies including local fields.
Note that the GSD-F models tend to have lower degeneracy because the inclusion of the fields breaks symmetries within energy levels.
The specific instances are referred to as GSD-\# and GSD-F-\# where the \# indicates the amount of ground state degeneracy.
The \rev{Hamiltonians} of each instance and a discussion of how they were designed is provided in Appendix \ref{app:gsd-details}.

\vspace{-0.65cm}
\subsubsection{Quantum Annealing Data Collection}
For each of the GSD models, $H_{\mathrm{Ising}}$, a family of hardware inputs is considered by sweeping $\alpha_{in}$ between 0.0 and 1.0 with step size of 0.0125 below 0.1, 0.025 between 0.1 and 0.5, and 0.1 above 0.5.
The increased step size for low scales is helpful as the output statistics tend to be more sensitive to $\alpha_{in}$ in this regime.
For each of these models and scales, $10^6$ samples are collected from the QA hardware. After every 100 samples, a gauge transform is applied in order to mitigate the effects of bias.
For each gauge transformation, a randomly generated vector $\vec{a} \in \left\{-1,1\right\}^V$ is used to redefine the Ising parameters via $h_i \to a_i h_i, J_{ij} \to a_i a_j J_{ij}$.
This transformation preserves the energy eigenvalues of $H(s)$, but it relabels the spin configurations $\sigma$ by $\sigma_i \to a_i \sigma_i$.
Using gauge transformations is a well-known technique for eliminating these field biases, which can add additional unwanted asymmetries \cite{Ronnow420,PhysRevA.91.042314,2012.08827}.
After the samples have been collected, the empirical probability of each state is recorded.
This data collection process is repeated for annealing times $t$ of 1 $\mu s$, 5 $\mu s$, 25 $\mu s$, and 125 $\mu s$.

\subsubsection{Estimating the Effective Output Temperature} \label{alphaout}

As discussed at the beginning of this section, we would like to identify the closest Gibbs distribution of $H_{\mathrm{Ising}}$ to the empirical distribution output by the QA hardware, i.e., \eqref{eq:alpha_out}.
However, the exact solution of the optimization task posed by \eqref{eq:alpha_out} presents a significant computational challenge because of difficulties in computing candidate Gibbs distributions.
The primary reason that this work has focused on the small-scale Ising models with only 16 spins is to make this optimization task computationally tractable.
In this work, the optimization problem \eqref{eq:alpha_out} is solved by a brute-force calculation of the \rev{total variation distance} for a discrete set of $\alpha$ values.
The $\alpha$ values are determined by calculating an optimistic estimate for $\alpha_{\mathrm{max}}$ and then scaling this value by the same step sizes that are used for $\alpha_{in}$.
The $\alpha_{\mathrm{max}}$ value is calculated by executing the \rev{single-qubit} protocol proposed in \cite{9465651} on each of the relevant spins and taking the average of the spin-by-spin effective temperatures recovered by that protocol. 
Therefore, the minimum of the range is $0.0$ and the maximum is $\alpha_{\mathrm{max}}$.
The $\alpha$ value from this discrete set that achieves the minimum \rev{total variation distance} between the quantum annealing data and the Gibbs distribution is selected as $\alpha_{out}$.
In the results, we observe that $\alpha_{out} \ll  \alpha_{\mathrm{max}}$ indicating that $\alpha_{\mathrm{max}}$ is a sufficiently optimistic value for this brute-force optimization procedure in practice.

\subsubsection{Bounding Finite Sampling Performance}

Because the empirical distribution produced by the QA hardware is calculated from finite samples, there is an unavoidable error caused by finite sampling (details are discussed in Appendix \ref{app:total-variation}).
To understand the impact of this finite sampling error, a lower bound is calculated by a simulation procedure 
that generates $10^6$ samples from the selected $\alpha_{out}$ Gibbs distribution and calculates the \rev{total variation distance} between this empirical distribution and the exact distribution.
Because the samples are generated from a known source distribution, the only source of error is a result of finite sampling.
Therefore, this value represents the best \rev{total variation distance} that is achievable given the number of samples that were used in these experiments $10^6$.
This lower bound is included in the result figures to provide a measure of the \rev{portion of the total variation distance} that can be attributed solely to the limitations of finite sampling.
In this work we find that this sampling issue is only significant when the input scale is low ($\alpha_{in} < 0.1$) as this regime reflects a distribution that is close to uniform and the probability mass is spread out among exponentially many states.
The presence of errors caused by finite sampling motivates our use of a large number of samples (i.e., $10^6$ per data point) and provides an explanation for some of the consistent features in the results that are observed in the low  $\alpha_{in}$ regime.

\subsection{Experiment Results}

\subsubsection{A Typical GSD Model}

\begin{figure}
    \includegraphics[width=\linewidth]{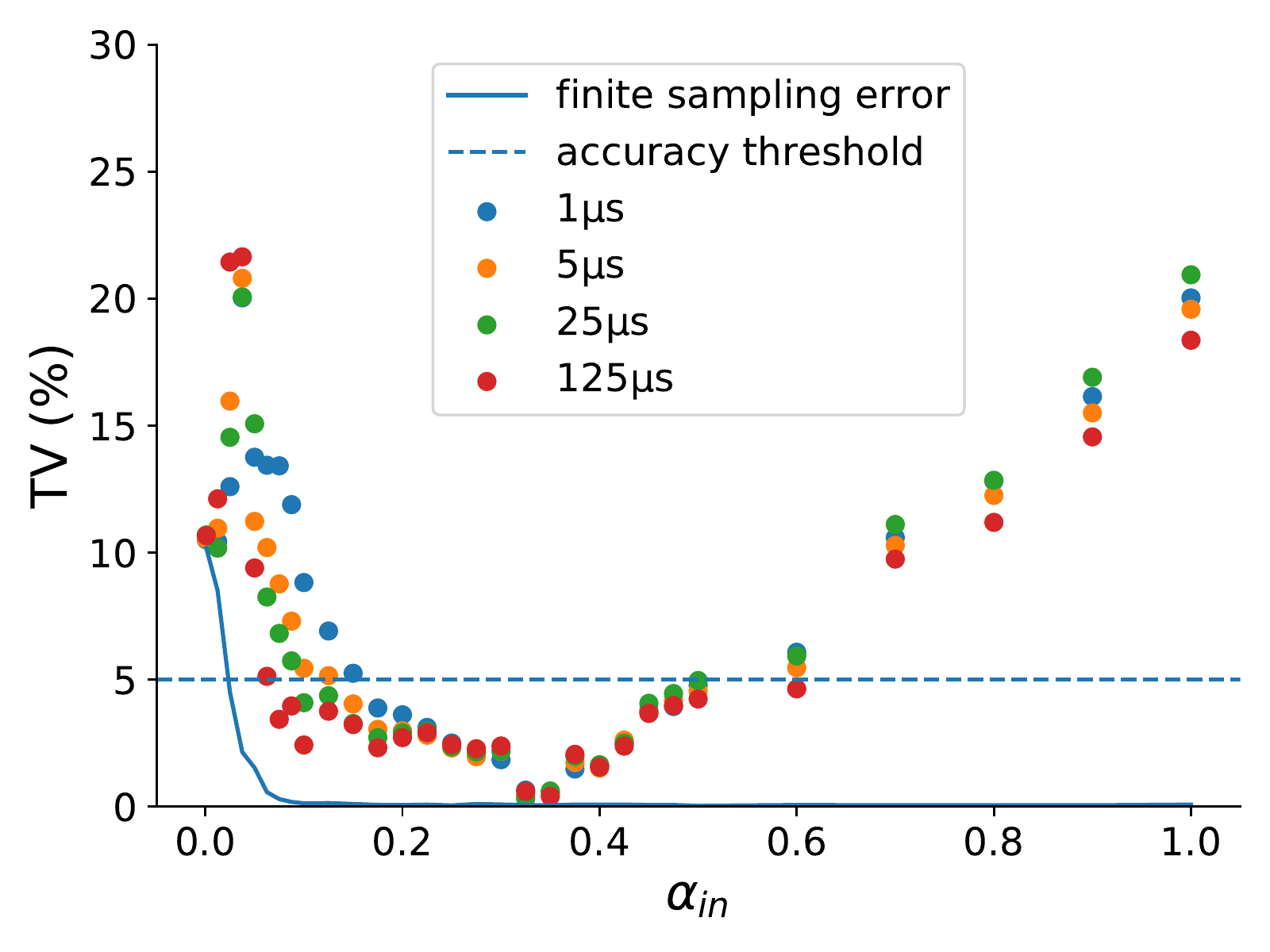}
    \caption{\rev{Total variation distance} between QA hardware output distribution for various scalings $\alpha_{in}$ and target Gibbs distribution with estimated $\alpha_{out}$ on the GSD-6 Hamiltonian. This model contains 16 spins and six ground states. Experiment is repeated for annealing times of 1 $\mu s$ (blue), 5 $\mu s$ (orange), 25 $\mu s$ (green), and 125 $\mu s$ (red). Finite sampling lower bound is indicated by blue curve. The {\em accuracy threshold} indicates the statistical requirement to be considered a high-quality Gibbs sampler in this work. The surprising finding is the band of $\alpha_{in}$ between approximately 0.2 and 0.4 where the hardware output does conform to a Gibbs distribution of the input Hamiltonian.}
    \label{fig:tv}
\end{figure}

We begin by presenting detailed results on a characteristic GSD model and then provide a summary of the results across the complete collection of models.
The \rev{total variation distance} results for the GSD-6 model are presented in Figure \ref{fig:tv}.
As the value of $\alpha_{in}$ is increased, the results show an up-down-up shape in the TV metric.
At very low $\alpha_{in} < 0.01$, the TV value is limited by finite sampling effects.
As $\alpha_{in}$ is increased, it quickly deviates from the family of $H_{\mathrm{Ising}}$ Gibbs distributions and gradually becomes more Gibbs-like.
At some point, while increasing $\alpha_{in}$, a minimum TV value is reached and it begins to increase monotonically until the maximum $\alpha_{in}$ value.
This up-down-up trend is replicated across a variety of the models considered in this work and plausible causes for it are discussed in Section \ref{sec:discussion}.
The central observation of this analysis is that there is a band of $\alpha_{in}$ between approximately 0.2 and 0.4, where the QA hardware's output distribution deviates from the target Gibbs distribution by less than $5\%$ in \rev{total variation distance}, which we designate as {\em high-quality}. 
Outside of this particular range of $\alpha_{in}$, the \rev{total variation distance} increases considerably, showing how the QA hardware's distribution dramatically shifts away from the family of $H_{\mathrm{Ising}}$ Gibbs distributions.
Interestingly, there does not seem to be any significant dependence of the \rev{total variation distance} on annealing time (i.e., each of the colored points in Figure \ref{fig:tv} follow similar trends).
These results suggest that this middle regime of $\alpha_{in}$ is favorable for using QA hardware as a Gibbs sampler at each of the annealing times considered.
Similar results on all of the GSD instances are provided in Appendix \ref{app:tv-details}.

\begin{figure*}
\centering
    \includegraphics[width=0.57\linewidth]{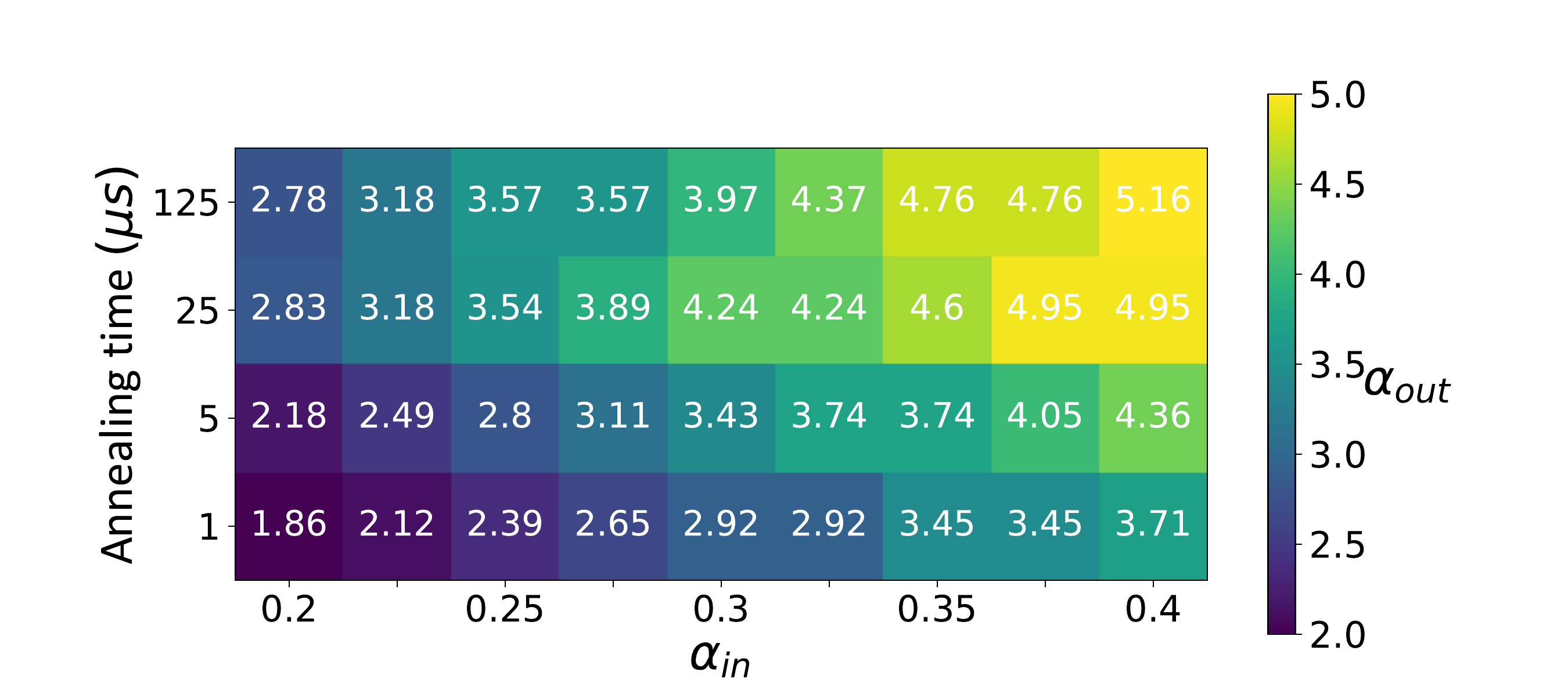}
    \caption{The values of $\alpha_{out}$ recovered by operating the QA hardware at different values of $\alpha_{in}$ and annealing times in the high-quality regime identified in Figure \ref{fig:tv} for the the GSD-6 Hamiltonian. Encouragingly, $\alpha_{out}$ increases monotonically (or nearly so) with both $\alpha_{in}$ and annealing time, allowing the hardware generate distributions at different effective temperatures, which is an essential feature for practical applications.}
    \label{fig:betaj}
\end{figure*}

The results from Figure \ref{fig:tv} suggest there are operating regimes where the QA hardware's is an accurate Gibbs sampler; however, the effective temperature (i.e. $\alpha_{out}$) of the that distribution is not presented.
Figure \ref{fig:betaj} explores this point by presenting the $\alpha_{out}$ values that were found for each of the high-quality input configurations considered (i.e., $0.2 \leq \alpha_{in} \leq 0.4$).
Given the prevailing theories of how QA hardware operates \cite{PhysRevA.92.052323,9465651}, one expects that increasing the energy scale of the input $\alpha_{in}$ or the annealing time will increase the effective temperature of the output distribution, $\alpha_{out}$.
Indeed, that trend is observed in Figure \ref{fig:betaj}, where the recovered $\alpha_{out}$ values increase monotonically (or nearly so) with both $\alpha_{in}$ and annealing time.
This monotonicity is a particularly useful result as it suggest a simple procedure for tuning the effective temperature of the distribution output by the QA hardware platform, which is an essential feature for most practical sampling applications.

Generating Gibbs samples in the $\alpha_{out}$ range of from 1.85 to 5.16 as shown in Figure \ref{fig:betaj} is an encouraging result as the critical points separating high- and low-temperature regimes of Ising spin glasses on lattices and random graphs typically occurs for values of $\alpha_{out} < 1.0$ \cite{lokhov2018optimal}.
Sampling from these low-temperature regimes is known to present significant computational challenges for classical algorithms, giving hope that QA hardware may be able to provide a performance enhancement on these tasks.
However, additional study is needed to determine if the results presented here can generalize to larger system sizes \rev{with different connectivity layouts} and to temperatures below the critical point of natively representable Ising spin glasses \cite{PhysRevX.4.021008}.

\subsubsection{Summary of the GSD Models}
\label{sec:other-problems}

\begin{figure}
     \includegraphics[width=\columnwidth]{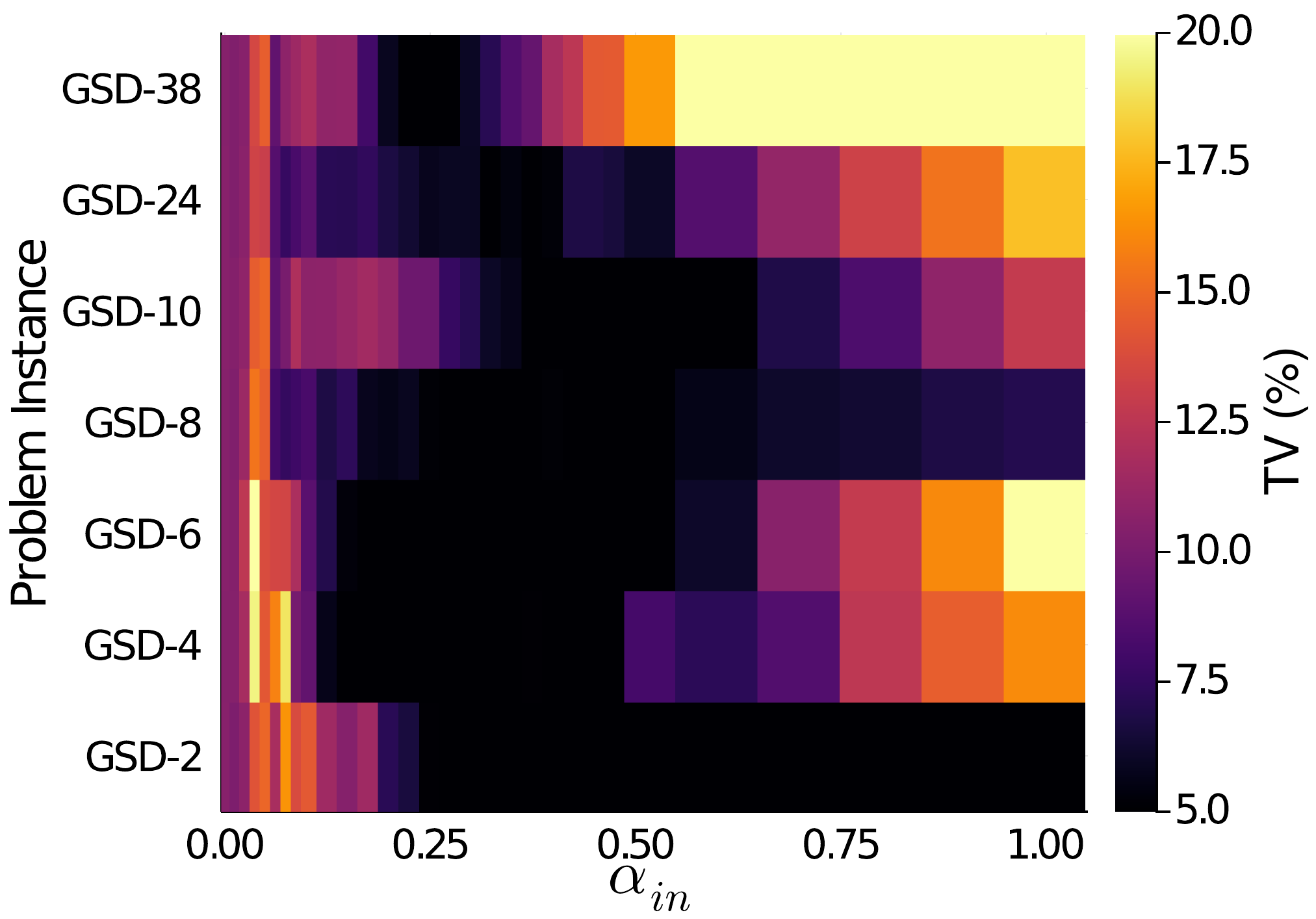}
     \includegraphics[width=\columnwidth]{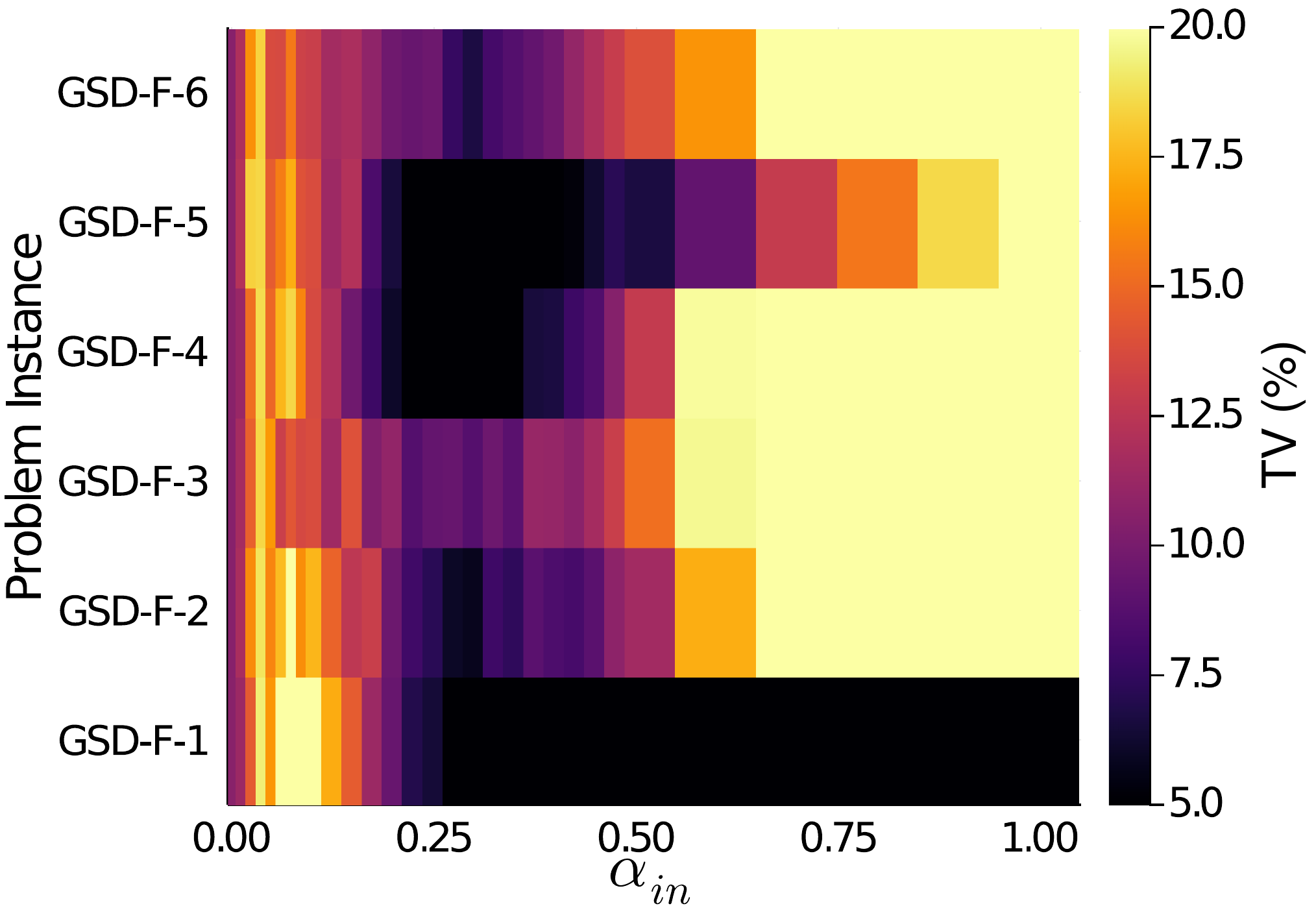}
    \caption{\rev{Total variation distance} between QA hardware output distribution for various scalings $\alpha_{in}$ and target Gibbs distribution with estimated $\alpha_{out}$ on the all GSD instances (top) and GSD-F instances bottom. A consistent high-quality Gibbs sampling band is observed for $0.2 \leq \alpha_{in} \leq 0.4$, with $\alpha_{in} \approx 0.3$ achieving a low \rev{total variation distance} for nearly all of the GSD models.}
    \label{fig:fulltv}
\end{figure}

To test if the observations on the GSD-6 instance generalize to a broader collection of models, we repeated the previous experiment on all 13 of the GSD models proposed in this work, seven GRD and six GSD-F cases.
Figure~\ref{fig:fulltv} displays the minimum \rev{total variation distance} with respect to the ideal Gibbs distribution for each scale of $\alpha_{in}$ and for each model. 
Only the 1 $\mu s$ data is presented, but results with other annealing times are available in Appendix \ref{app:tv-details}.
Although one can observe a variety of distinct features across these instances, the encouraging finding is that the range of $\alpha_{in}$ within 0.2 and 0.4 consistently yields statistics with low \rev{total variation distance}, suggesting the observations on the GSD-6 model do generalize to a broader class of models.
In particular, $\alpha_{in} \approx 0.3$ achieves a low \rev{total variation distance} for nearly all of the models considered. 

It is important to briefly discuss the GSD-2 and GSD-F-1 models, as these are the only ones that do not follow the up-down-up trajectory shown in Figure \ref{fig:tv}.
In these cases, the \rev{total variation distance} is high for low $\alpha_{in}$ and then drops below 5\% for $\alpha_{in}$ greater than 0.2 as expected.
However, the \rev{total variation distance} remains low even for $\alpha_{in}$ greater than 0.4.
This result is because when $\alpha_{in}$ is large, the probability distribution is highly concentrated in the ground states and these instances do not exhibit ground state degeneracy breaking.
As there is only one ground state in the GSD-F case and two symmetrical ground states in the GSD case, the sampling task is relatively easy and effectively reduces to a ground state identification task.
The QA hardware achieves low TV in these cases by simply outputting the ground states with high probability.
Although one can consider these two cases to represent \emph{easy} sampling tasks, we include them to highlight the increased challenge that instances with more ground state degeneracy pose to using QA for sampling applications.

The results from Figure \ref{fig:fulltv} indicate that the proposed high-quality Gibbs sampling regime generalizes to a variety of Ising models; however, the stability of the effective temperatures (i.e., $\alpha_{out}$) of those distributions is an important question.
Table \ref{tab:betaj} provides a summary of this information by presenting the minimum and maximum $\alpha_{out}$ values that can are achieved on each model by varying both $\alpha_{in}$ and the annealing time (see Appendix \ref{app:alpha-out-details} for more detailed information).
Although there is some variability in the largest $\alpha_{out}$ values, the results indicate that the range of effective temperatures output by the hardware remains relatively stable and the hardware is suitable for sampling from all models between 1.85 and 3.97.
These results also suggest that the the observations on the GSD-6 model generalizes to a broader class of models.

\begin{table}
 \centering
 \begin{tabular}{|l||r|r|r|}
 \hline
 {\bf Instance} & {\bf min $\alpha_{out}$} & {\bf max $\alpha_{out}$} \\
 \hline
%
 \hline
 GSD-2 & 1.32 & 3.97 \\
 \hline
 GSD-4  & 1.85 & 4.95 \\
 \hline
 GSD-6 & 1.85 & 5.16 \\
 \hline
 GSD-8 & 1.59 & 4.37 \\
 \hline
 GSD-10 & 1.32 & 4.76  \\
 \hline
 GSD-24 & 1.59 & 4.58 \\
 \hline
 GSD-38 & 1.59 & 4.76 \\
 \hline
 \hline
 GSD-F-1 & 1.32 & 3.97 \\
 \hline
 GSD-F-2 & 1.32 & 4.37 \\
 \hline
 GSD-F-3 & 1.59 & 4.37 \\
 \hline
 GSD-F-4 & 1.59 & 4.58 \\
 \hline
 GSD-F-5 & 1.59 & 4.76 \\
 \hline
 GSD-F-6 & 1.59 & 4.37 \\
 \hline
 \end{tabular}
 \caption{A summary of the smallest and largest $\alpha_{out}$ values recovered by operating the QA hardware at different values of $\alpha_{in}$ and annealing times in the high-quality regime across all GSD models.
 Although there is some variation in the largest $\alpha_{out}$ values, the QA hardware has a fairly consistent behavior across the GSD models and suggests that samples can be generated effectively in the range of effective temperatures from at least 1.85 to 3.97 for a variety of Hamiltonians.}
 \label{tab:betaj}
 \end{table}

\section{Discussion}
\label{sec:discussion}


Using quantum annealing for Gibbs sampling has been proposed as early as 2010, \cite{dwave_boltzmann}.
Since then, several studies have indicated that the output distribution of a quantum annealer does not sample from the input Hamiltonian's Gibbs distribution \cite{Perdomo_Ortiz_2018,PhysRevA.94.022308,Zhang2017,PhysRevA.100.030303}.
However, many of these works use an input scaling that is outside of the range identified by this work and are therefore subject to the distortions that we see in Figures~\ref{fig:tv} and \ref{fig:fulltv}.
These figures show that if $\alpha_{in}$ is too small ($< 0.2$) or too large ($> 0.4$) then QA hardware's output distribution will differ greatly from that of a corresponding Gibbs distribution.
Additionally, observing that the QA operating protocol proposed by this work yields high-quality Gibbs samples in a variety of Ising models suggests a systematic phenomenon that produces this behavior.

To investigate what might account for the distortions in these Gibbs distributions and the unique features that occur in the $0.2 \leq \alpha_{in} \leq 0.4$ range, we conduct a detailed study of the output statistics of a very small Ising Hamiltonian, i.e., a three-spin Ising chain.
With such a small system, we are able to create a theoretical model that reproduces the experimental data.
This theoretical model suggests local field noise and residual transverse field effects as potential explanations for the observed distortions.
The theoretical model found that the three-spin system sampled from an Ising Hamiltonian with extraneous couplings when $\alpha_{in}$ was in the low- or high-scaling regime.
Only when $\alpha_{in}$ remained in the range identified by this work did the system produce the desired Gibbs distribution.
This result provides additional evidence that this restricted scaling regime is optimal for conducting Gibbs sampling with QA hardware.

\vspace{0.50cm}
\subsection{Three-spin Ising Chain Statistics}

\begin{figure}
    \includegraphics[width=0.5\linewidth]{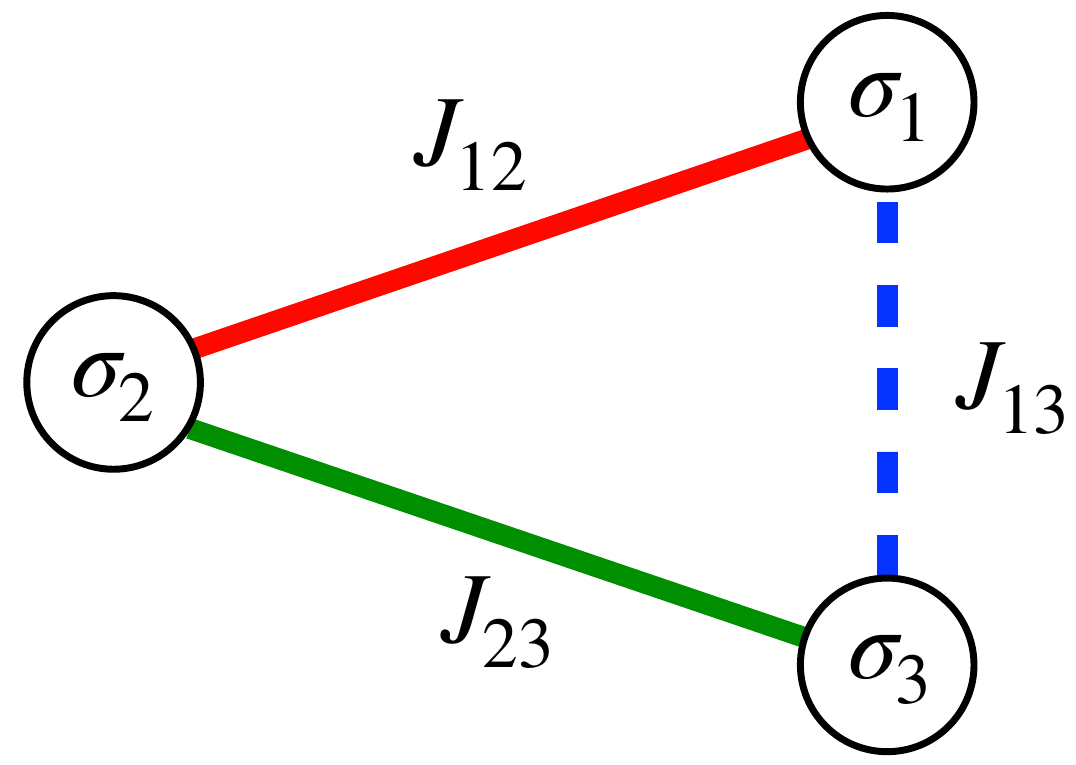}
    \caption{Diagram of three-spin Ising chain experiment. Solid lines for $J_{12}$ and $J_{23}$ indicate couplings that are programmed in the input Hamiltonian. Dashed line for $J_{13}$ indicates that this coupling is not programmed in the input Hamiltonian even though it is reconstructed from the output statistics and is thus a ``spurious coupling."}
    \label{fig:3spin_diagram}
\end{figure}

In this experiment, three spins denoted by $\sigma_1$, $\sigma_2$, and $\sigma_3$ are linked together in a ferromagnetic chain with $\sigma_1$ coupled to $\sigma_2$ and $\sigma_2$ coupled to $\sigma_3$ as shown in Figure~\ref{fig:3spin_diagram}. Notice that $\sigma_3$ is not coupled to $\sigma_1$. Formally, the input Hamiltonian is defined as follows: 
\begin{equation}
H_{\text{Ising}} = -J_{in} (\widehat{\sigma}^{z}_1 \widehat{\sigma}^{z}_2 + \widehat{\sigma}^{z}_2 \widehat{\sigma}^{z}_3),
\end{equation}
where $J_{in}$ is the value of the coupling.
\rev{Note that $J_{in}$ is equivalent to $\alpha_{in}$ from the 16-spin experiments.}
$J_{in}$ is swept between 0.0 and 1.0 with step-sizes matching the discretization of $\alpha_{in}$.\footnote{Note when programming a D-Wave system, negative values of $J_{in}$ are used to follow the ferromagnetic sign convention of that platform.}
For each value of $J_{in}$, $5 \times 10^6$ samples are collected from the QA hardware and the output statistics are recorded. Once again, in order to mitigate bias a spin-reversal transform is applied after every 100 samples. 

\begin{figure}
    \begin{subfigure}{0.95\linewidth}
    \includegraphics[width=\linewidth]{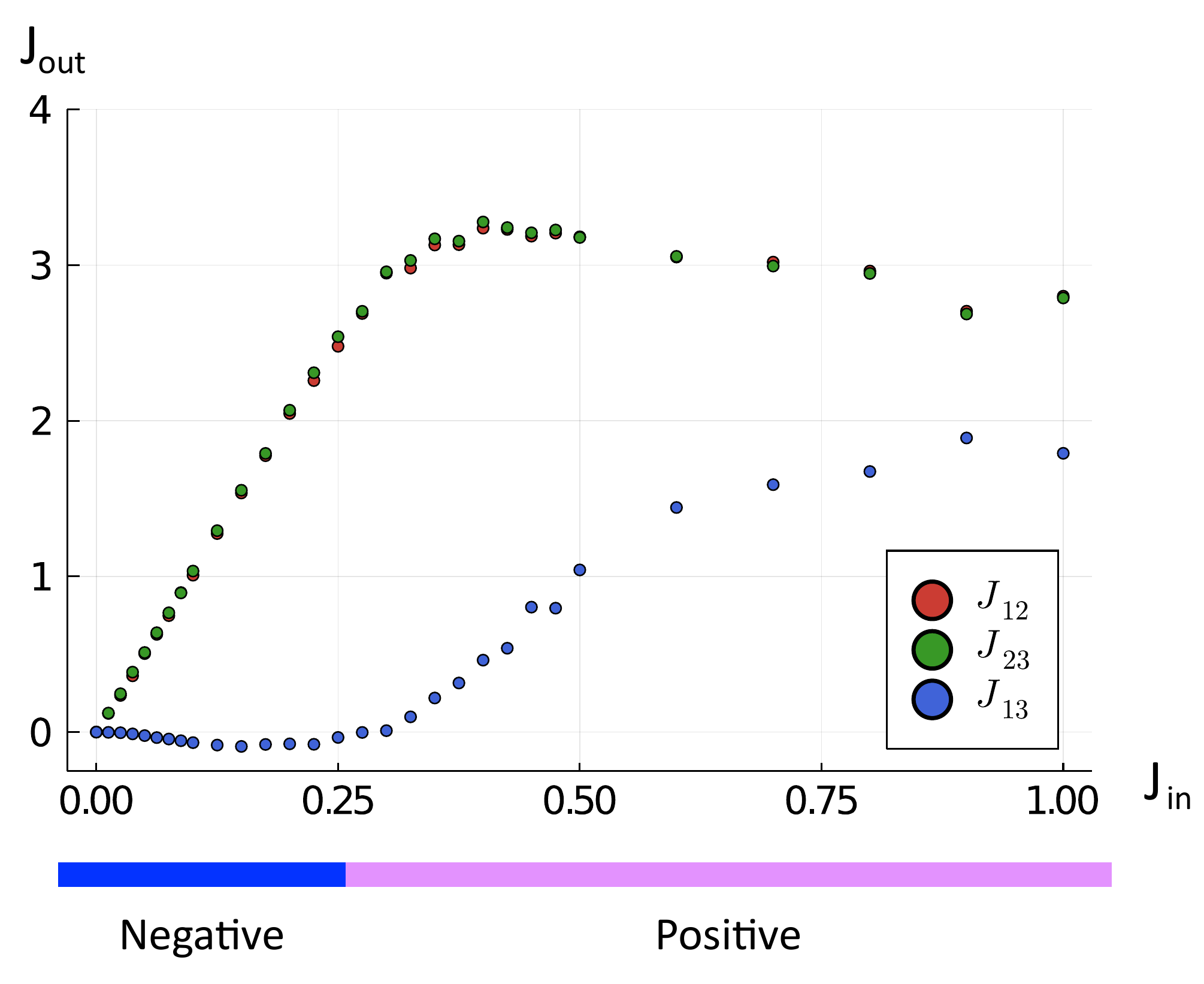}
    \caption{Reconstructed coupling values for three-spin Ising chain sampled at various coupling strengths. Only $J_{12}$ (coupling between $\sigma_1$ and $\sigma_2$ represented by green in figure) and $J_{23}$ (coupling between $\sigma_2$ and $\sigma_3$ represented by red in figure) are programmed in the input Hamiltonian. $J_{13}$ (blue) is not included in the input Hamiltonian \rev{and represents a spurious coupling between $\sigma_1$ and $\sigma_3$. Blue and purple bars represent the intervals of the input coupling strength $J_{in}$ that produce negative and positive spurious couplings, respectively.}}
    \label{fig:3spin-inv}
    \end{subfigure}
     \begin{subfigure}{0.95\linewidth}
    \includegraphics[width=\linewidth]{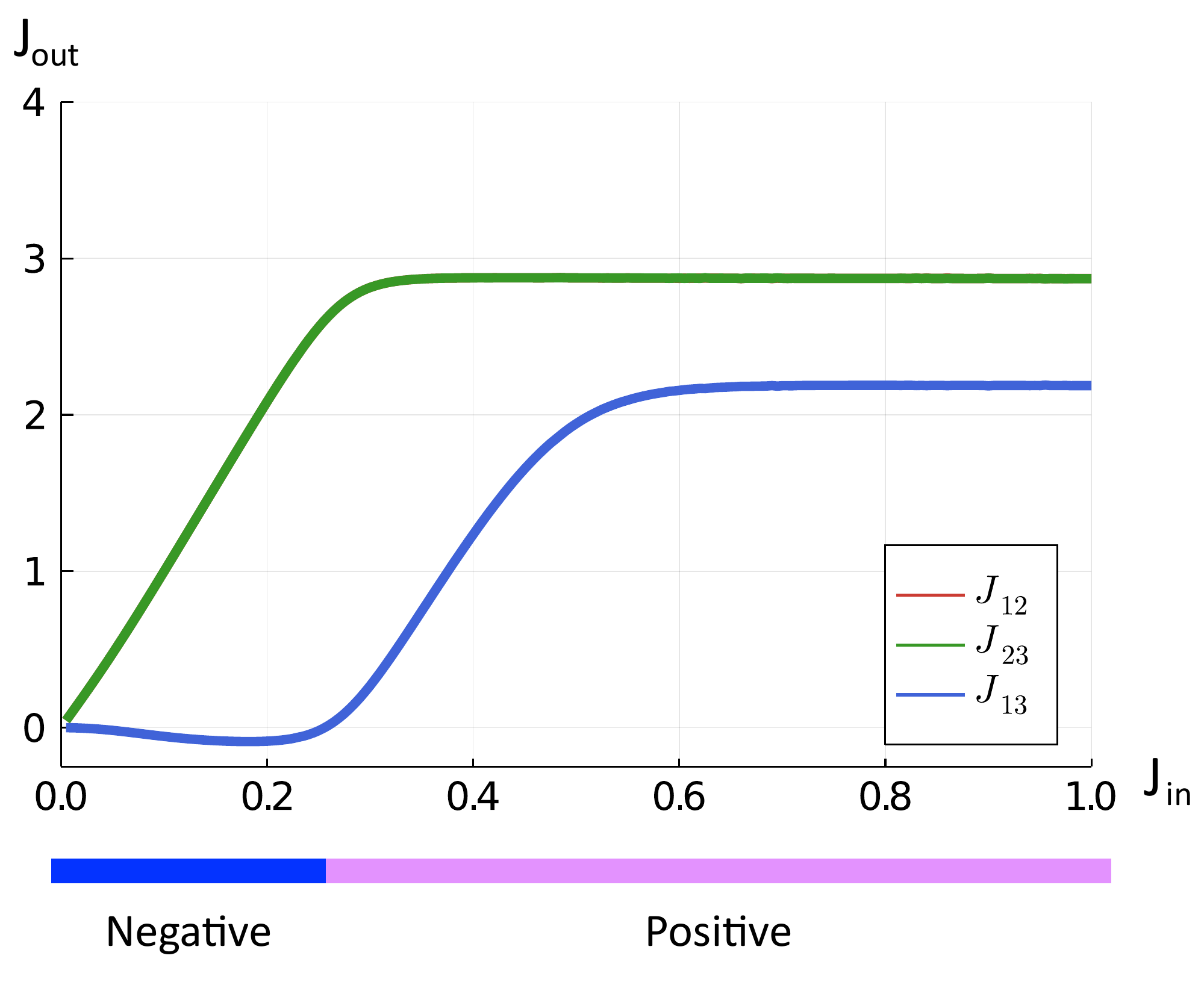}
    \caption{Results from model simulation that replicates three-spin Ising chain experiment. $J_{12}$ and $J_{23}$ are identical and only the green line is visible. $J_{13}$ is spurious coupling and represented by blue. \rev{Blue and purple bars represent the intervals of the input coupling strength $J_{in}$ that produce negative and positive spurious couplings, respectively.}}
    \label{fig:3spin-model}
    \end{subfigure}
    \caption{Notice how $J_{13}$ is reconstructed as negative for low-input couplings and then transitions to positive as input coupling increases beyond 0.275. As this coupling is not programmed in the input Hamiltonian, this non-zero reconstruction marks a disruption to the desired distribution. The negative and positive regions indicate the scaling regimes where noise and quantum effects distort the output distribution, respectively. We observe that the optimal scaling regime to use for Gibbs sampling is when $J_{13}$ is small relative to $J_{12}$ and $J_{23}$.}
    \label{fig:3spin}
\end{figure}

The output distribution is analyzed by solving the inverse-Ising problem using the Interaction screening estimation algorithm \cite{vuffray2016interaction, lokhov2018optimal}.
This algorithm takes the empirical distribution produced by the hardware and estimates the Ising Hamiltonian that most likely produced the given output statistics.
The estimated values for each of the couplings are denoted as $J_{out}$ and more specifically as $J_{12}$, $J_{23}$, and $J_{13}$ to indicate the specific edge in the three-spin chain. These estimated values are compared to the corresponding input values $J_{in}$ in Figure \ref{fig:3spin-inv}.
This procedure provides an understanding of the effective Hamiltonian output by the QA hardware, which may differ from the Ising Hamiltonian that has been programmed.
The entire protocol is repeated for annealing times of 1 $\mu s$, 5 $\mu s$, 25 $\mu s$, and 125 $\mu s$ with similar results.
Only the data for a 1 $\mu s$ annealing time is presented in Figure~\ref{fig:3spin-inv}, see Appendix \ref{app:ising-chain-at} for results for the other annealing times.

Because $\sigma_1$ and $\sigma_3$ are not coupled together in the input Hamiltonian, one expects $J_{13}$ from the reconstruction to be close to zero. However, as shown in Figure \ref{fig:3spin-inv}, $J_{13}$ appears to be negative and, hence, anti-ferromagnetic below $J_{in} = 0.275$ and positive above that point. In \cite{2012.08827}, this effect is referred to as a ``spurious coupling" because it appears in the output distribution despite not being programmed in the input Hamiltonian. It is striking that this spurious coupling exactly cancels out when $J_{in} = 0.275$, which coincides with the optimal input scaling for $\alpha_{in}$ for which almost all GSD models in Figure~\ref{fig:fulltv} are shown to have low \rev{total variation distance}.
Furthermore, the regime where the spurious coupling strength is low relative to the intended coupling strengths can be expanded to approximately include $J_{in}$ between 0.2 to 0.35, which again closely matches the optimal range for $\alpha_{in}$ that was found for the 16 spin experiments.
This provides additional support to the observation that the QA hardware considered here is an effective Gibbs sampler in this specific scaling range.

\subsection{A Three-Qubit Ising Chain Effective Model}

To further explore a connection between these experimental observations and the spurious couplings observed in the hardware's output (i.e., Figure \ref{fig:3spin-inv}), we propose an extension of the effective single-qubit quantum model developed in \cite{2012.08827} to the three-qubit context.
The single-qubit model considered in \cite{2012.08827,9465651} includes a transverse field with an intensity proportional to the input local field parameter, $h$.
This transverse field component was able to reproduce an observed saturation of the output field for large input values. Another important feature proposed in \cite{2012.08827} was qubit noise perturbing the input parameters of the model, which explained the spurious couplings in the regime of low input parameters.
Building on these characteristics, we consider the following toy model on three qubits controlled by a single parameter $J_{in}$ in the absence of local fields. We assume the output distribution is a noise-averaged thermal distribution:

\begin{align}
        \rho &= \frac{1}{8}\sum_{s_1, s_2, s_3=\pm 1}\frac{\exp\left(-\beta H\right)}{\Tr{\exp\left(-\beta H\right)}},
        \label{eq:mixture_density_matrix}
\end{align}

where $H$ is a three-qubit Hamiltonian with independent binary qubit noise realized through binary random variables $s_i$:

\begin{align}
        H \!= -J_{in} (\widehat{\sigma}^{z}_1 \widehat{\sigma}^{z}_2 + \widehat{\sigma}^{z}_2 \widehat{\sigma}^{z}_3) - \! \sum_{i=1}^{3} \gamma_i J_{in} \widehat{\sigma}^{x}_i - \! \sum_{i=1}^{3} \eta_i s_i \widehat{\sigma}^{z}_i.
        \label{eq:mixture_Hamiltonian}
\end{align}

Equation~\eqref{eq:mixture_Hamiltonian} describes a three-qubit Ising chain, $J_{in}$ is a single input parameter controlling the strength of interactions, while $\gamma_i$ and $\eta_i$ are constants re-scaling the strength of the transverse field and qubit noise.
Leveraging the parameters estimated in \cite{2012.08827}, we select $\beta =11$, $\gamma_i = 0.013$ and $\eta_i = 0.04$ as typical values for these parameters. We would like to highlight that the model from Equation~\eqref{eq:mixture_Hamiltonian} is very different quantitatively and qualitatively from what is known as the Background Suceptibility model as discussed in Appendix~\ref{app:bs-model}.

It was shown in \cite{2012.08827} that for small values of the input parameter $J_{in}$, the negative spurious coupling can be explained by field noise on the involved qubits.
In fact, it is information-theoretically impossible to distinguish between a model with field noise and a model with the corresponding anti-ferromagnetic coupling.
As seen in Figure \ref{fig:3spin-model}, the toy model in Eq.~\eqref{eq:mixture_density_matrix} indeed predicts that when $J_{in}$ is small and thus field noise is significant relative to $J_{in}$, a negative spurious coupling is preset.
We therefore designate this low-scaling regime where $J_{in} < 0.2$ as the noisy regime.

When $J_{in}$ increases, the field noise becomes less pronounced relative to $J_{in}$. In this regime of higher input values, the model \eqref{eq:mixture_density_matrix} predicts an emergence of a positive spurious coupling -- an effect that is also observed in the experimental data. This suggests that a $J$-dependent residual transverse field on each of the qubits can account for the positive response in spurious link and the saturation of that link for large inputs. Similarly to how we designated the low-scaling regime as noise-dominated, we can designate the high-scaling regime as dominated by the effect of residual transverse fields.

An interesting intermediate regime predicted by the toy model as shown in Figure~\ref{fig:3spin} happens around $J_{in} = 0.275$, where noise and transverse field influences cancel each other, leading to an effective absence of spurious couplings, which also occurs experimentally.
This observation suggests that the optimal sampling parameter range for $\alpha_{in}$, referred to as the {\em sweet-spot} previously in this work, may be explained by the interacting effects of noise and residual fields in the system.

\section{Conclusion}
\label{sec:conclusion}

Inspired by recent work indicating that quantum annealing hardware behaves as a noisy Gibbs sampler in very low energy scales \cite{2012.08827}, this work identified an approach for mitigating the impacts of noise and conducting high-quality Gibbs sampling in a range of effective temperatures for a class of small hardware-native Ising models.
This approach to using quantum annealing hardware for Gibbs sampling opens new opportunities for applications in machine learning and exploring the physics of condensed matter systems.

More broadly, the computational task of Gibbs sampling and the protocol developed in this work could provide an avenue for exploring the potential for quantum advantage in quantum annealing hardware.
To that end, two follow-on works would be required.
First, a class of Ising models needs to be identified that are challenging to sample from with classical algorithms (e.g., spin-glasses) that also adhere to the criteria proposed in this work, i.e., naively representable on the hardware with $J,h \in \{-1,0,1\}$.
Such examples seem unlikely in the Chimera hardware architecture \cite{PhysRevX.4.021008,PhysRevX.5.019901}; however, the new Pegasus hardware architecture \cite{2020arXiv200300133B} will likely provide new opportunities for identifying such models. 
Second, significant additional research is required to verify that the protocol proposed by this work will scale to larger systems, ideally with 100's to 1000's of qubits, so that more of the quantum annealing hardware can be used.
This type of verification requires an amazing amount of computation and tuning of sophisticated Monte Carlo methods as there are no known efficient algorithms for generating Gibbs samples of the target distributions that would be required as a baseline of comparison.
Another practical challenge is that scaling to larger systems may also yield unexpected side effects in practice, such as amplifying the role of hardware programming errors \cite{PhysRevA.93.012317,Albash_2019}, putting an implicit limit on sampling accuracy at larger scales.
Although significant follow-on investigation is required to more deeply understand the potential for performing thermal Gibbs sampling with quantum annealing hardware, this work provides a foundation for maximizing the performance available hardware platforms when conducting these computational tasks.

\section{Acknowledgements}

This work was supported by the U.S. DOE through the Laboratory Directed Research and Development (LDRD) program of LANL under project number 20210114ER and the Center for NonLinear Studies (CNLS). This research used computing resources provided by the LANL Institutional Computing Program, which is supported by the U.S. Department of Energy National Nuclear Security Administration under Contract No. 89233218CNA000001.
This material is also based upon work supported by the National Science Foundation the Quantum Leap Big Idea under Grant No. OMA-1936388.

\bibliography{main}
\vspace{0.25cm}
LA-UR-21-28692

\clearpage
\appendix

\section{Comparing Distributions with Total Variation Distance}
\label{app:total-variation}

Given two distributions $\mu$ and $\nu$ over a binary string of size $N$, i.e. $\sigma \in \{-1, 1\}^N$, the \rev{total variation distance} between them is defined as,
\begin{align}
\operatorname{TV}(\mu,\nu) &= \frac{1}{2} \sum_{\sigma\in \{-1, 1\}^N} | \mu(\sigma) - \nu(\sigma) |,
\label{eq:tv}
\end{align}
that is, the absolute difference between the probability of each state \rev{is added together and} divided by 2, to normalize the metric to the range of 0 to 1.
Given that the metric is in the range of 0 to 1, it is often presented as a percentage between 0 and 100.

The TV distance can be interpreted as the maximum discrepancy between the probability of an event computed with the distribution $\mu$ instead of $\nu$, that is,
\begin{align}
\sup_A |\mathbb{P}_{\mu}\left(A\right) - \mathbb{P}_{\nu}\left(A\right)| = \operatorname{TV}(\mu,\nu),
\end{align}
where $A\in \mathcal{P}\left(\{-1, 1\}^N \right)$ is an element of the power set of $\{-1, 1\}^N$. Therefore, it comes as the distance of choice to measure the error that can be made by estimating probabilities using a surrogate distribution $\nu$ of a target distribution $\mu$. However, note that this sets a very strong quality criteria for estimating the reliability of a surrogate distribution as the TV is equal to the worse-case estimation error.
It is important to note that if $\mu$ or $\nu$ are only accessible through a set of finite samples, there is an unavoidable error caused by sampling that will be captured by the TV distance. To avoid finite sampling error in the TV estimate, one needs to use a number of samples that is on the order of the typical support of $\mu$ and $\nu$. This is required to estimate the absolute difference in Eq.~\eqref{eq:tv} accurately for each state $\sigma$. Typically, the number of samples required grows exponentially in $N$ and is maximal for distribution $\mu$ and $\nu$ that are close to a uniform distribution where it is in the order of $2^N$.

It is fairly common to consider KL divergence as an alternative to the TV distance, especially when the non-convexity of the latter becomes a computational bottleneck, taking advantage of Pinsker's inequality, i.e., $\operatorname{TV}(\mu, \nu) \leq \sqrt{1/2 \operatorname{KL}(\mu ||\nu)}$. However, the KL divergence does not share the same meaning as the TV distance: the KL divergence estimates the inefficiency in compressing a source $\mu$ using a code optimized for $\nu$. In general, it is not suitable for measuring the quality of probability estimates between a target distribution and its surrogate. A simple situation that illustrates this last point is when the distribution $\nu$ is almost identical to $\mu$ except that it has an infinitesimally small probability mass that lies outside of the support of $\mu$. The TV distance between these distribution is infinitesimally small, while the KL divergence becomes infinite.

\section{The Ground State Degeneracy Models}
\label{app:gsd-details}

The Ising model instances used in this work were generated by the following procedure. 
First, 16 spins (2~unit cells) on the D-Wave 2000Q Quantum Annealer \texttt{DW\_2000Q\_LANL} hardware graph were selected with qubits 296 to 303 making up one unit cell and qubits 304 to 311 making up an adjacent unit cell in the Chimera graph architecture.
Then the values for $J_{ij}$ incident to these qubits were assigned randomly from the set $\{-1,1\}$ for each coupler in the hardware graph.
Second, a brute-force calculation was used to compute the ground-state degeneracy for a given instance.\footnote{This approach is only computationally feasible for small systems spins, such the ones considered here.}
This procedure was repeated many times and a subset of cases were selected to highlight a smooth range of ground-state degeneracies.
This yielded a total of seven GSD (ground-state degeneracy) cases with degeneracy values of 2, 4, 6, 8, 10, 24, and 38, which are presented in Table \ref{tab:gsd}.
The same procedure was repeated while also assigning random values to the local fields $h_i$ to $\{-1,1\}$, yielding six additional \rev{independently generated} GSD-F models with 1, 2, 3, 4, 5, and 6 ground state degeneracies, which are presented in Table \ref{tab:gsd-f}.
The specific instances are referred to as GSD-\# and GSD-F-\#, where the \# indicates the amount of ground state degeneracy.
Note that the GSD-F models tend to have lower degeneracy because the inclusion of the fields breaks symmetries within energy levels.

It is worth noting that these models are reminiscent of the RAN and RANF models that have been used to benchmark earlier generations of QA processors \cite{Boi2014,10.1007/978-3-030-19212-9_11}. 
The RAN model class on Chimera graphs has been shown to exhibit a zero temperature phase transition \cite{PhysRevX.4.021008,PhysRevX.5.019901}, indicating that sampling from this model class at finite temperature is not expected to be computationally hard. 
Here, however, instances are hand-picked from this class to exhibit varying degrees of ground state degeneracy.
In this sense, we do not expect that the instances of this work to be representative samples from the RAN class, and instead expect \rev{the high degeneracy instances} to be challenging for a quantum annealer to sample from \rev{due to the degeneracy breaking effect of the transverse field.}

\section{\rev{Total Variation Distance} at Different Scales and Annealing Times}
\label{app:tv-details}

\begin{figure*}[h]
\centering
    \begin{subfigure}{0.4\linewidth}
        \includegraphics[width=0.9\linewidth]{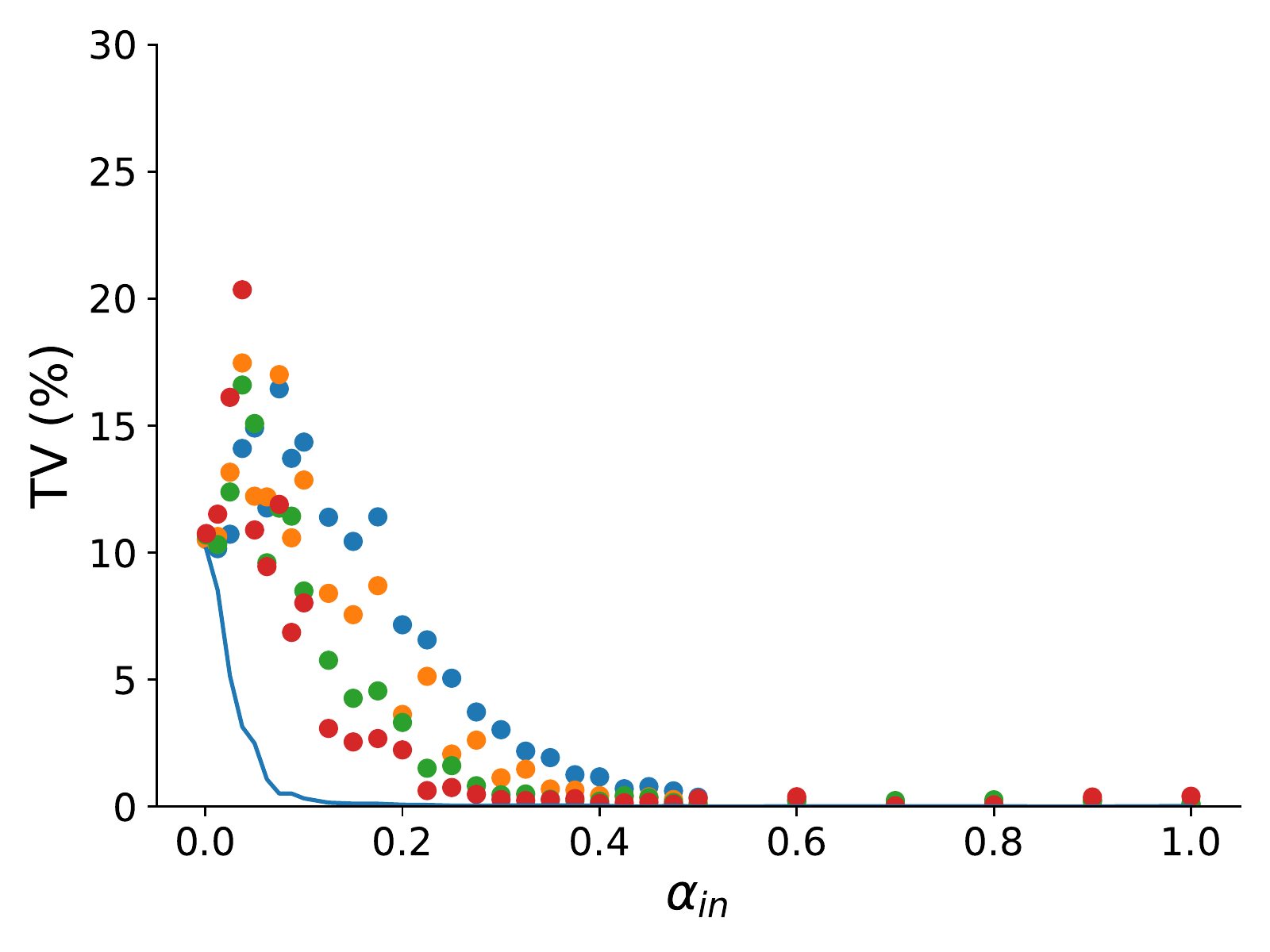}
        \caption{GSD-2}
    \end{subfigure}
    \begin{subfigure}{0.4\linewidth}
        \includegraphics[width=0.9\linewidth]{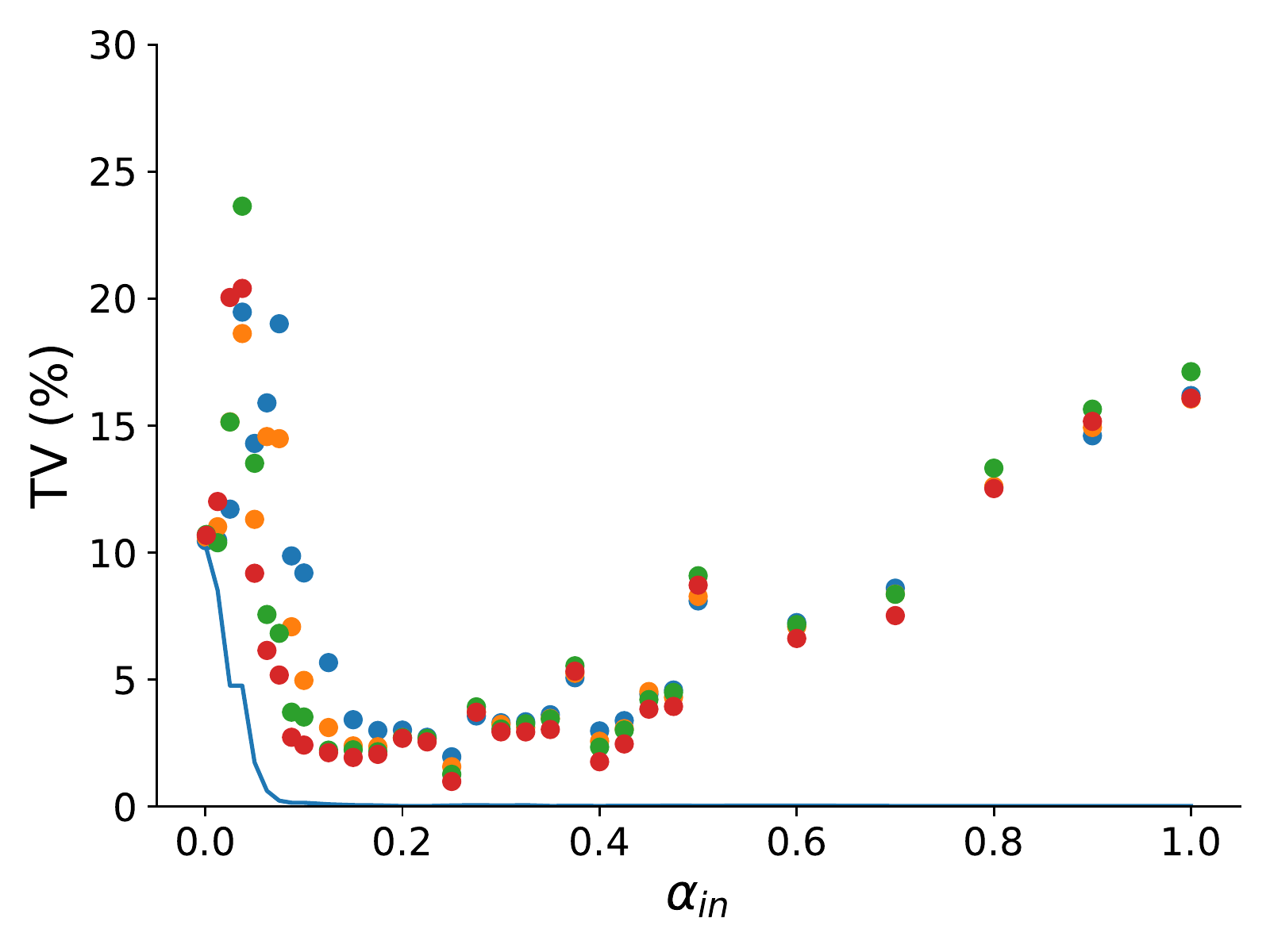}
        \caption{GSD-4}
    \end{subfigure}
    \begin{subfigure}{0.4\linewidth}
        \includegraphics[width=0.9\linewidth]{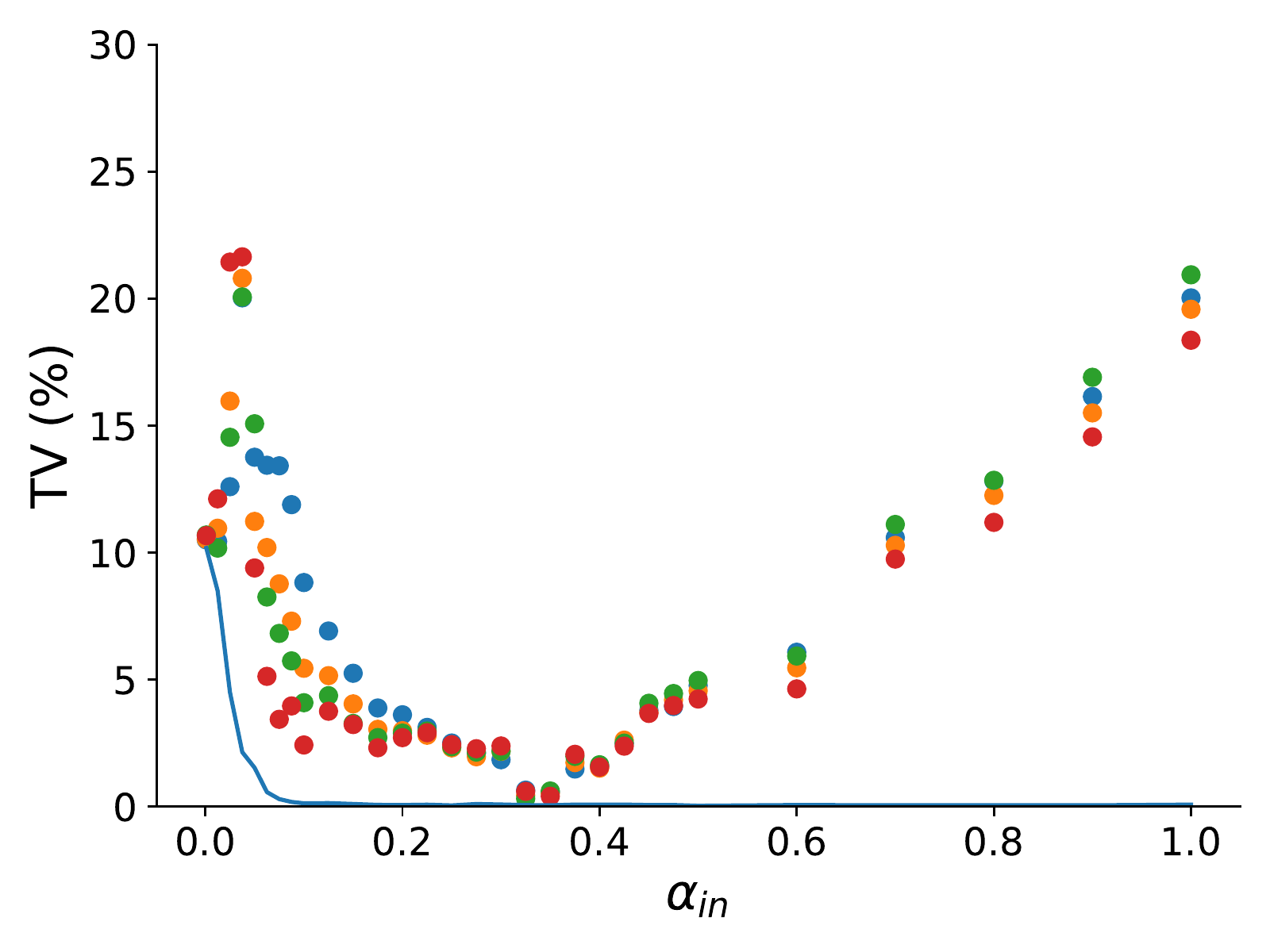}
        \caption{GSD-6}
    \end{subfigure}
    \begin{subfigure}{0.4\linewidth}
       \includegraphics[width=0.9\linewidth]{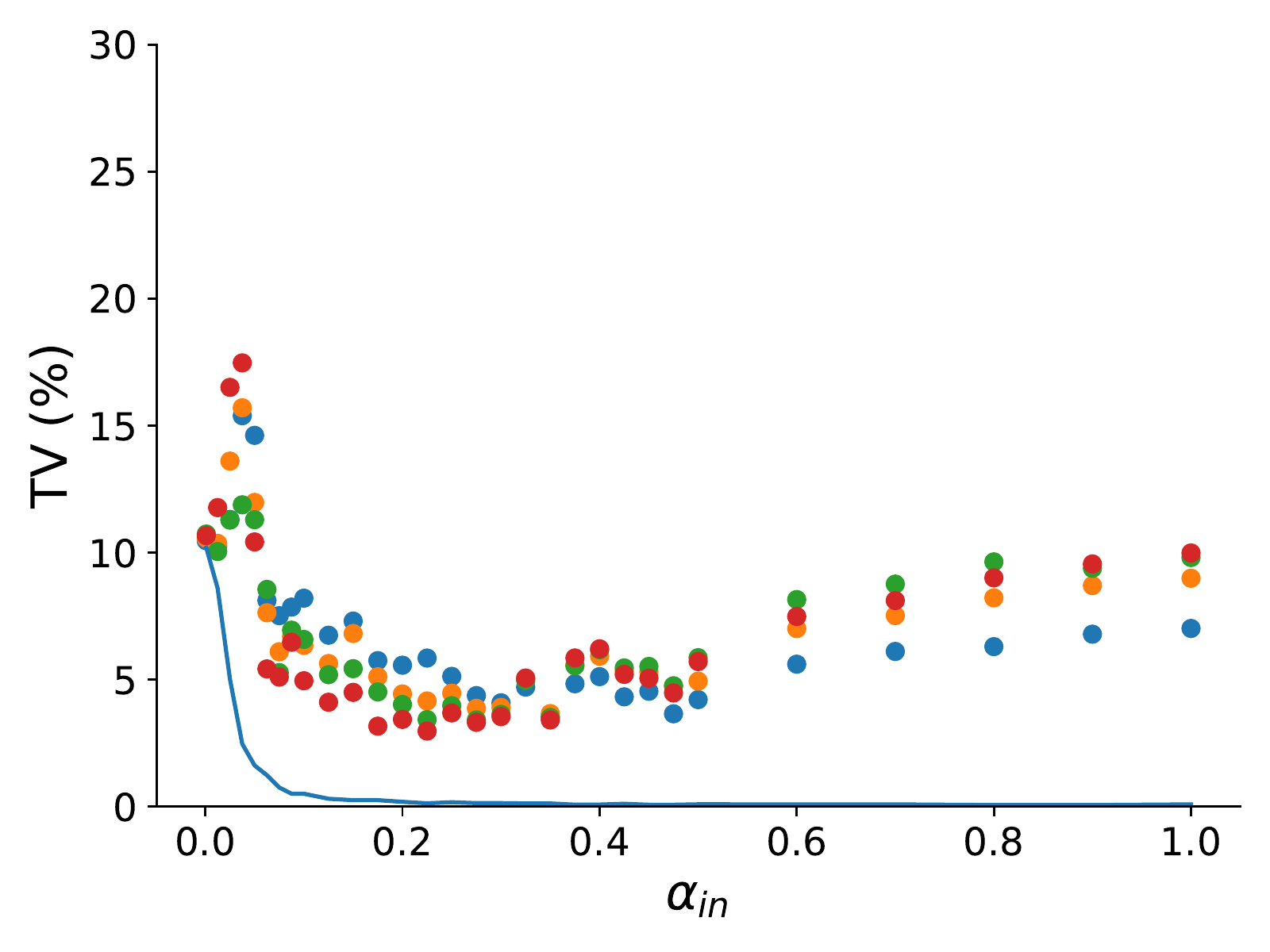}
        \caption{GSD-8}
    \end{subfigure}
    \begin{subfigure}{0.4\linewidth}
        \includegraphics[width=0.9\linewidth]{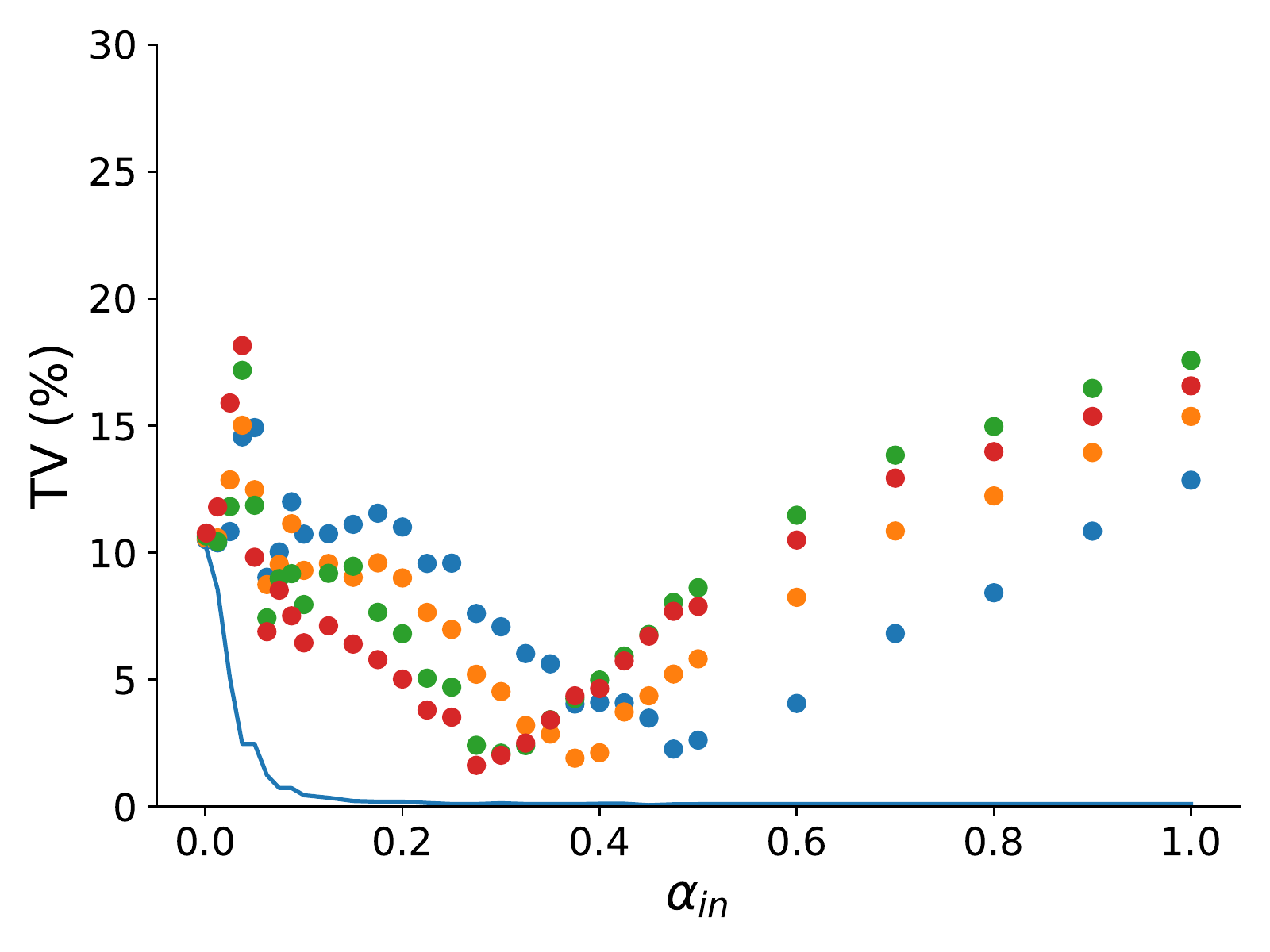}
        \caption{GSD-10}
    \end{subfigure}
    \begin{subfigure}{0.4\linewidth}
        \includegraphics[width=0.9\linewidth]{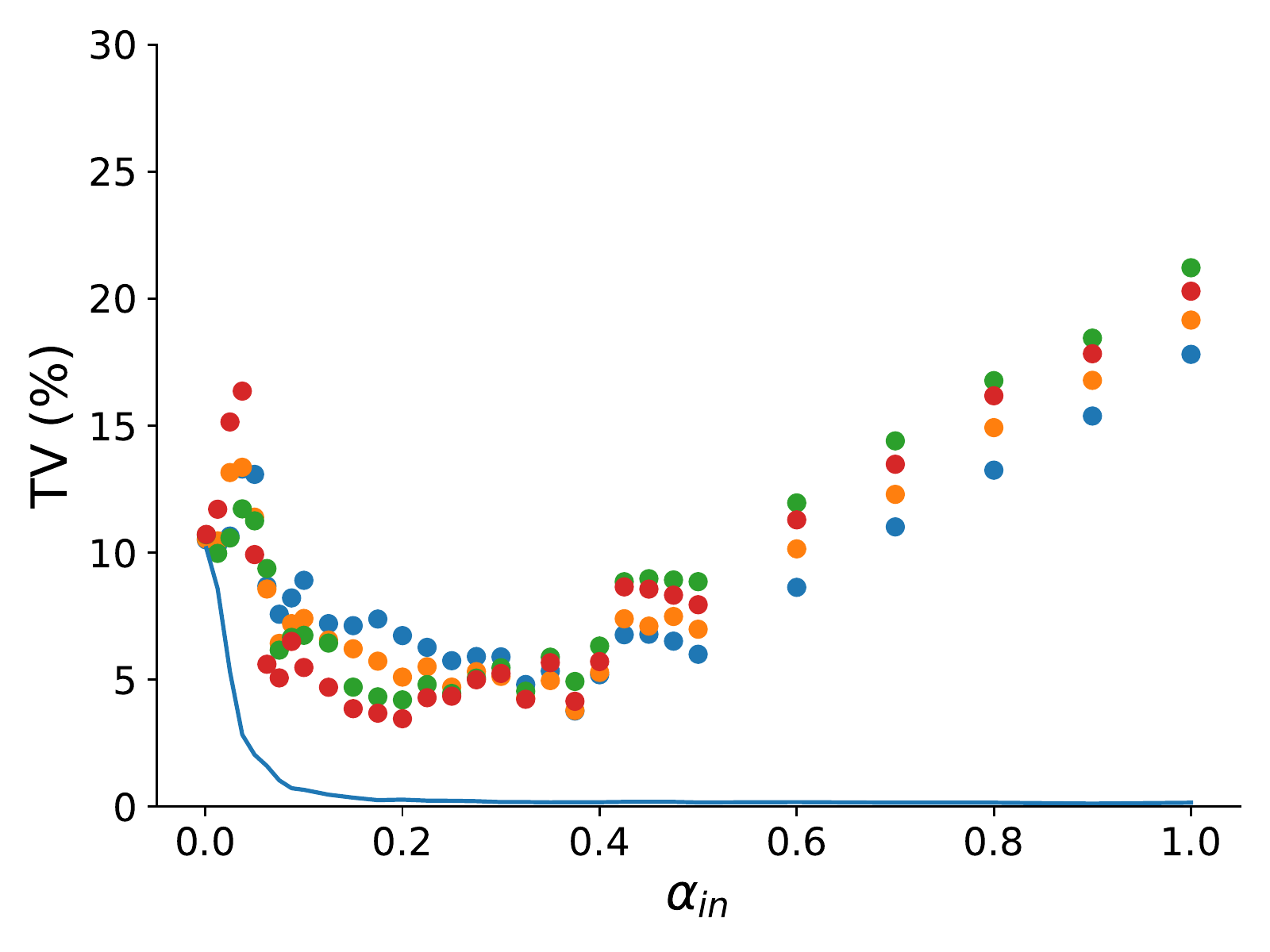}
        \caption{GSD-24}
    \end{subfigure}
    \begin{subfigure}{0.4\linewidth}
        \includegraphics[width=0.9\linewidth]{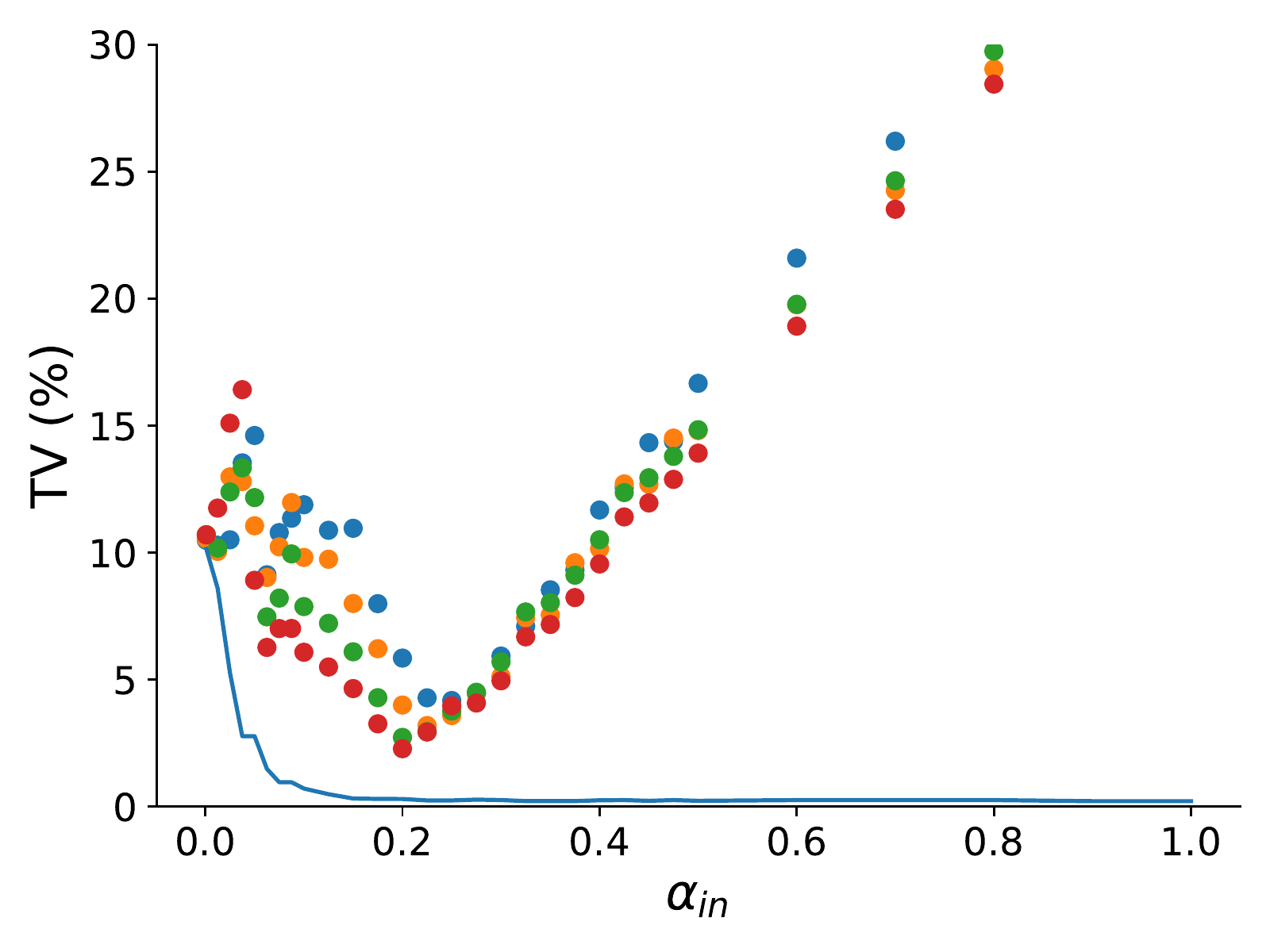}
        \caption{GSD-38}
    \end{subfigure}
    \caption{Dependence of \rev{total variation distance} on $\alpha_{in}$ for GSD models. Recall from Figure \ref{fig:tv} the color notation for various anneal times (blue: 1 $\mu s$, orange: 5 $\mu s$, green: 25 $\mu s$, and red: 125 $\mu s$). The TV lower bound is a result of finite sampling is indicated by the blue line.}
    \label{fig:app-tv-gsd}
\end{figure*}

\begin{figure*}[h]
\centering
    \begin{subfigure}{0.49\linewidth}
        \includegraphics[width=0.9\linewidth]{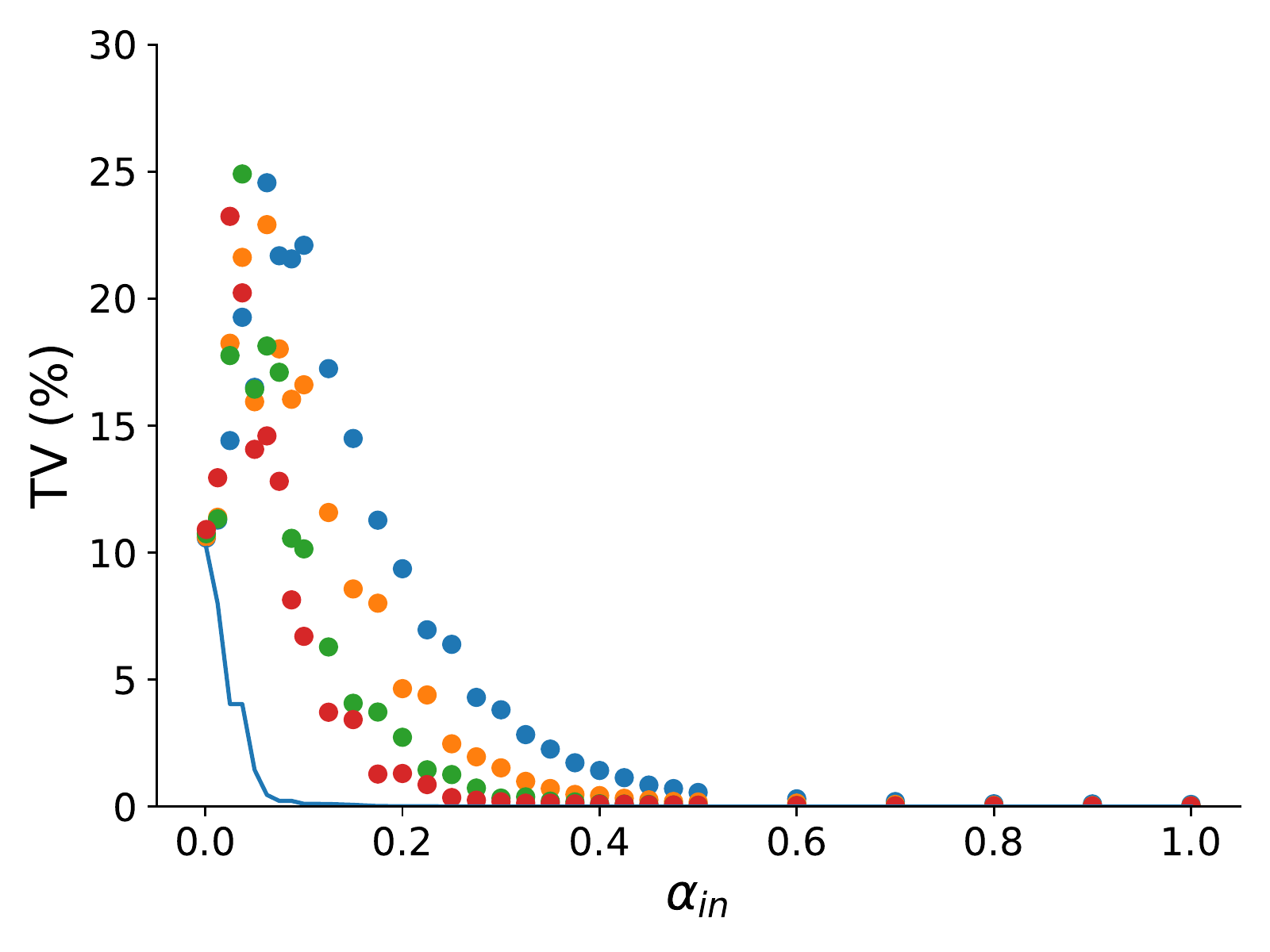}
        \caption{GSD-F-1}
    \end{subfigure}
    \begin{subfigure}{0.49\linewidth}
        \includegraphics[width=0.9\linewidth]{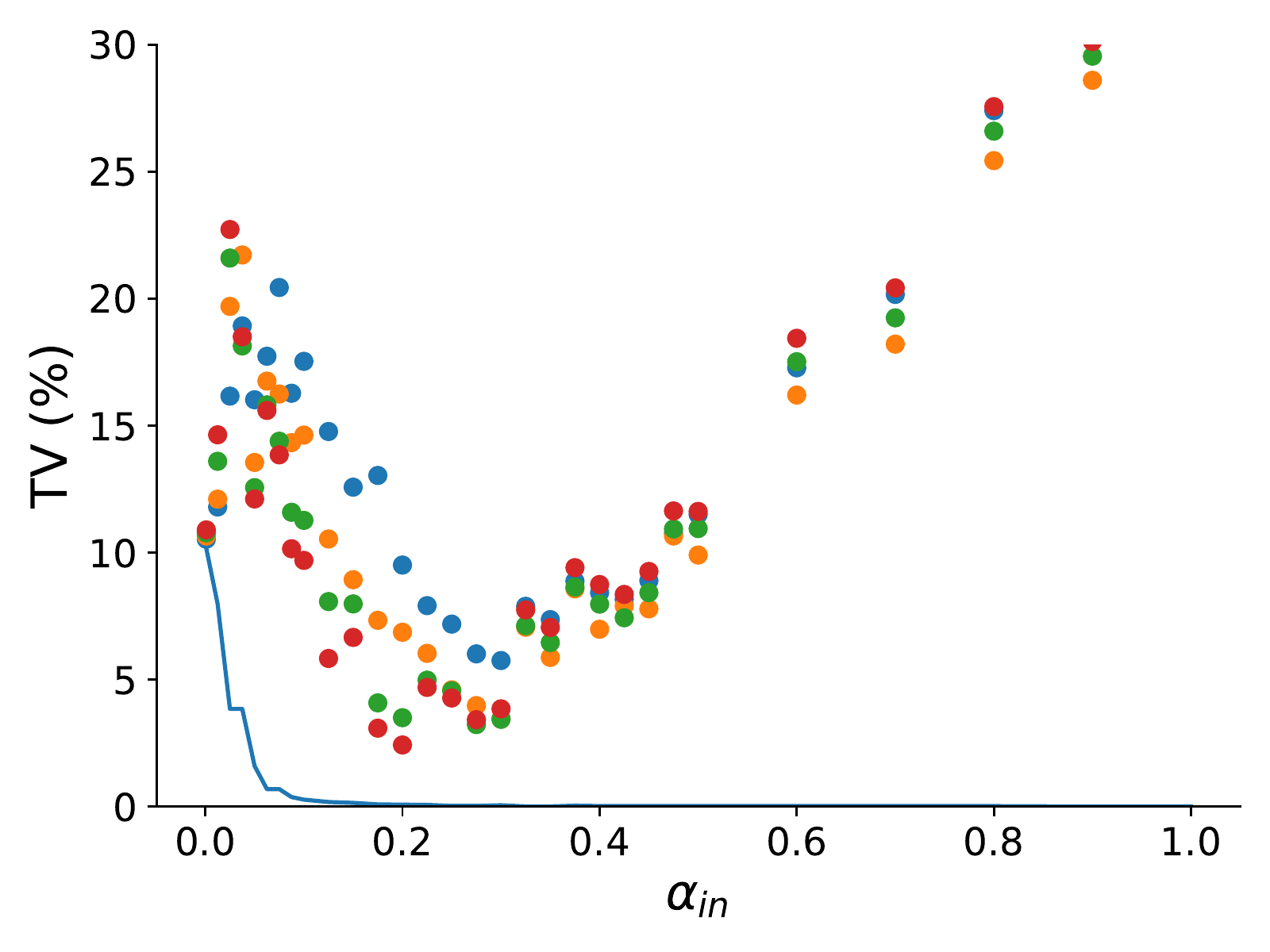}
        \caption{GSD-F-2}
    \end{subfigure}
    \begin{subfigure}{0.49\linewidth}
        \includegraphics[width=0.9\linewidth]{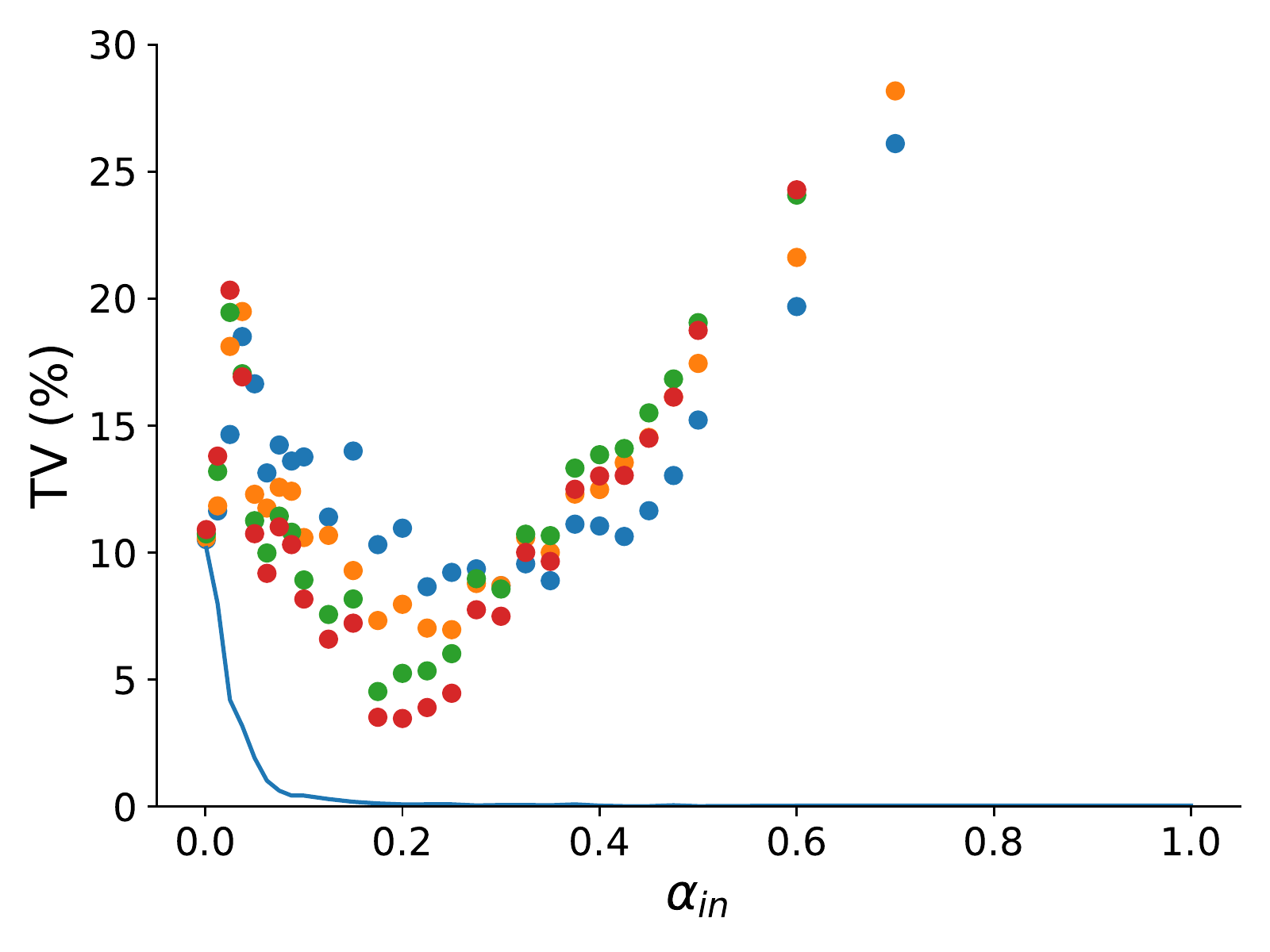}
        \caption{GSD-F-3}
    \end{subfigure}
    \begin{subfigure}{0.49\linewidth}
        \includegraphics[width=0.9\linewidth]{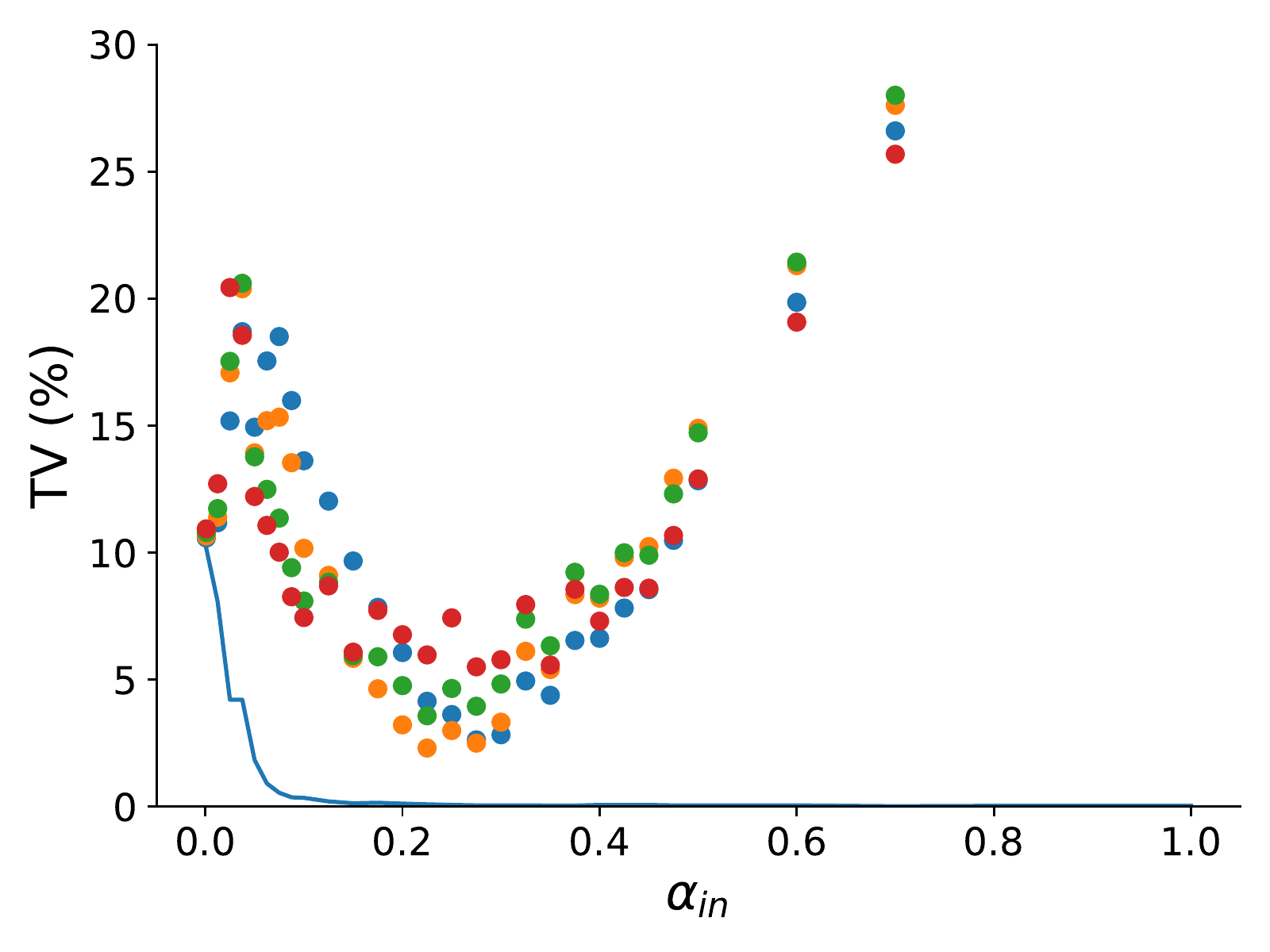}
        \caption{GSD-F-4}
    \end{subfigure}
    \begin{subfigure}{0.49\linewidth}
        \includegraphics[width=0.9\linewidth]{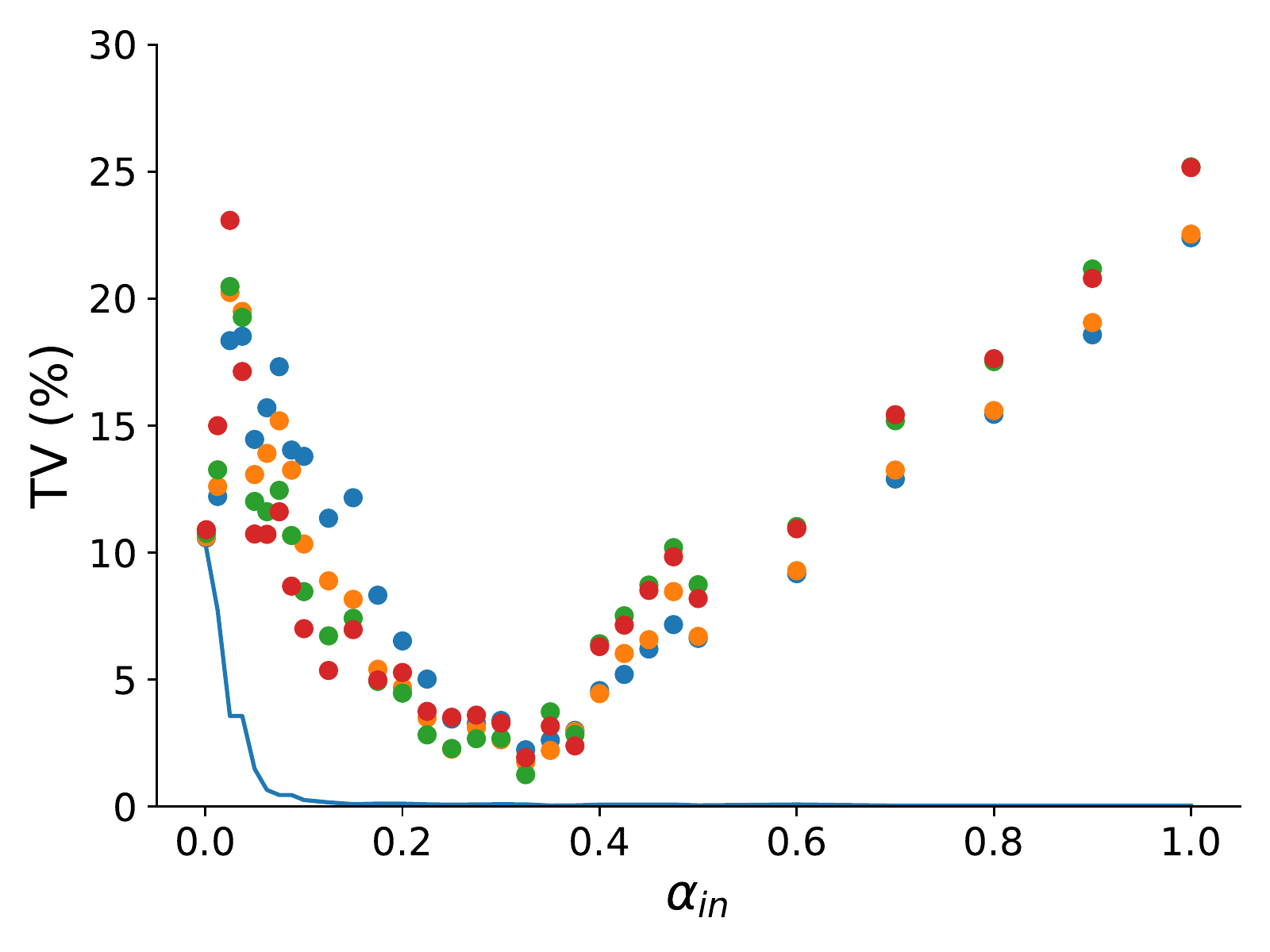}
        \caption{GSD-F-5}
    \end{subfigure}
    \begin{subfigure}{0.49\linewidth}
        \includegraphics[width=0.9\linewidth]{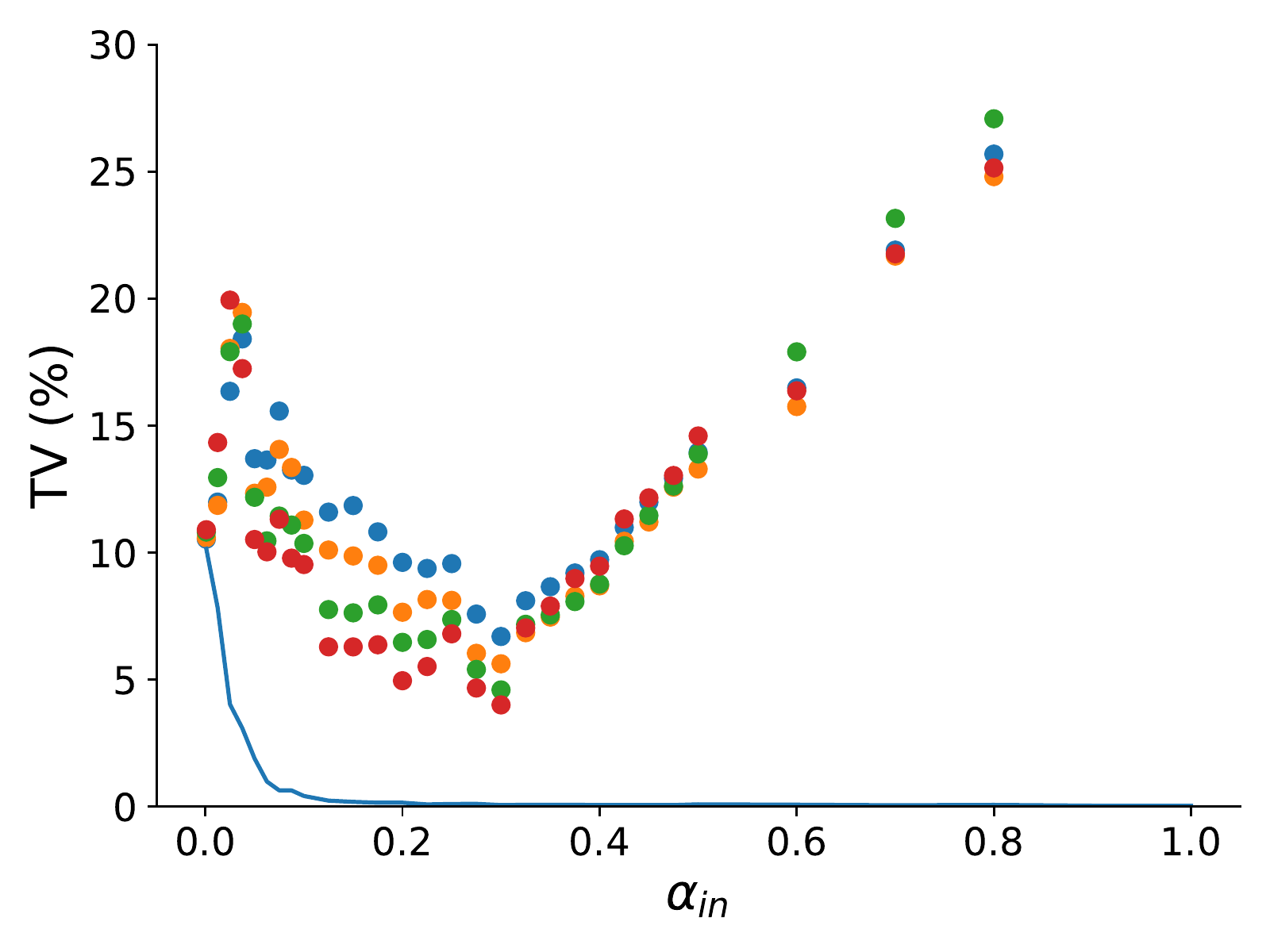}
        \caption{GSD-F-6}
    \end{subfigure}
    \caption{Dependence of \rev{total variation distance} on $\alpha_{in}$ for GSD-F models. Recall from Figure \ref{fig:tv} and \ref{fig:app-tv-gsd} the color notation for various anneal times (blue: 1 $\mu s$, orange: 5 $\mu s$, green: 25 $\mu s$, and red: 125 $\mu s$). The TV lower bound resulting from finite sampling is indicated by the blue line.}
    \label{fig:app-tv-gsd-f}
\end{figure*}

In this section, we present supplementary figures for each of the experiments in the main text. 
Figures \ref{fig:app-tv-gsd} and \ref{fig:app-tv-gsd-f} display the \rev{total variation distance} results for every GSD model as a function of $\alpha_{in}$. As in Figure \ref{fig:tv} where only GSD-6 is shown, the results for each of the annealing times are overlaid with different colors. Note how GSD-2 and GSD-F-1 are the only instances where \rev{total variation distance} decreases as $\alpha_{in}$ increases. This is because models with low degeneracy are easier to sample from in the high scale regime, which was previously explained in Section \ref{sec:other-problems}.  For the rest of the instances, the consistent U-shaped trend supports our claim that the most effective $\alpha_{in}$ for achieving high accuracy is in the medium scale regime. This optimal band between $\alpha_{in}$ of 0.2 and 0.4 remains consistent across all GSD models. In addition, these figures show that varying annealing time does not shift the optimal region by a significant amount, with the exception of GSD-10. However, there is a slight yet noticeable shift to the left in almost all instances as anneal times increase.

Figure \ref{fig:app-betaj-gsd} communicates the same information as Figures \ref{fig:app-tv-gsd} and \ref{fig:app-tv-gsd-f}, but more directly compares various GSD models by stacking their \rev{total variation distance} values in a heatmap.
Only the $1 \mu s$ plot was presented in Figure \ref{fig:fulltv} in the main text, but here the results for all anneal times are shown. Once again, it is evident that the optimal region for $\alpha_{in}$ remains the same for various anneal times.

\section{Effective Temperatures at Different Scales and Annealing Times}
\label{app:alpha-out-details}

\begin{figure*}[h]
\centering
    \begin{subfigure}{0.49\linewidth}
        \includegraphics[width=0.9\linewidth]{fig/heatmap_16_spin_1ms_v2_medium_font_no_clb_title_v2.pdf}
        \caption{$1 \mu s$}
    \end{subfigure}
    \begin{subfigure}{0.49\linewidth}
        \includegraphics[width=0.9\linewidth]{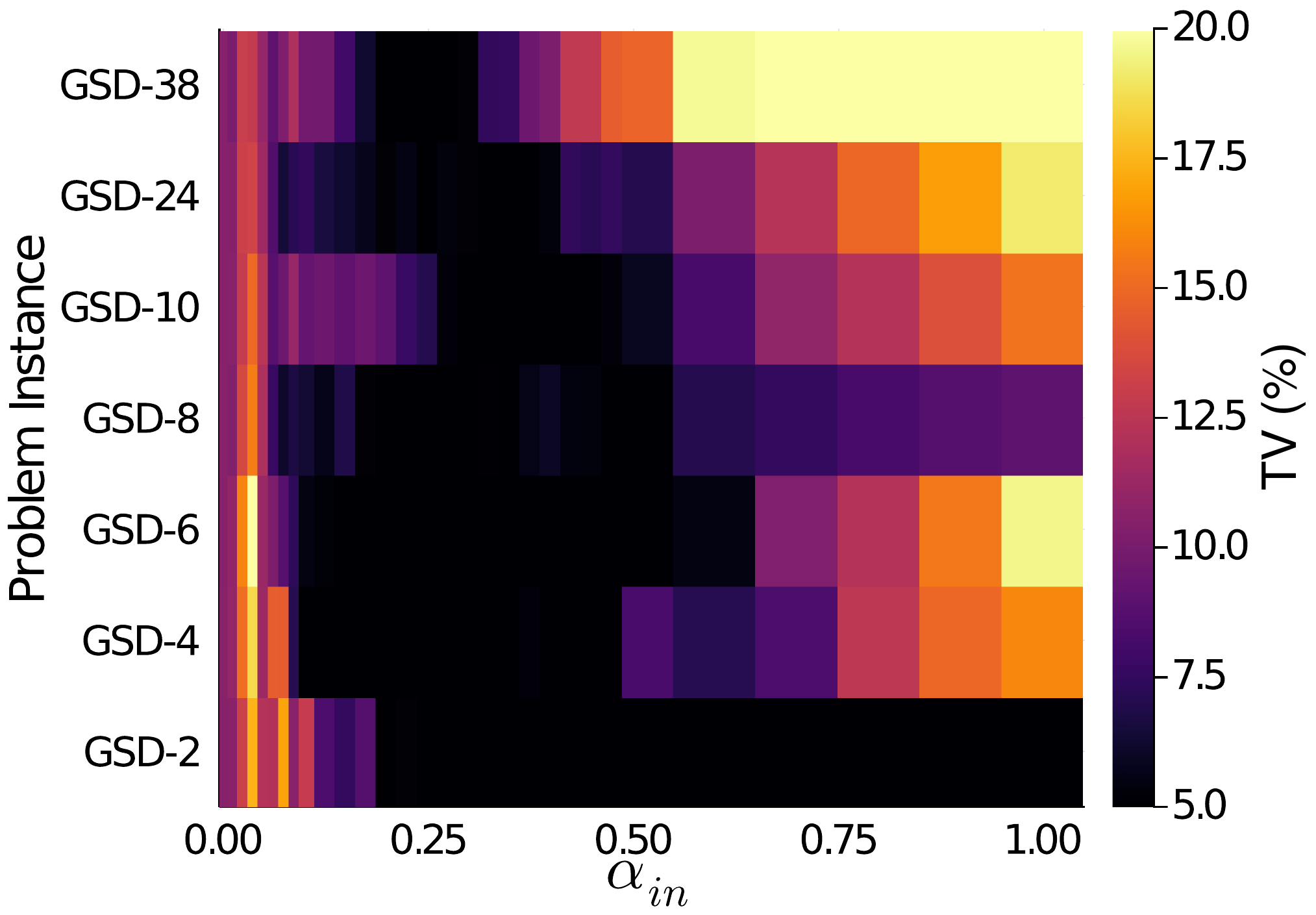}
        \caption{$5 \mu s$}
    \end{subfigure}
    \begin{subfigure}{0.49\linewidth}
        \includegraphics[width=0.9\linewidth]{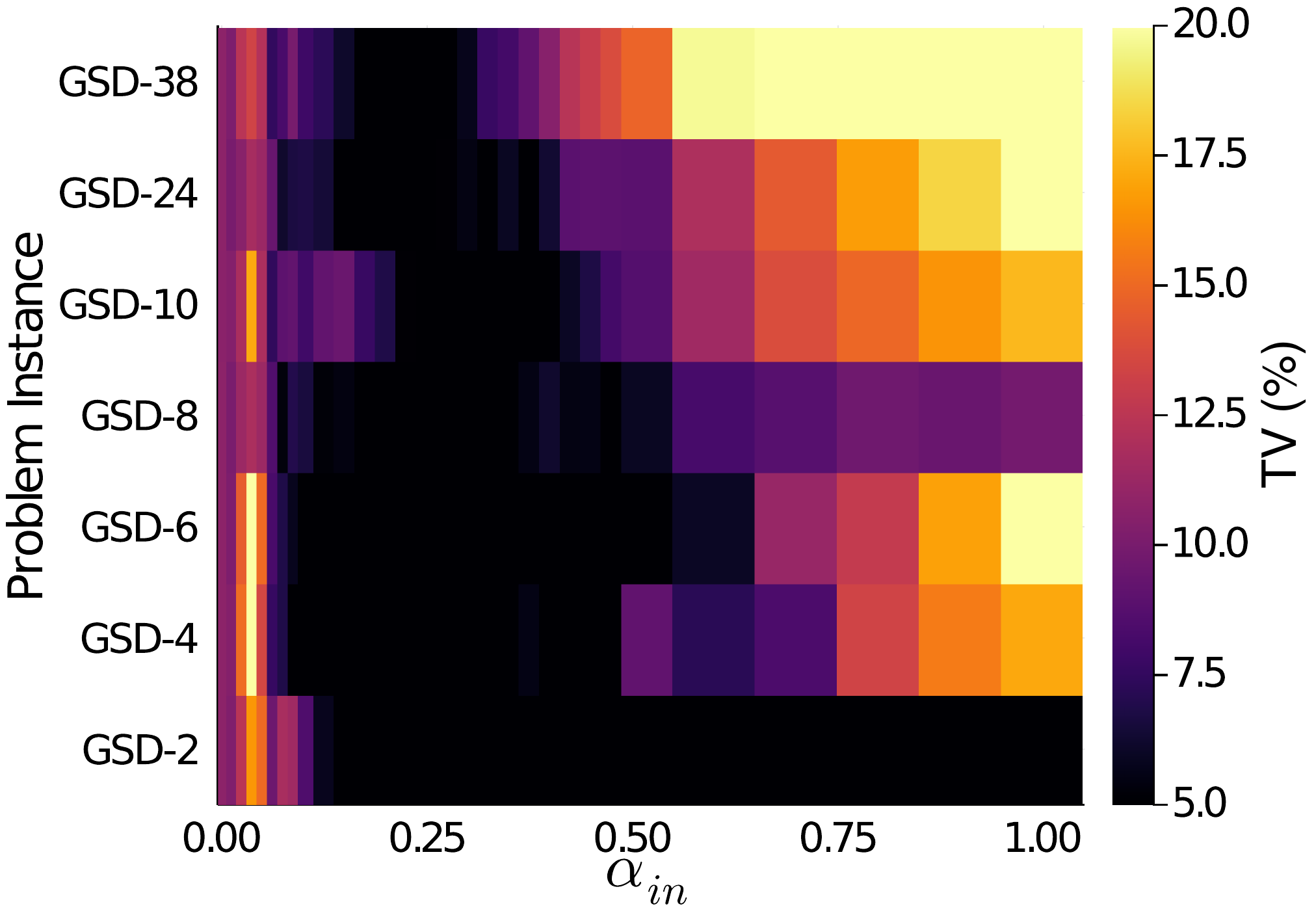}
        \caption{$25 \mu s$}
    \end{subfigure}
    \begin{subfigure}{0.49\linewidth}
       \includegraphics[width=0.9\linewidth]{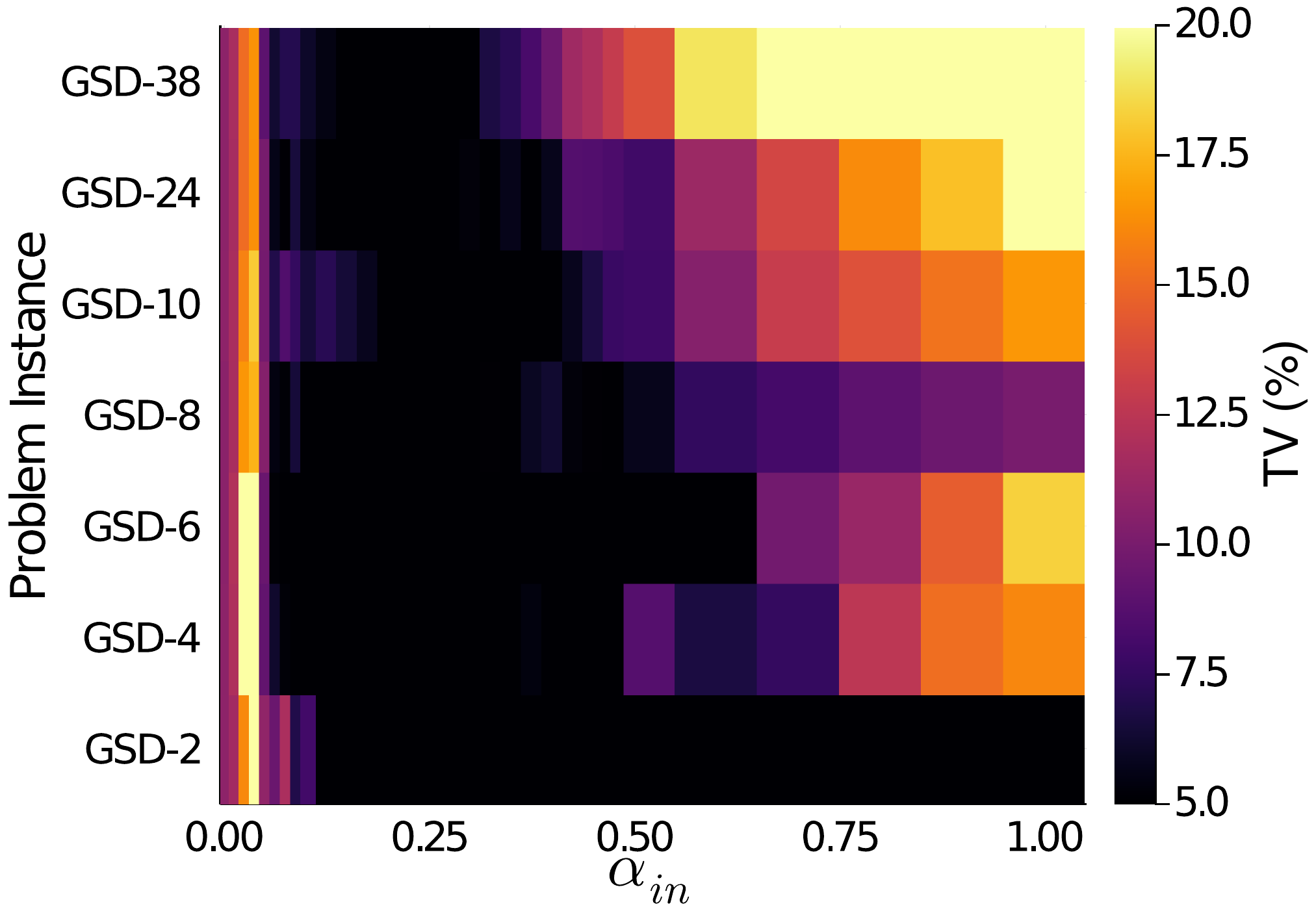}
        \caption{$125 \mu s$}
    \end{subfigure}
    \caption{A heatmap summary of the dependence of \rev{total variation distance} on $\alpha_{in}$ and annealing time for all GSD models.}
    \label{fig:app-betaj-gsd}
\end{figure*}

\begin{figure*}[h]
\centering
    \begin{subfigure}{0.49\linewidth}
        \includegraphics[width=0.9\linewidth]{fig/heatmap_16_spin_field_1ms_v2_medium_font_no_clb_title_v2.pdf}
        \caption{$1 \mu s$}
    \end{subfigure}
    \begin{subfigure}{0.49\linewidth}
        \includegraphics[width=0.9\linewidth]{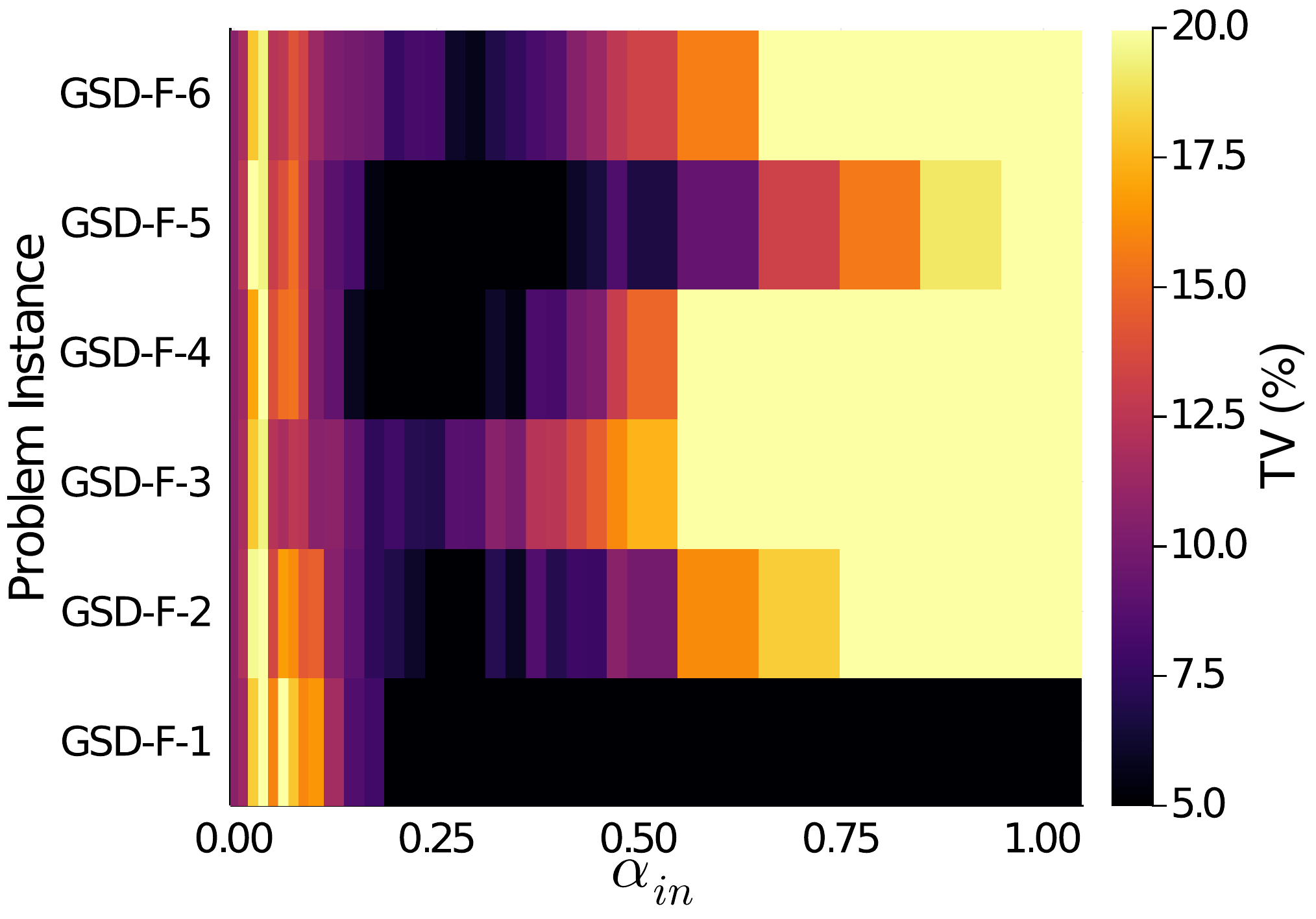}
        \caption{$5 \mu s$}
    \end{subfigure}
    \begin{subfigure}{0.49\linewidth}
        \includegraphics[width=0.9\linewidth]{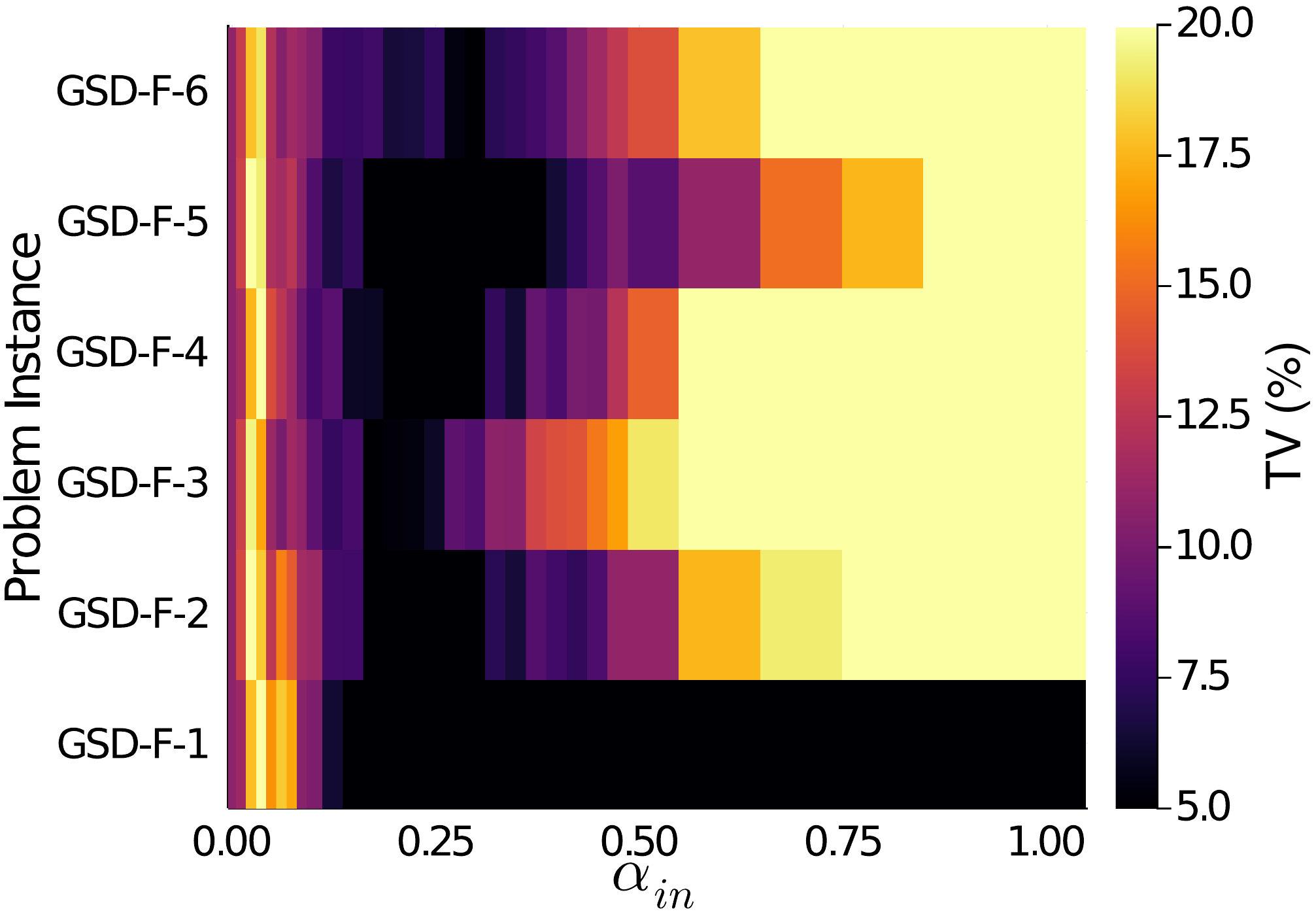}
        \caption{$25 \mu s$}
    \end{subfigure}
    \begin{subfigure}{0.49\linewidth}
       \includegraphics[width=0.9\linewidth]{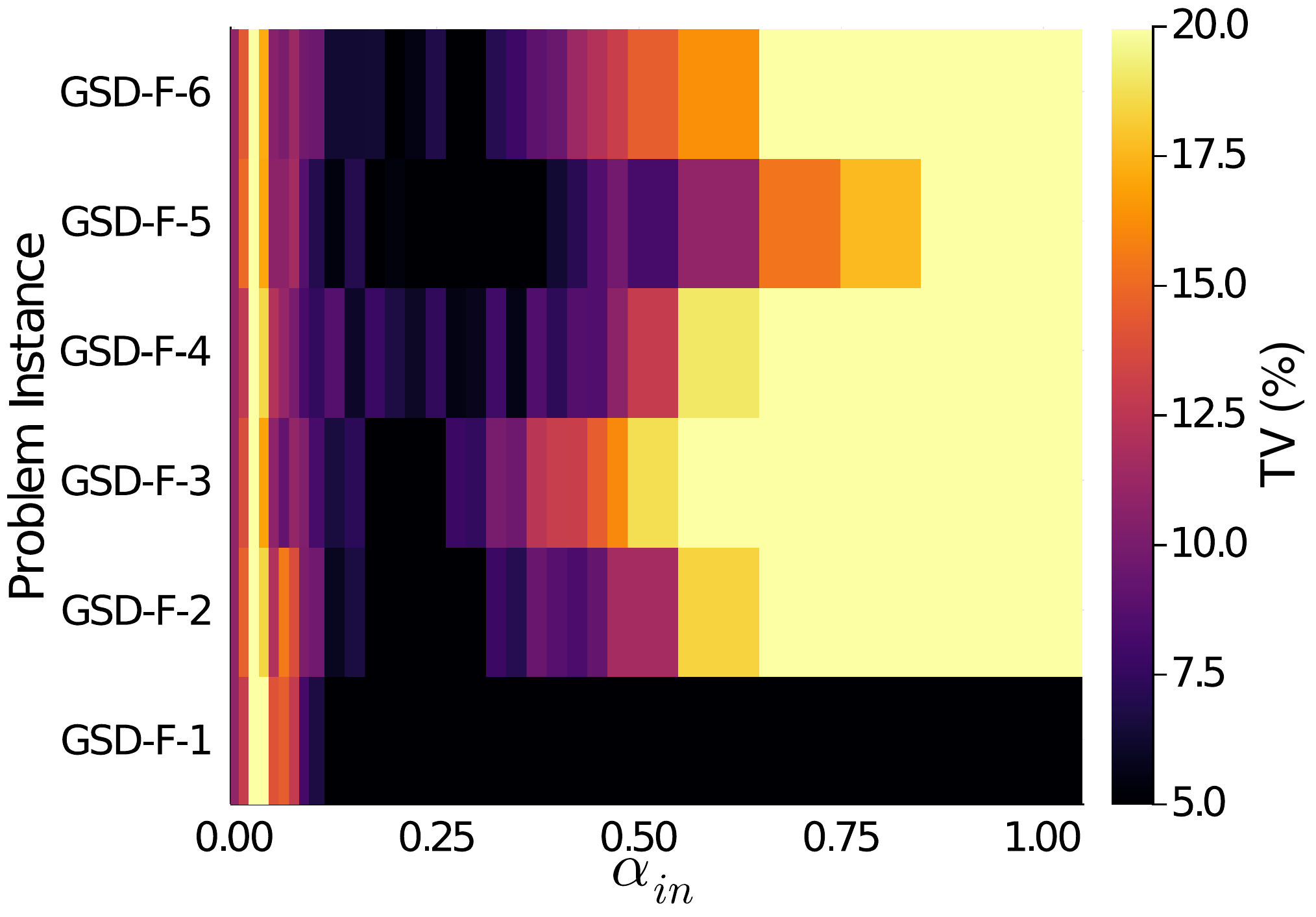}
        \caption{$125 \mu s$}
    \end{subfigure}
    \caption{A heatmap summary of the dependence of \rev{total variation distance} on $\alpha_{in}$ and annealing time for all GSD-F models.}
    \label{fig:app-betaj-gsd-f}
\end{figure*}

Figures \ref{fig:app-betaj-gsd} and \ref{fig:app-betaj-gsd-f} present the $\alpha_{out}$ values that correspond to the optimal $\alpha_{in}$ range between 0.2 and 0.4. The minimum and maximum value in each heatmap was used to construct Table \ref{tab:betaj}. These figures show how $\alpha_{in}$ and anneal time can be used to smoothly tune $\alpha_{out}$ and that the effect is consistent across all GSD models. Although the exact $\alpha_{out}$ values vary slightly from model to model, the trend is predictable and so these figures suggest a straitforward approach for producing Gibbs samples at at desired $\alpha_{out}$ value.

\section{Three-Spin Ising Chain Annealing Times}
\label{app:ising-chain-at}

\begin{figure*}[h!]
\centering
    \begin{subfigure}{\columnwidth}
        \includegraphics[width=\columnwidth]{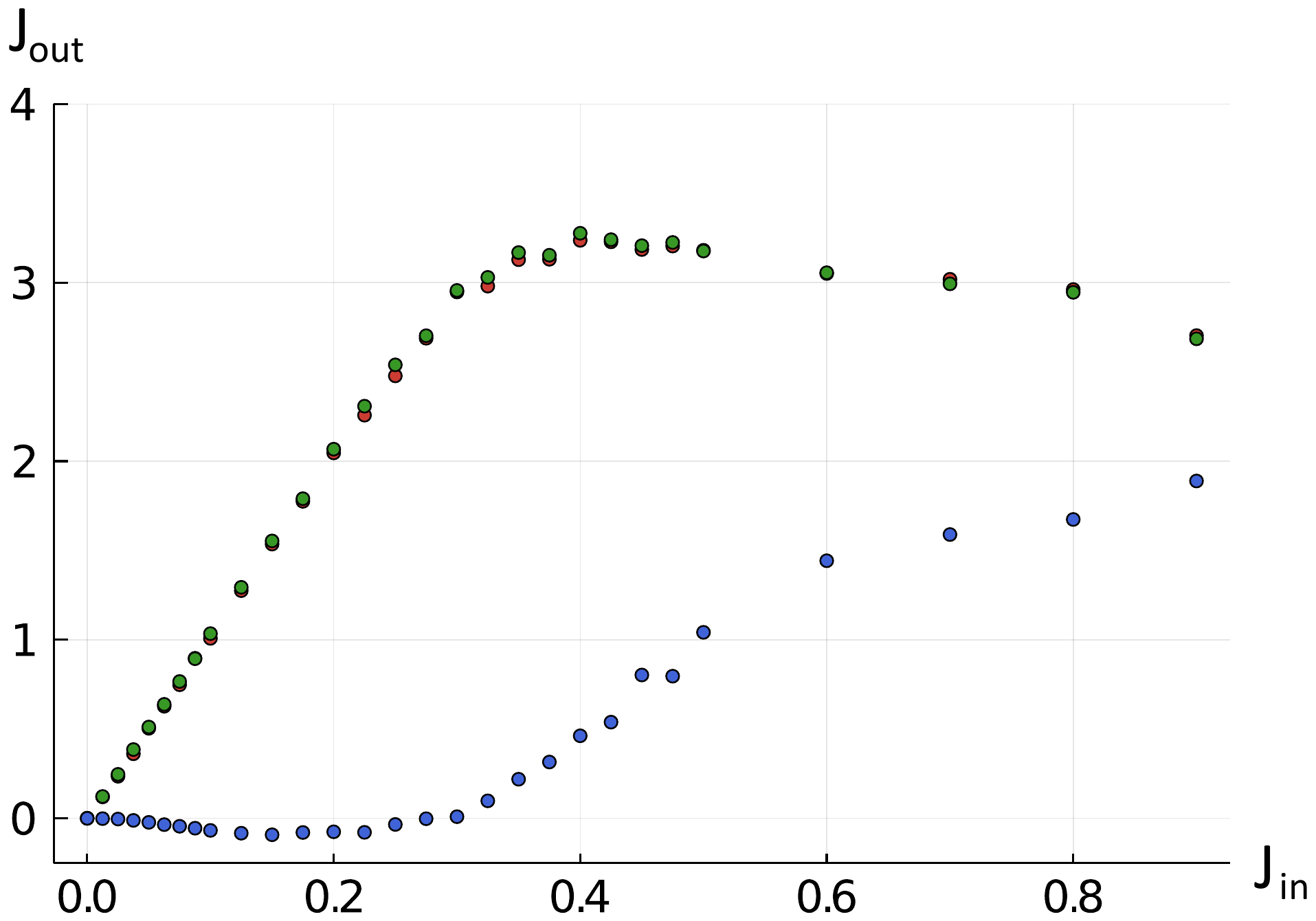}
        \caption{1 $\mu s$}
    \end{subfigure}
    \begin{subfigure}{\columnwidth}
        \includegraphics[width=\columnwidth]{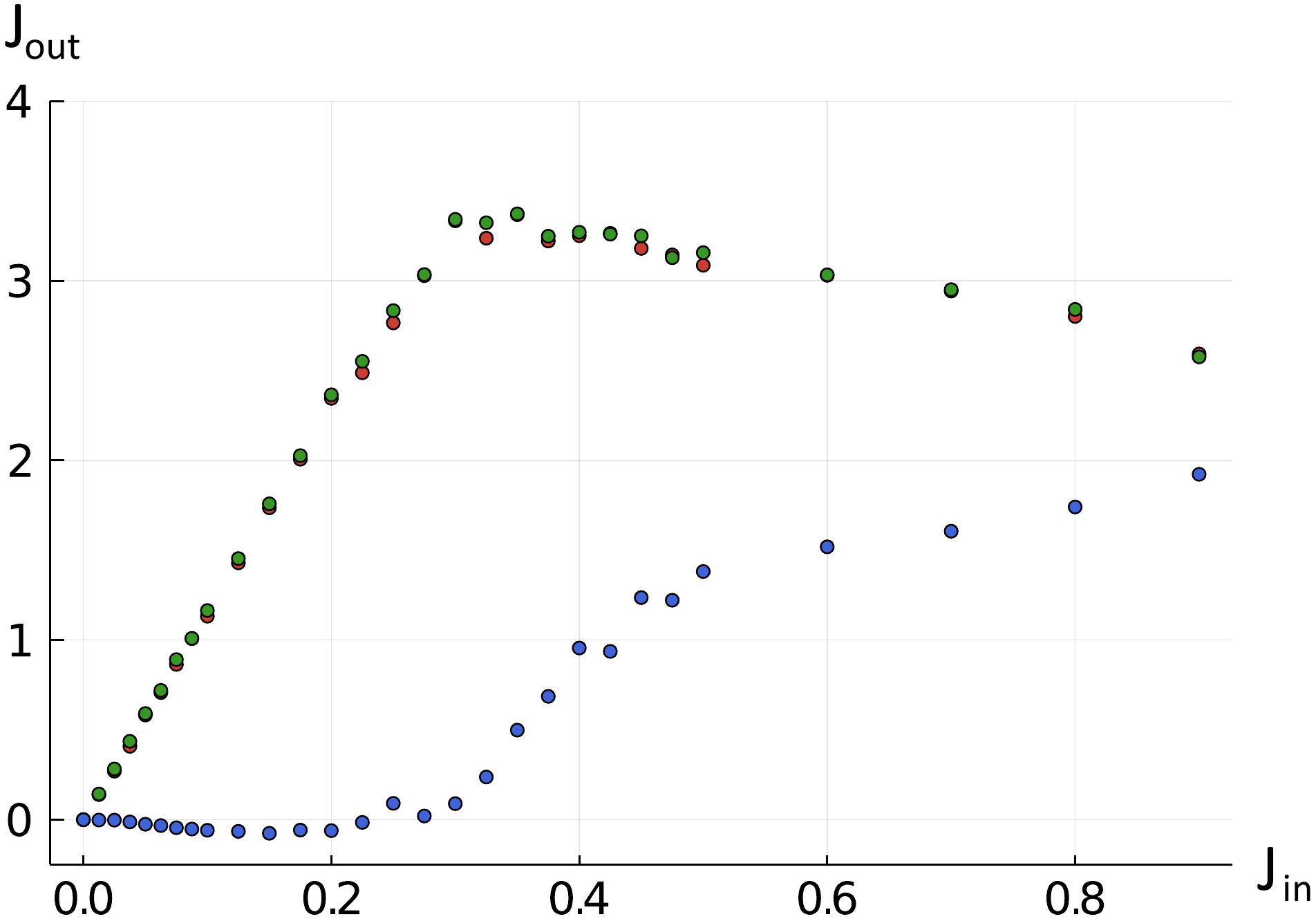}
        \caption{5 $\mu s$}
    \end{subfigure}
    \begin{subfigure}{\columnwidth}
        \includegraphics[width=\columnwidth]{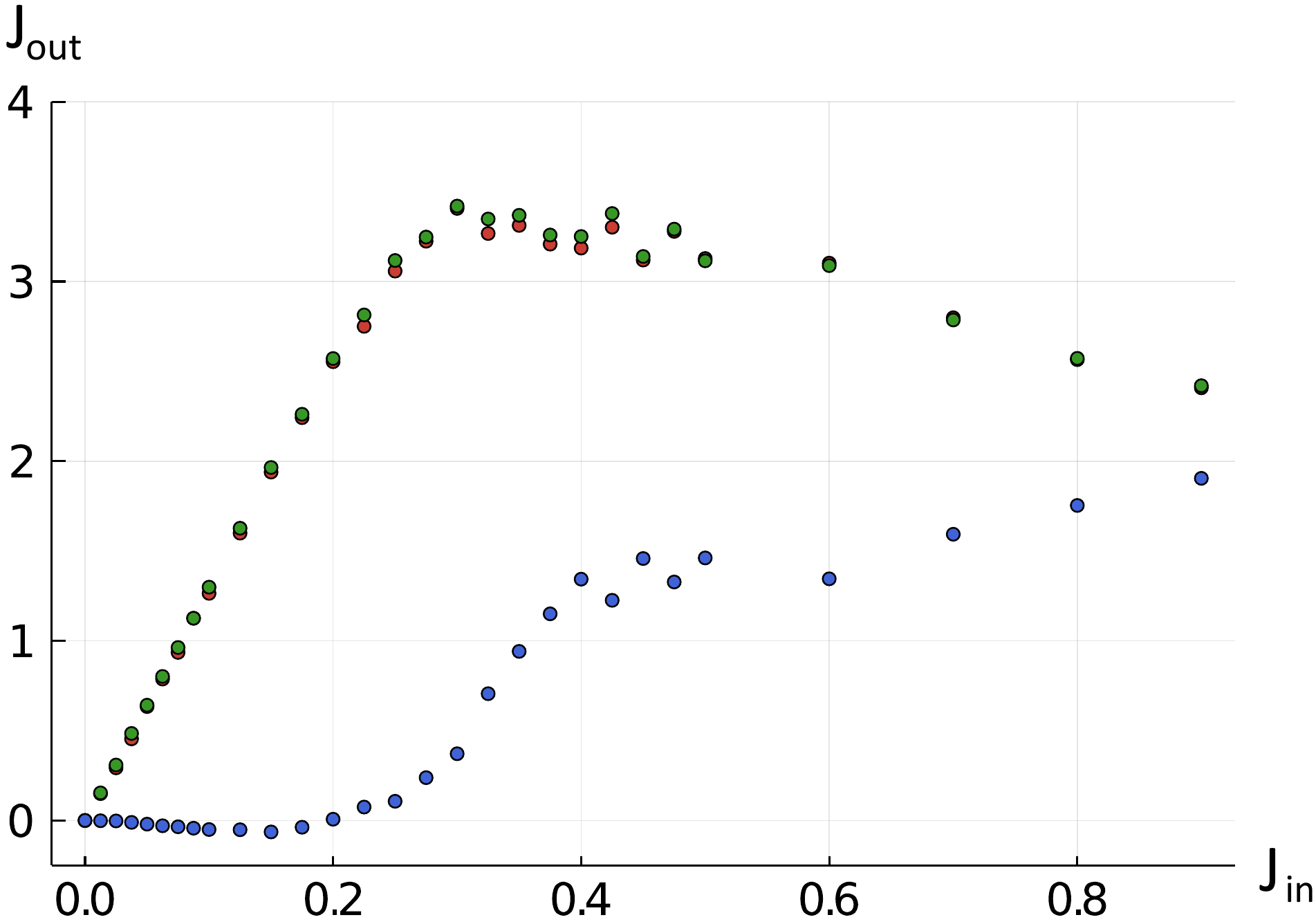}
        \caption{25 $\mu s$}
    \end{subfigure}
    \begin{subfigure}{\columnwidth}
       \includegraphics[width=\columnwidth]{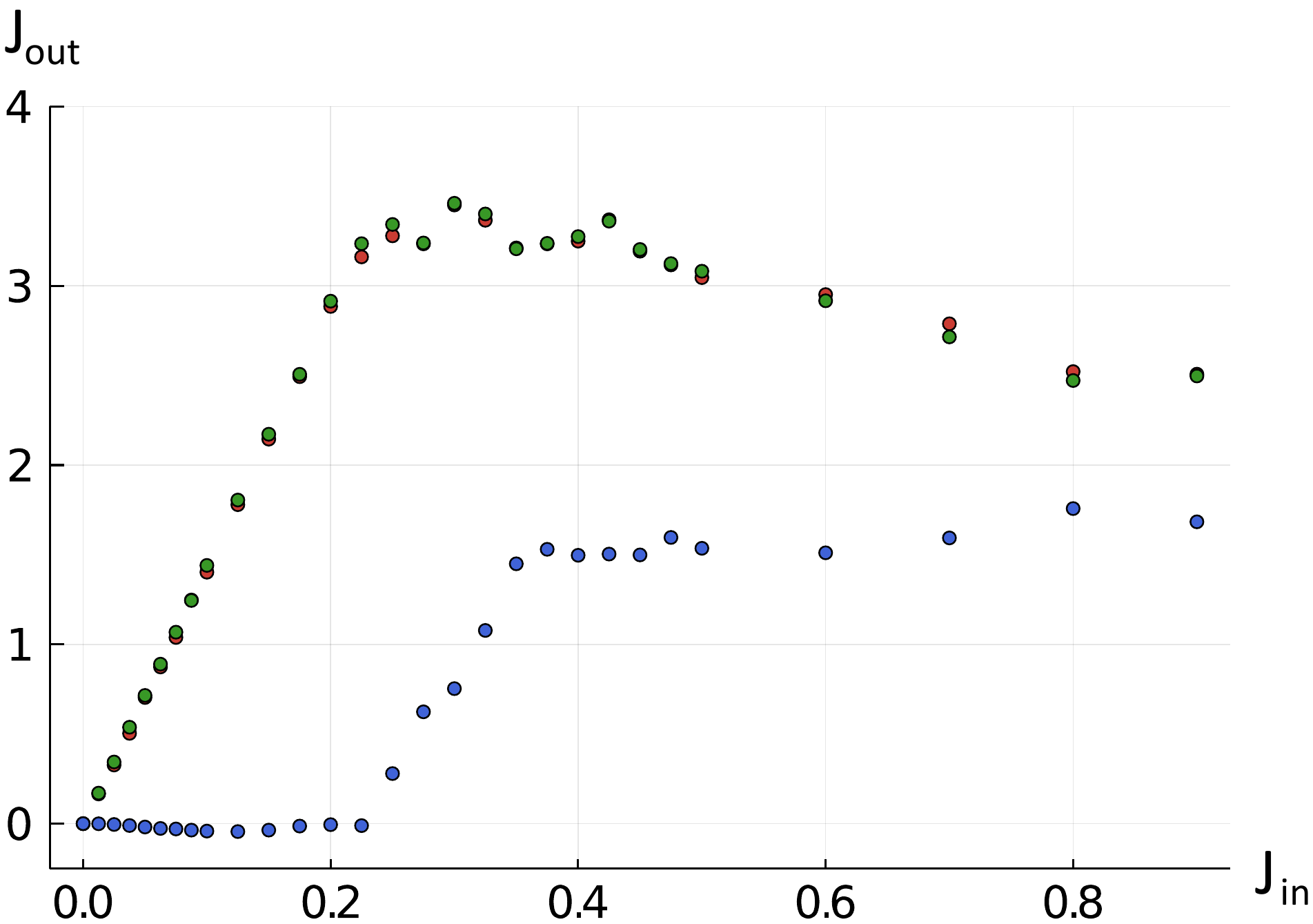}
        \caption{125 $\mu s$}
    \end{subfigure}
    \caption{Three Spin Experiments for Various Annealing Times. Green and red denote the programmed edges $J_{12}$ and $J_{23}$, respectively. Blue represents the spurious coupling $J_{13}$.}
    \label{fig:app-3spin}
\end{figure*}

Finally, Figure \ref{fig:app-3spin} shows results for the three-spin Ising chain experiment for various anneal times. The most important feature of these plots is the point at which the spurious coupling reduces to zero, as we hypothesize that this is the optimal input scale for sampling. From the plots, this optimal $J_{in}$ value slightly decreases from 0.275 to a low of 0.2 as anneal time increases. However, this value still remains within the optimal region seen in the 16-spin scaling experiments. As anneal time increases, both programmed and spurious couplings appear to increase more rapidly with the increase of $J_{in}$. This increase results in the spurious coupling almost saturating by the time $J_{in}$ reaches 0.4 for anneal times of 25 $\mu s$ and 125 $\mu s$. However, the majority of our claimed optimal region between 0.2 and 0.4 remains in the regime where the spurious coupling is small compared to the programmed couplings.
\\

\begin{figure}[t!]
    \centering
    \includegraphics[width=1\linewidth]{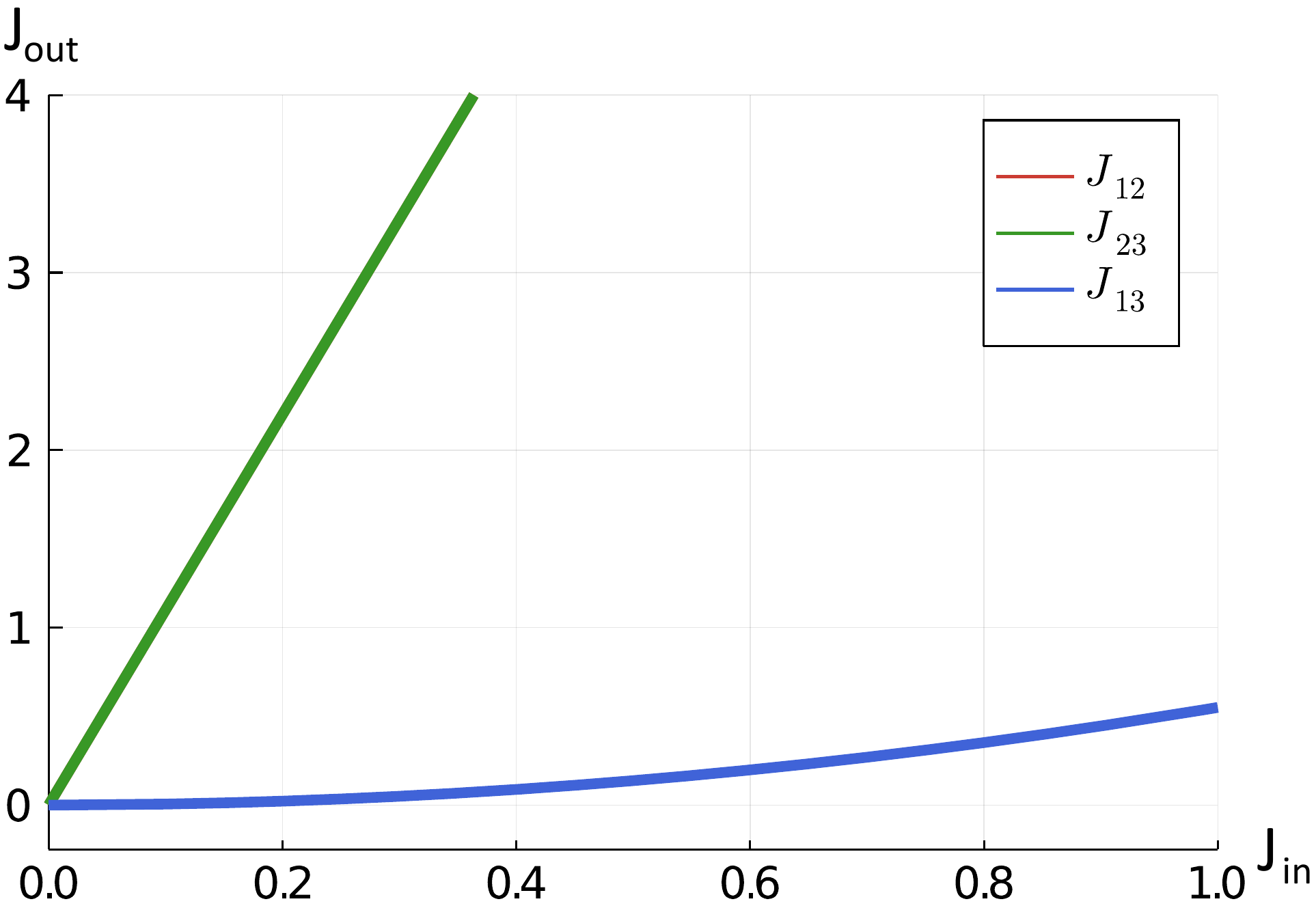}
    \caption{Input and spurious coupling values as a function of the input coupling intensity according to the Background susceptibility model.}
    \label{fig:BS_hamiltonian}
\end{figure}

\section{Background Susceptibility Model for the Three-Spin Ising Chain}
\label{app:bs-model}

It is believed that physically neighboring qubits are not perfectly isolated from each other and give rise to uncontrolled interactions. These spurious couplings are described by the Background Susceptibility model~\cite{dwave_docs} and take the form of an interaction that is proportional to the coupling intensities emanating from neighboring qubits. For the three-spin Ising chain discussed in Section~\ref{sec:discussion}, it results in the following Hamiltonian with background susceptibility,
\begin{align}
        H_{BS} =  -J_{in} \widehat{\sigma}^{z}_1 \widehat{\sigma}^{z}_2 - J_{in} \widehat{\sigma}^{z}_2 \widehat{\sigma}^{z}_3 - \chi J_{in}^2 \widehat{\sigma}^{z}_1 \widehat{\sigma}^{z}_3,
        \label{eq:BS_Hamiltonian}
\end{align}
where $\chi > 0$ is the background susceptibility constant between qubits one and three (Note in~\cite{dwave_docs}, $\chi$ is a negative quantity following the D-Wave programming convention that ferromagnetic couplings are negative.). Notice that the intensity of the spurious link between $\sigma_1$ and $\sigma_3$ is quadratic in the input coupling intensity $J_{\textrm{in}}$ and is of ferromagnetic nature when $J_{\textrm{in}}$ is positive. 
Because the Hamiltonian in Eq.~\eqref{eq:BS_Hamiltonian} is diagonal, we immediately see that this model behaves very differently than the effective Hamiltonian found experimentally (see Figure~\ref{fig:3spin}). The background susceptibility model predicts no saturation in the input coupling or spurious coupling intensity, as the earlier grow linearly and the later grows quadratically. Moreover, the spurious coupling of the background susceptibility model remains ferromagnetic unlike what is measured experimentally for small values of $J_{\textrm{in}}$. This behavior is depicted in Figure~\rev{\ref{fig:BS_hamiltonian}} for the typically encountered values of $\beta = 11$ and $\chi = 0.05$.

\begin{table*}[h!]
 \centering
 \begin{tabular}{|c||r|r|r|r|r|r|r|}
 \hline
  & GSD-2 & GSD-4 & GSD-6 & GSD-8 & GSD-10 & GSD-24 & GSD-38\\
 \hline
 \hline
 $J_{296, 300}$ & 1.0 & -1.0 & 1.0 & 1.0 & -1.0 & 1.0 & 1.0 \\ 
\hline 
$J_{296, 301}$ & 1.0 & 1.0 & -1.0 & -1.0 & 1.0 & 1.0 & 1.0 \\ 
\hline 
$J_{296, 302}$ & -1.0 & 1.0 & -1.0 & -1.0 & -1.0 & -1.0 & -1.0 \\ 
\hline 
$J_{296, 303}$ & -1.0 & -1.0 & 1.0 & -1.0 & 1.0 & -1.0 & 1.0 \\ 
\hline 
$J_{297, 300}$ & 1.0 & 1.0 & 1.0 & -1.0 & -1.0 & -1.0 & 1.0 \\ 
\hline 
$J_{297, 301}$ & 1.0 & 1.0 & 1.0 & 1.0 & -1.0 & 1.0 & 1.0 \\ 
\hline 
$J_{297, 302}$ & -1.0 & -1.0 & 1.0 & -1.0 & 1.0 & -1.0 & 1.0 \\ 
\hline 
$J_{297, 303}$ & 1.0 & -1.0 & -1.0 & 1.0 & -1.0 & -1.0 & -1.0 \\ 
\hline 
$J_{298, 300}$ & -1.0 & 1.0 & -1.0 & 1.0 & -1.0 & -1.0 & -1.0 \\ 
\hline 
$J_{298, 301}$ & 1.0 & -1.0 & -1.0 & -1.0 & -1.0 & 1.0 & 1.0 \\ 
\hline 
$J_{298, 302}$ & 1.0 & 1.0 & -1.0 & 1.0 & 1.0 & -1.0 & -1.0 \\ 
\hline 
$J_{298, 303}$ & 1.0 & 1.0 & 1.0 & -1.0 & -1.0 & 1.0 & -1.0 \\ 
\hline 
$J_{299, 300}$ & 1.0 & 1.0 & 1.0 & -1.0 & -1.0 & 1.0 & -1.0 \\ 
\hline 
$J_{299, 301}$ & 1.0 & -1.0 & -1.0 & -1.0 & 1.0 & 1.0 & -1.0 \\ 
\hline 
$J_{299, 302}$ & -1.0 & 1.0 & -1.0 & -1.0 & 1.0 & -1.0 & 1.0 \\ 
\hline 
$J_{299, 303}$ & -1.0 & 1.0 & 1.0 & -1.0 & 1.0 & -1.0 & -1.0 \\ 
\hline 
$J_{300, 308}$ & 1.0 & -1.0 & -1.0 & -1.0 & -1.0 & 1.0 & 1.0 \\ 
\hline 
$J_{301, 309}$ & 1.0 & 1.0 & -1.0 & -1.0 & -1.0 & 1.0 & 1.0 \\ 
\hline 
$J_{302, 310}$ & 1.0 & 1.0 & 1.0 & 1.0 & 1.0 & -1.0 & -1.0 \\ 
\hline 
$J_{303, 311}$ & -1.0 & 1.0 & 1.0 & -1.0 & -1.0 & -1.0 & 1.0 \\ 
\hline 
$J_{304, 308}$ & 1.0 & -1.0 & 1.0 & -1.0 & 1.0 & 1.0 & 1.0 \\ 
\hline 
$J_{304, 309}$ & 1.0 & -1.0 & 1.0 & -1.0 & -1.0 & -1.0 & -1.0 \\ 
\hline 
$J_{304, 310}$ & 1.0 & 1.0 & -1.0 & 1.0 & -1.0 & 1.0 & -1.0 \\ 
\hline 
$J_{304, 311}$ & -1.0 & 1.0 & 1.0 & -1.0 & -1.0 & 1.0 & 1.0 \\ 
\hline 
$J_{305, 308}$ & -1.0 & 1.0 & 1.0 & 1.0 & -1.0 & -1.0 & -1.0 \\ 
\hline 
$J_{305, 309}$ & -1.0 & -1.0 & 1.0 & 1.0 & -1.0 & -1.0 & -1.0 \\ 
\hline 
$J_{305, 310}$ & -1.0 & 1.0 & -1.0 & -1.0 & 1.0 & 1.0 & -1.0 \\ 
\hline 
$J_{305, 311}$ & -1.0 & -1.0 & 1.0 & 1.0 & 1.0 & -1.0 & -1.0 \\ 
\hline 
$J_{306, 308}$ & 1.0 & 1.0 & -1.0 & -1.0 & -1.0 & -1.0 & 1.0 \\ 
\hline 
$J_{306, 309}$ & -1.0 & -1.0 & 1.0 & 1.0 & -1.0 & 1.0 & 1.0 \\ 
\hline 
$J_{306, 310}$ & 1.0 & 1.0 & -1.0 & 1.0 & -1.0 & -1.0 & 1.0 \\ 
\hline 
$J_{306, 311}$ & 1.0 & 1.0 & 1.0 & 1.0 & 1.0 & 1.0 & 1.0 \\ 
\hline 
$J_{307, 308}$ & 1.0 & 1.0 & -1.0 & -1.0 & -1.0 & -1.0 & 1.0 \\ 
\hline 
$J_{307, 309}$ & 1.0 & 1.0 & 1.0 & 1.0 & -1.0 & 1.0 & -1.0 \\ 
\hline 
$J_{307, 310}$ & -1.0 & -1.0 & -1.0 & 1.0 & -1.0 & 1.0 & -1.0 \\ 
\hline 
$J_{307, 311}$ & 1.0 & -1.0 & -1.0 & -1.0 & 1.0 & -1.0 & 1.0 \\ 
\hline 

 \end{tabular}
 \caption{Input coupling values for GSD models before scaling by $\alpha_{in}$. The D-Wave programming convention is used where negative couplings indicate ferromagnetic and positive couplings indicate antiferromagnetic.}
 \label{tab:gsd}
\end{table*}

\begin{table*}
 \centering
\begin{tabular}{|c||r|r|r|r|r|r|}
 \hline
  & GSD-F-1 & GSD-F-2 & GSD-F-3 & GSD-F-4 & GSD-F-5 & GSD-F-6 \\
\hline
\hline 
$h_{296}$ & 1.0 & 1.0 & 1.0 & -1.0 & 1.0 & -1.0 \\ 
\hline 
$h_{297}$ & 1.0 & -1.0 & 1.0 & 1.0 & -1.0 & -1.0 \\ 
\hline 
$h_{298}$ & 1.0 & 1.0 & -1.0 & 1.0 & -1.0 & 1.0 \\ 
\hline 
$h_{299}$ & 1.0 & -1.0 & -1.0 & 1.0 & -1.0 & 1.0 \\ 
\hline 
$h_{300}$ & -1.0 & -1.0 & -1.0 & -1.0 & -1.0 & 1.0 \\ 
\hline 
$h_{301}$ & -1.0 & 1.0 & -1.0 & 1.0 & 1.0 & 1.0 \\ 
\hline 
$h_{302}$ & 1.0 & 1.0 & 1.0 & 1.0 & -1.0 & 1.0 \\ 
\hline 
$h_{303}$ & -1.0 & 1.0 & 1.0 & -1.0 & 1.0 & -1.0 \\ 
\hline 
$h_{304}$ & -1.0 & -1.0 & 1.0 & -1.0 & -1.0 & 1.0 \\ 
\hline 
$h_{305}$ & -1.0 & 1.0 & -1.0 & 1.0 & 1.0 & 1.0 \\ 
\hline 
$h_{306}$ & 1.0 & -1.0 & -1.0 & -1.0 & -1.0 & 1.0 \\ 
\hline 
$h_{307}$ & 1.0 & 1.0 & -1.0 & 1.0 & 1.0 & -1.0 \\ 
\hline 
$h_{308}$ & -1.0 & 1.0 & -1.0 & -1.0 & 1.0 & 1.0 \\ 
\hline 
$h_{309}$ & 1.0 & -1.0 & 1.0 & -1.0 & 1.0 & 1.0 \\ 
\hline 
$h_{310}$ & -1.0 & 1.0 & -1.0 & -1.0 & -1.0 & -1.0 \\ 
\hline 
$h_{311}$ & -1.0 & -1.0 & 1.0 & -1.0 & -1.0 & -1.0 \\ 
\hline 
\hline 
$J_{296, 301}$ & -1.0 & 1.0 & 1.0 & -1.0 & -1.0 & 1.0 \\ 
\hline 
$J_{296, 302}$ & 1.0 & -1.0 & -1.0 & 1.0 & 1.0 & 1.0 \\ 
\hline 
$J_{296, 303}$ & -1.0 & -1.0 & -1.0 & 1.0 & -1.0 & -1.0 \\ 
\hline 
$J_{297, 300}$ & 1.0 & 1.0 & 1.0 & -1.0 & 1.0 & -1.0 \\ 
\hline 
$J_{297, 301}$ & -1.0 & -1.0 & -1.0 & 1.0 & -1.0 & 1.0 \\ 
\hline 
$J_{297, 302}$ & 1.0 & 1.0 & 1.0 & 1.0 & 1.0 & -1.0 \\ 
\hline 
$J_{297, 303}$ & -1.0 & 1.0 & -1.0 & -1.0 & 1.0 & -1.0 \\ 
\hline 
$J_{298, 300}$ & 1.0 & 1.0 & 1.0 & -1.0 & -1.0 & -1.0 \\ 
\hline 
$J_{298, 301}$ & -1.0 & 1.0 & -1.0 & -1.0 & -1.0 & 1.0 \\ 
\hline 
$J_{298, 302}$ & -1.0 & -1.0 & -1.0 & 1.0 & -1.0 & -1.0 \\ 
\hline 
$J_{298, 303}$ & -1.0 & -1.0 & 1.0 & -1.0 & -1.0 & -1.0 \\ 
\hline 
$J_{299, 300}$ & -1.0 & 1.0 & 1.0 & 1.0 & 1.0 & -1.0 \\ 
\hline 
$J_{299, 301}$ & -1.0 & 1.0 & -1.0 & -1.0 & 1.0 & 1.0 \\ 
\hline 
$J_{299, 302}$ & 1.0 & -1.0 & -1.0 & -1.0 & -1.0 & 1.0 \\ 
\hline 
$J_{299, 303}$ & 1.0 & 1.0 & 1.0 & 1.0 & 1.0 & 1.0 \\ 
\hline 
$J_{300, 308}$ & -1.0 & -1.0 & 1.0 & -1.0 & 1.0 & 1.0 \\ 
\hline 
$J_{301, 309}$ & -1.0 & -1.0 & -1.0 & -1.0 & 1.0 & 1.0 \\ 
\hline 
$J_{302, 310}$ & 1.0 & 1.0 & 1.0 & 1.0 & -1.0 & -1.0 \\ 
\hline 
$J_{303, 311}$ & -1.0 & -1.0 & -1.0 & -1.0 & -1.0 & 1.0 \\ 
\hline 
$J_{304, 308}$ & -1.0 & 1.0 & 1.0 & 1.0 & -1.0 & -1.0 \\ 
\hline 
$J_{304, 309}$ & -1.0 & 1.0 & 1.0 & 1.0 & 1.0 & -1.0 \\ 
\hline 
$J_{304, 310}$ & 1.0 & 1.0 & 1.0 & -1.0 & -1.0 & 1.0 \\ 
\hline 
$J_{304, 311}$ & 1.0 & 1.0 & 1.0 & -1.0 & -1.0 & 1.0 \\ 
\hline 
$J_{305, 308}$ & -1.0 & -1.0 & 1.0 & -1.0 & 1.0 & -1.0 \\ 
\hline 
$J_{305, 309}$ & 1.0 & -1.0 & -1.0 & 1.0 & -1.0 & -1.0 \\ 
\hline 
$J_{305, 310}$ & -1.0 & -1.0 & -1.0 & -1.0 & 1.0 & 1.0 \\ 
\hline 
$J_{305, 311}$ & 1.0 & -1.0 & 1.0 & 1.0 & -1.0 & -1.0 \\ 
\hline 
$J_{306, 308}$ & -1.0 & 1.0 & 1.0 & -1.0 & 1.0 & -1.0 \\ 
\hline 
$J_{306, 309}$ & 1.0 & 1.0 & -1.0 & -1.0 & -1.0 & 1.0 \\ 
\hline 
$J_{306, 310}$ & 1.0 & 1.0 & 1.0 & 1.0 & 1.0 & -1.0 \\ 
\hline 
$J_{306, 311}$ & 1.0 & -1.0 & -1.0 & 1.0 & 1.0 & 1.0 \\ 
\hline 
$J_{307, 308}$ & 1.0 & 1.0 & 1.0 & -1.0 & -1.0 & 1.0 \\ 
\hline 
$J_{307, 309}$ & -1.0 & -1.0 & -1.0 & -1.0 & -1.0 & 1.0 \\ 
\hline 
$J_{307, 310}$ & -1.0 & 1.0 & 1.0 & -1.0 & 1.0 & 1.0 \\ 
\hline 
$J_{307, 311}$ & -1.0 & -1.0 & -1.0 & -1.0 & 1.0 & -1.0 \\ 
\hline 

        \end{tabular}
        \caption{Input coupling and field values for GSD-F models before scaling by $\alpha_{in}$. The D-Wave programming convention is used where negative couplings indicate ferromagnetic and positive couplings indicate antiferromagnetic.}
        \label{tab:gsd-f}
    \end{table*}

\begin{figure*}[h]
\centering
    \begin{subfigure}{0.49\linewidth}
        \includegraphics[width=0.9\linewidth]{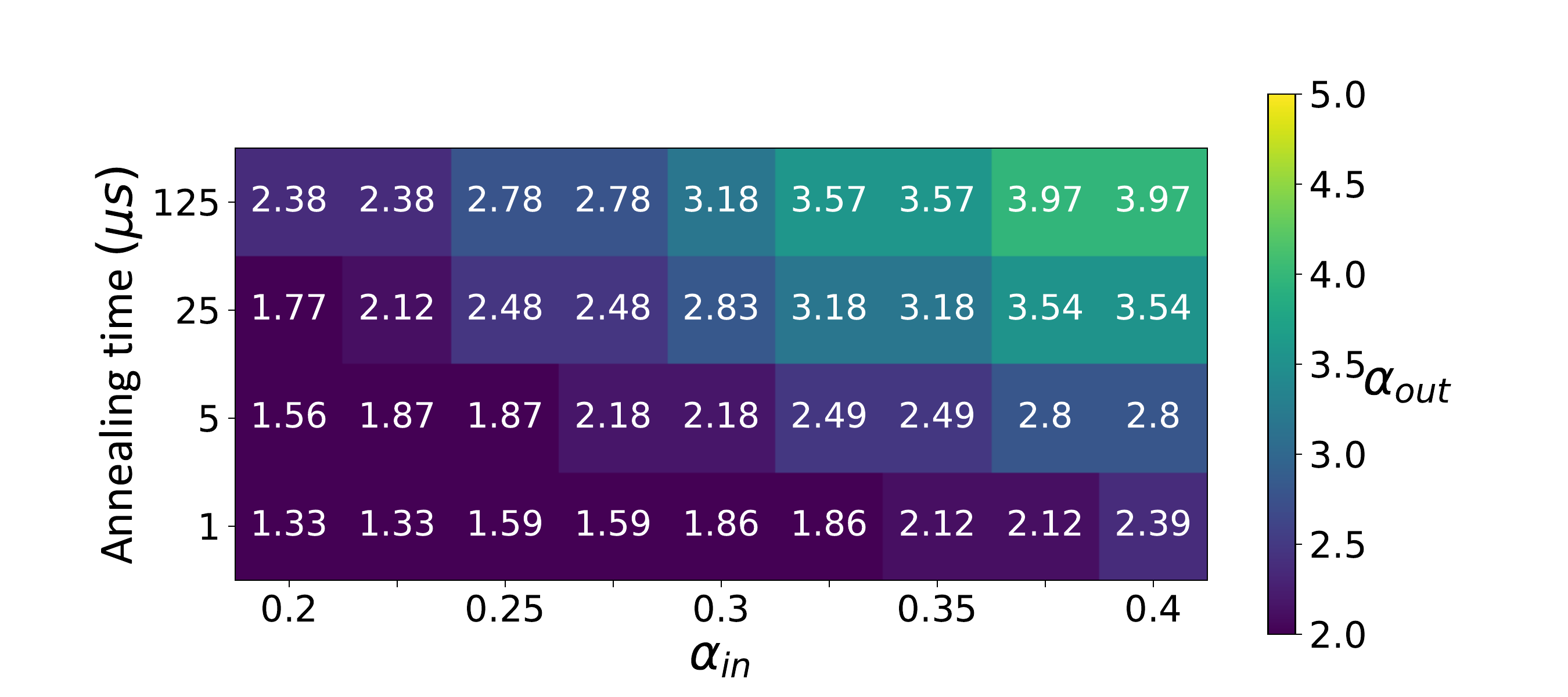}
        \caption{GSD-2}
    \end{subfigure}
    \begin{subfigure}{0.49\linewidth}
        \includegraphics[width=0.9\linewidth]{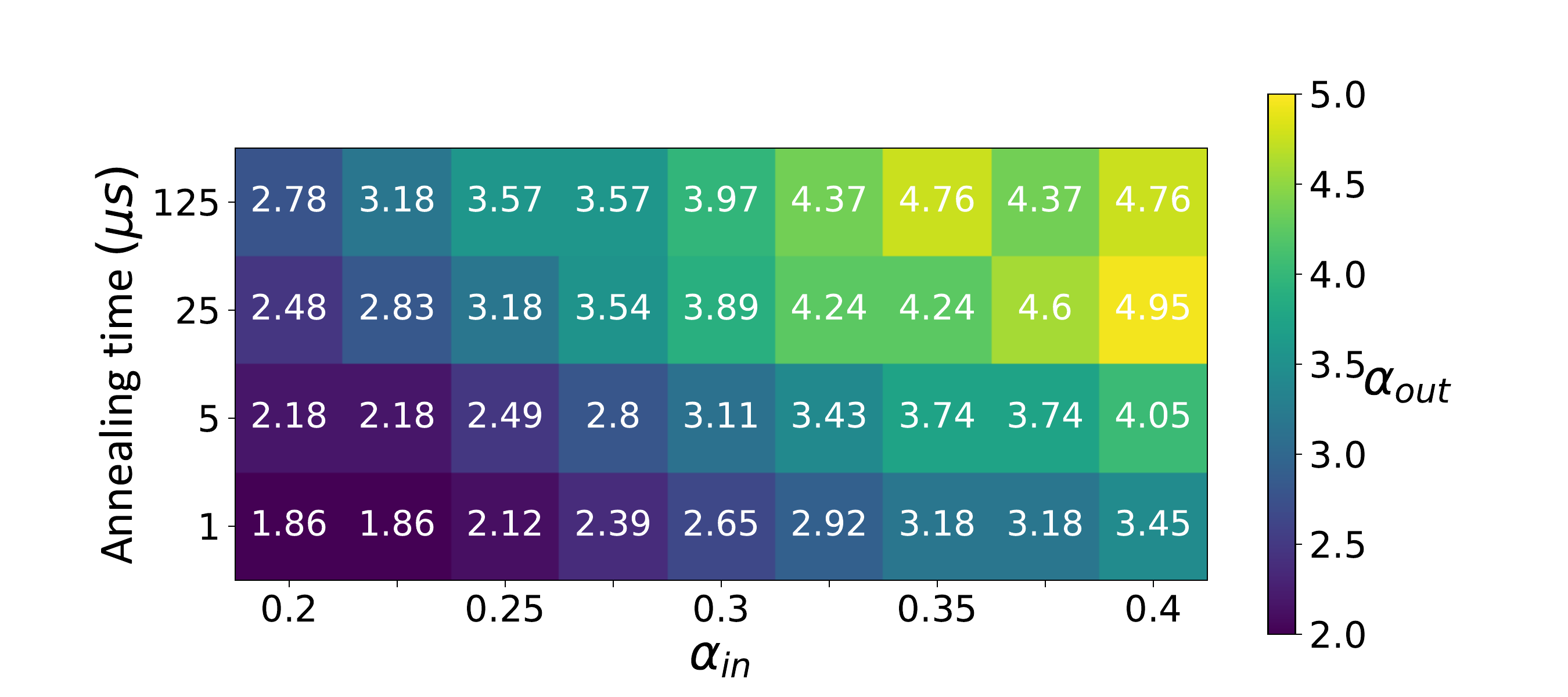}
        \caption{GSD-4}
    \end{subfigure}
    \begin{subfigure}{0.49\linewidth}
        \includegraphics[width=0.9\linewidth]{fig/ran_degen_6_j_sweep_16_spin_beta_j_matrix_font_increased.pdf}
        \caption{GSD-6}
    \end{subfigure}
    \begin{subfigure}{0.49\linewidth}
       \includegraphics[width=0.9\linewidth]{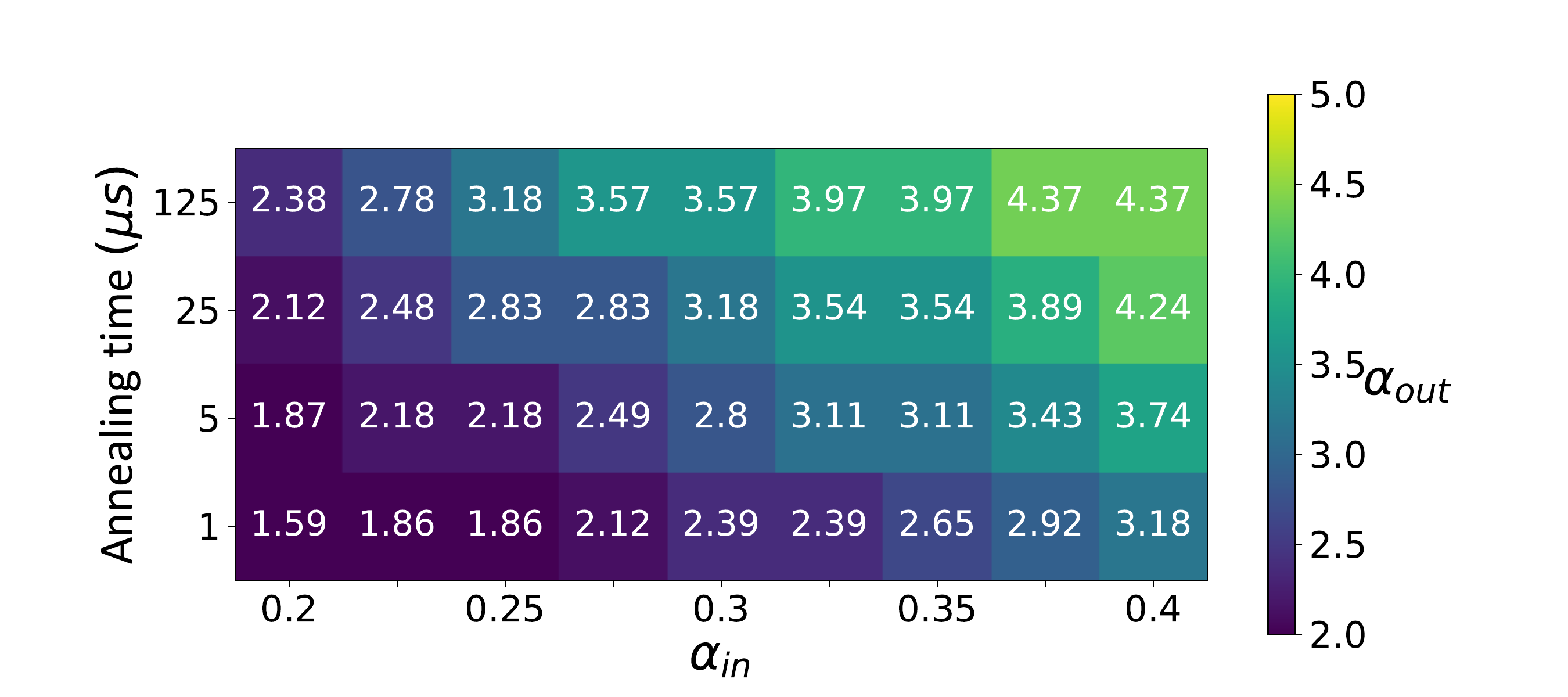}
        \caption{GSD-8}
    \end{subfigure}
    \begin{subfigure}{0.49\linewidth}
        \includegraphics[width=0.9\linewidth]{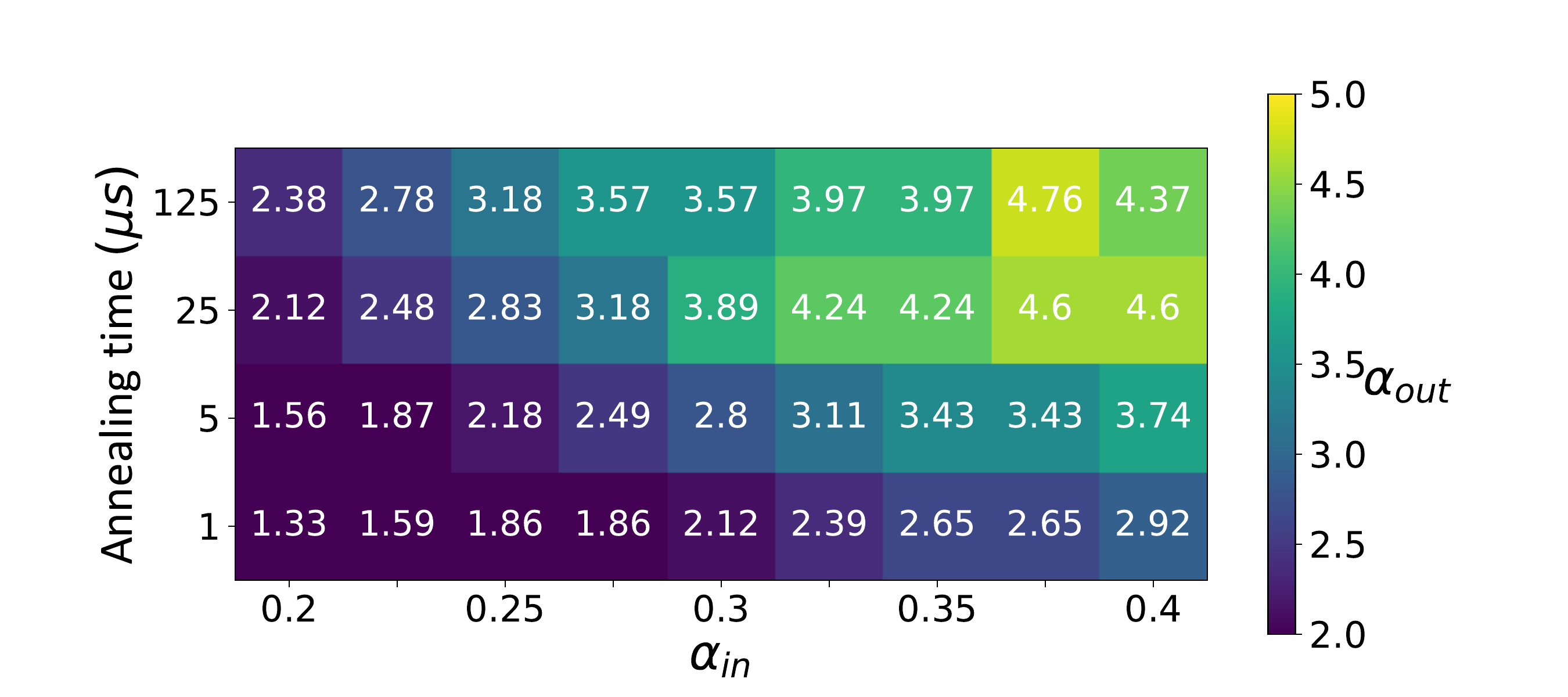}
        \caption{GSD-10}
    \end{subfigure}
    \begin{subfigure}{0.49\linewidth}
        \includegraphics[width=0.9\linewidth]{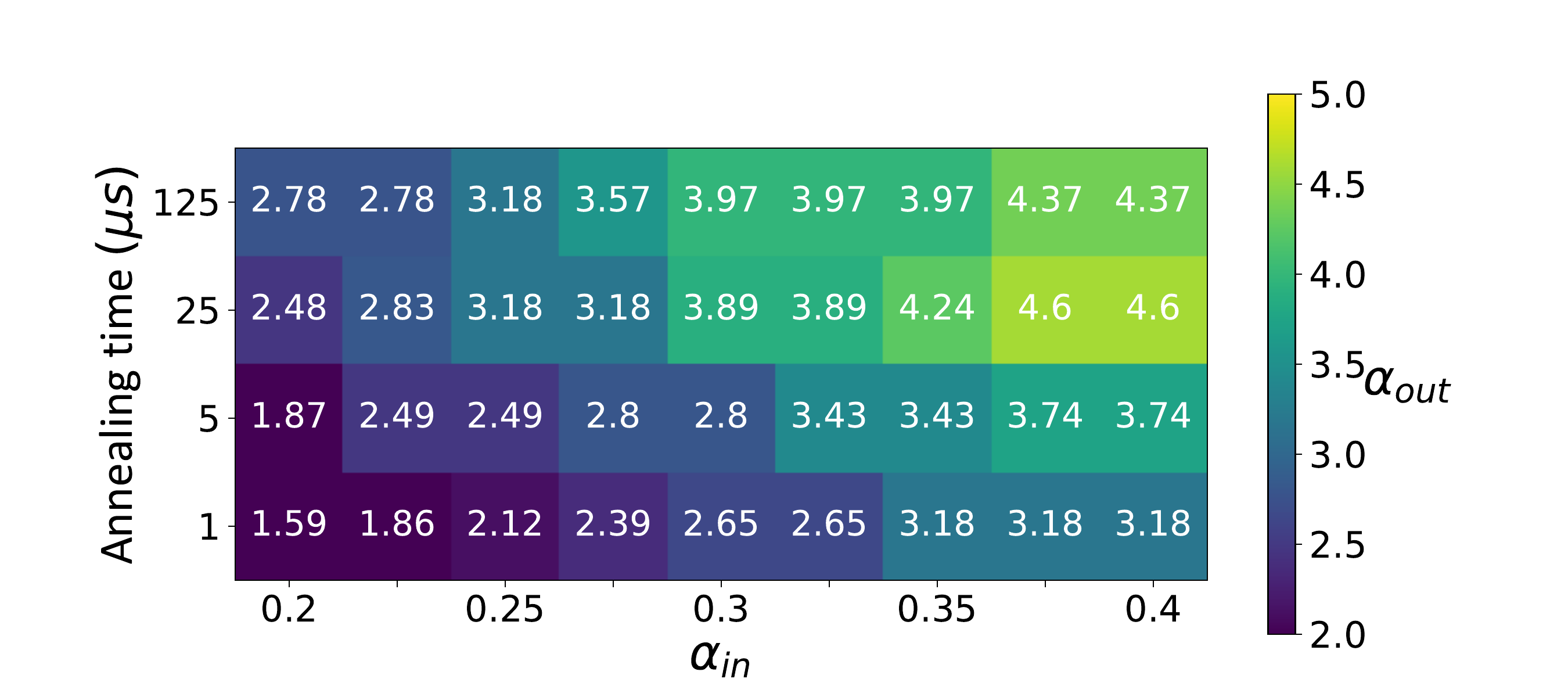}
        \caption{GSD-24}
    \end{subfigure}
    \begin{subfigure}{0.49\linewidth}
        \includegraphics[width=0.9\linewidth]{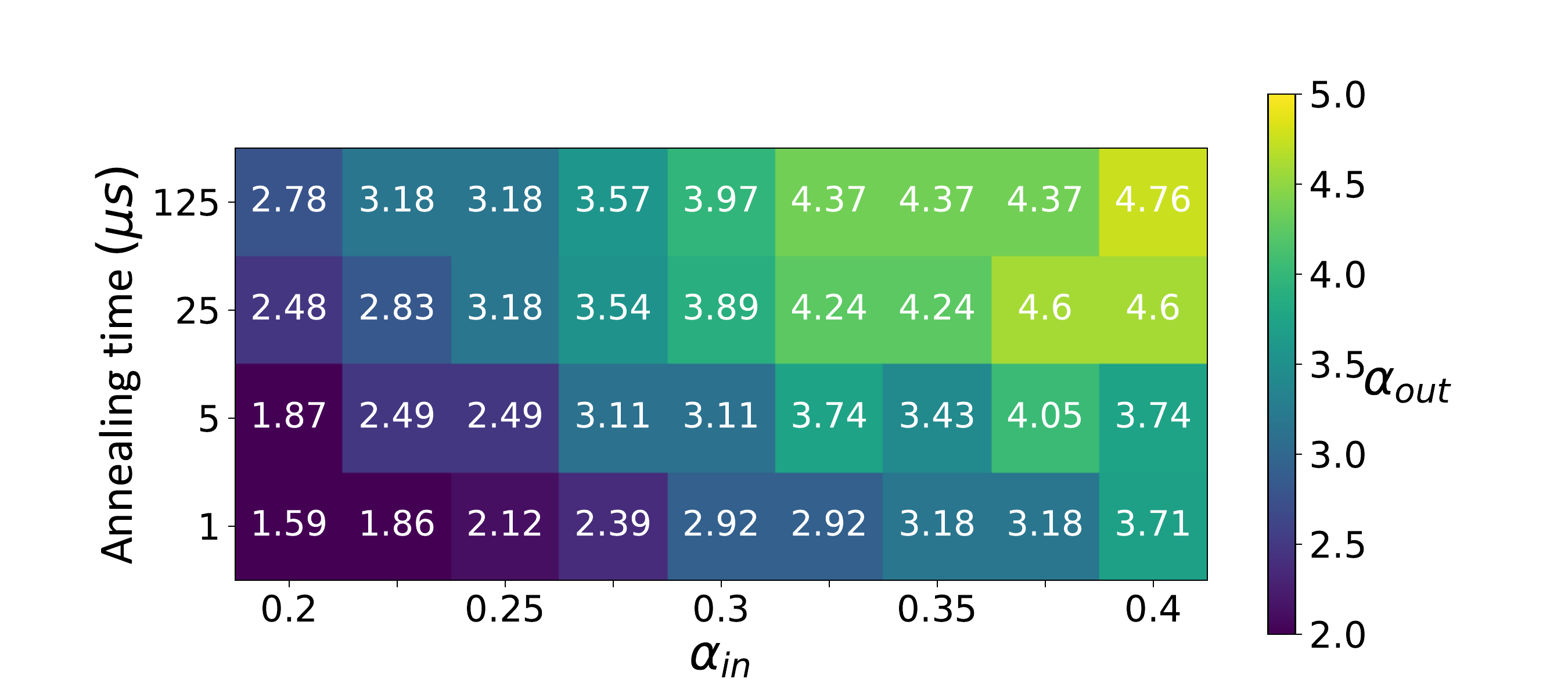}
        \caption{GSD-38}
    \end{subfigure}
    \caption{The dependence of effective temperature $\alpha_{out}$ on $\alpha_{in}$ and annealing time for all GSD models.}
\end{figure*}

\begin{figure*}[h]
\centering
    \begin{subfigure}{0.49\linewidth}
        \includegraphics[width=0.9\linewidth]{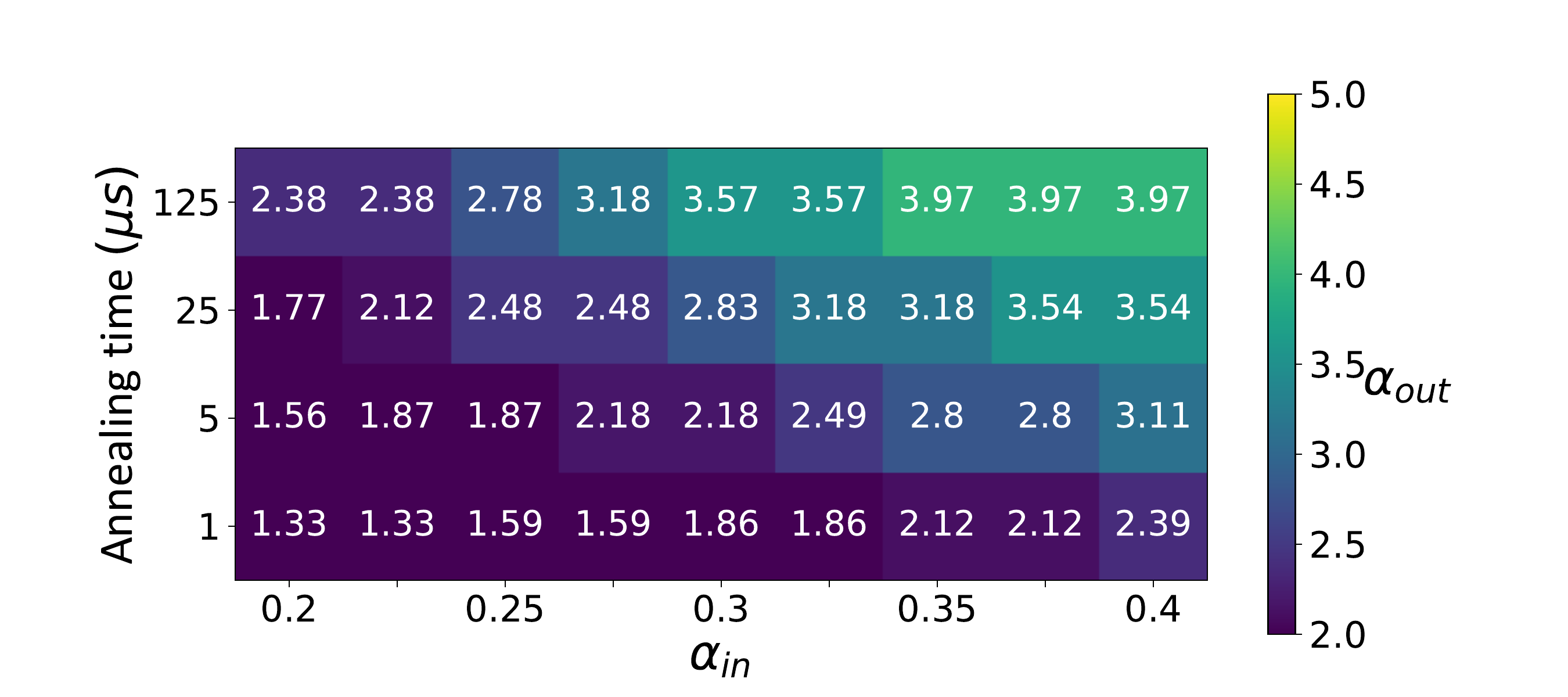}
        \caption{GSD-F-1}
    \end{subfigure}
    \begin{subfigure}{0.49\linewidth}
        \includegraphics[width=0.9\linewidth]{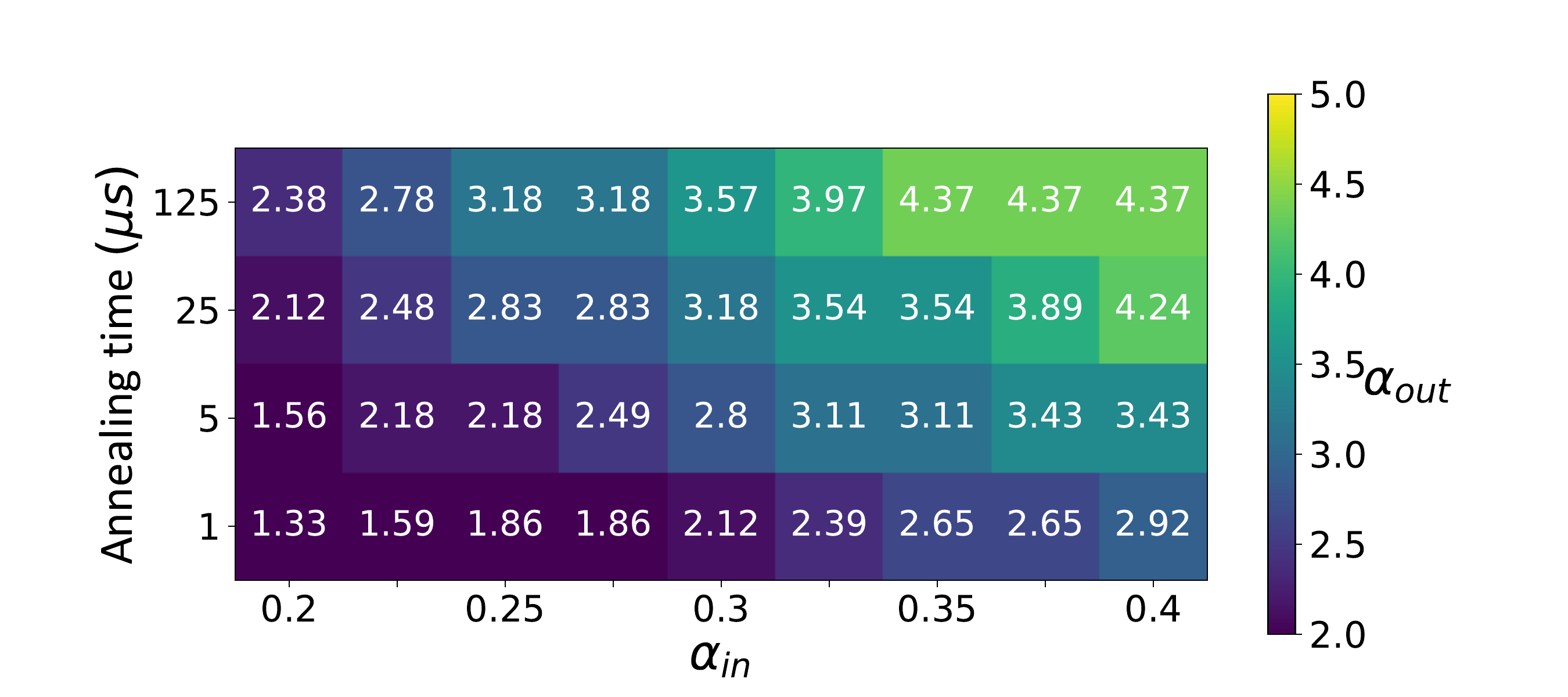}
        \caption{GSD-F-2}
    \end{subfigure}
    \begin{subfigure}{0.49\linewidth}
        \includegraphics[width=0.9\linewidth]{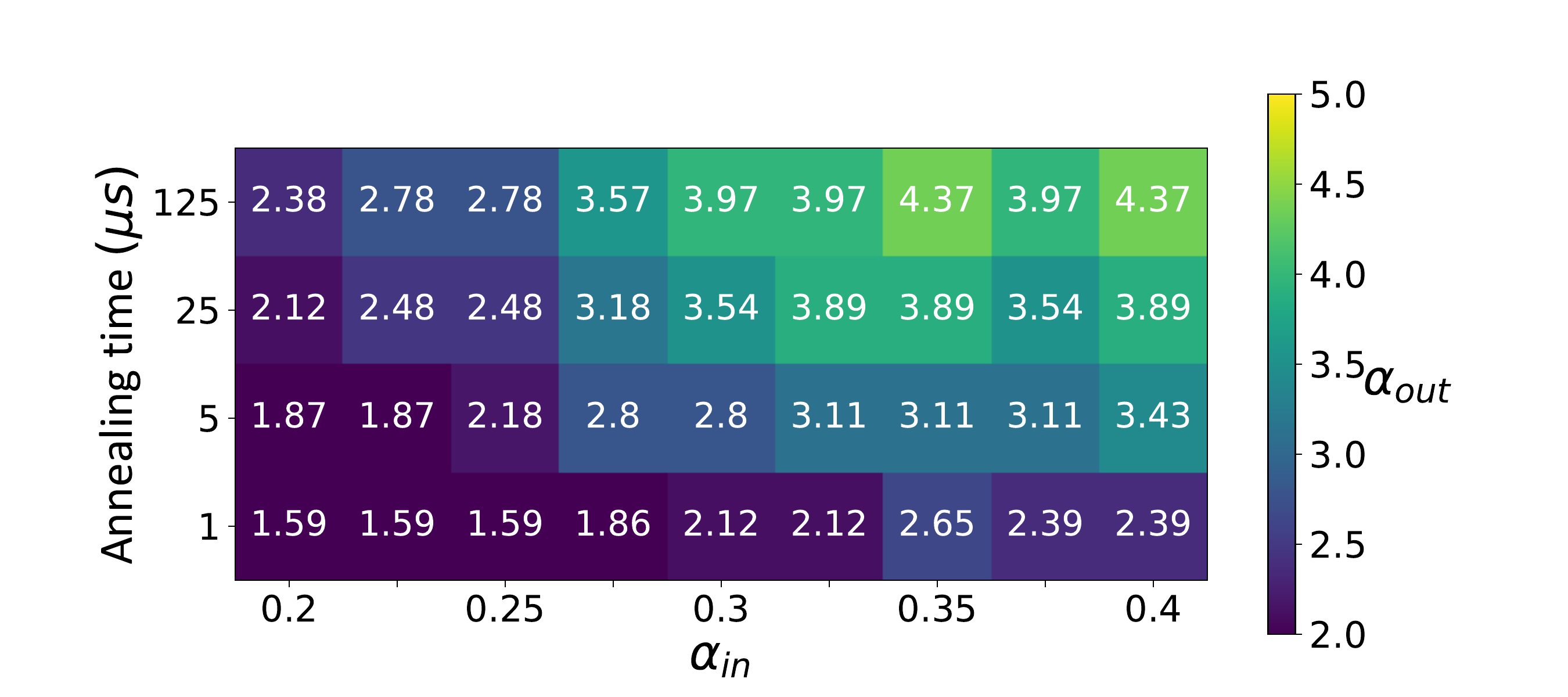}
        \caption{GSD-F-3}
    \end{subfigure}
    \begin{subfigure}{0.49\linewidth}
        \includegraphics[width=0.9\linewidth]{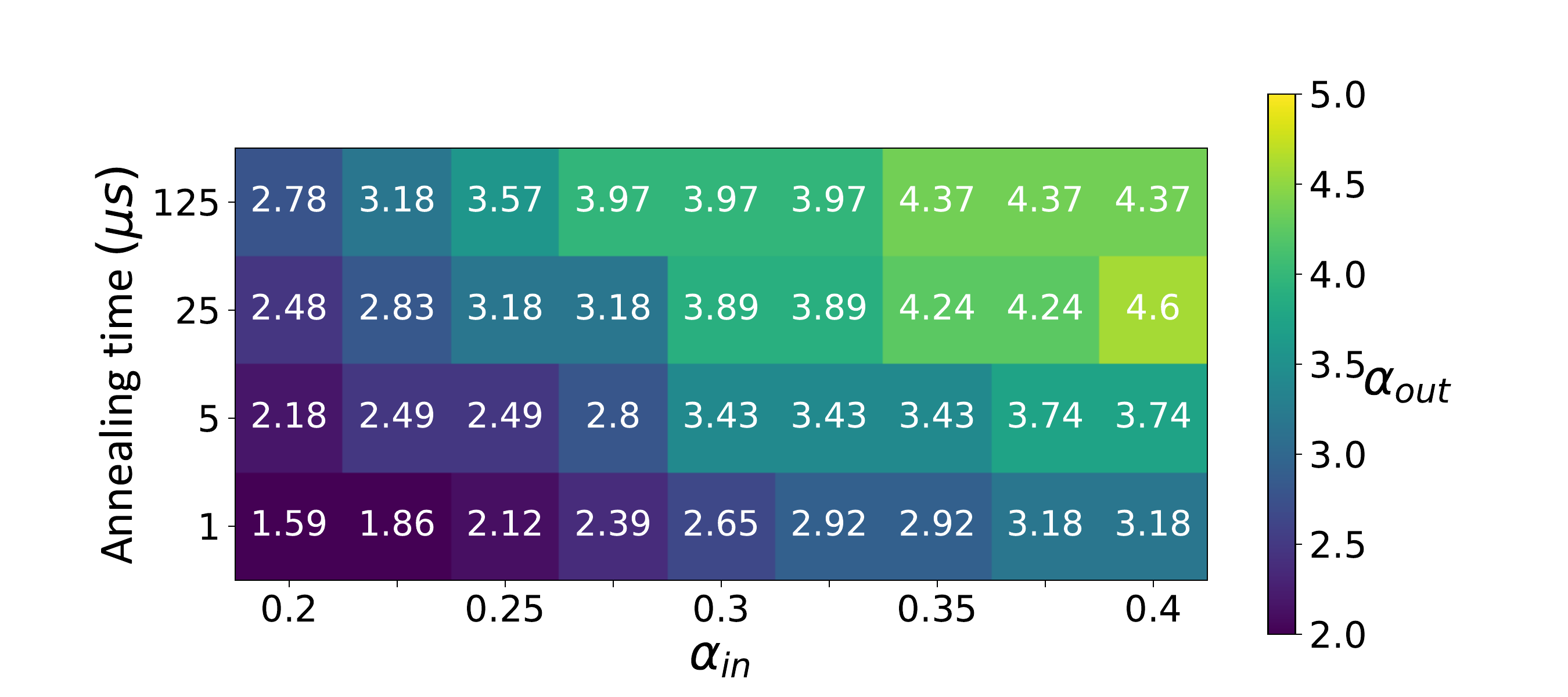}
        \caption{GSD-F-4}
    \end{subfigure}
    \begin{subfigure}{0.49\linewidth}
        \includegraphics[width=0.9\linewidth]{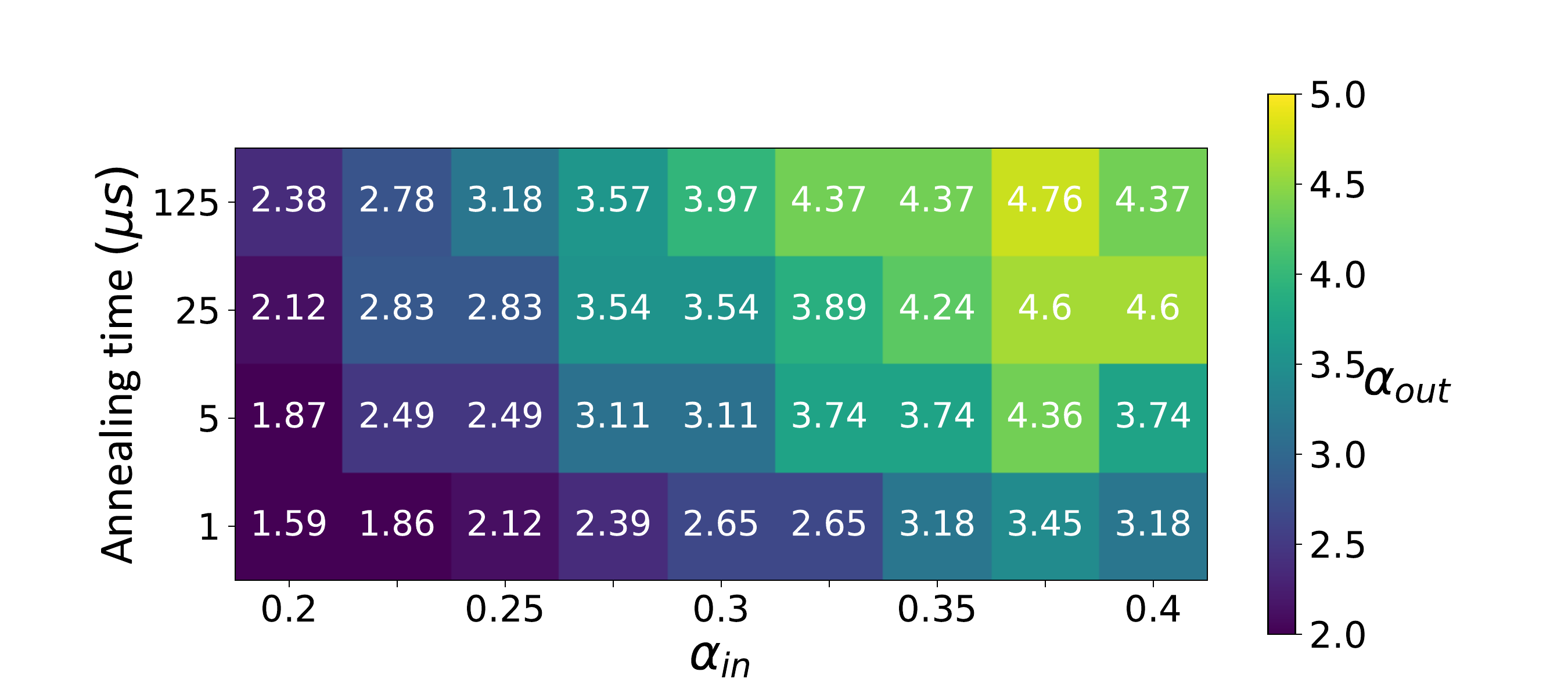}
        \caption{GSD-F-5}
    \end{subfigure}
    \begin{subfigure}{0.49\linewidth}
        \includegraphics[width=0.9\linewidth]{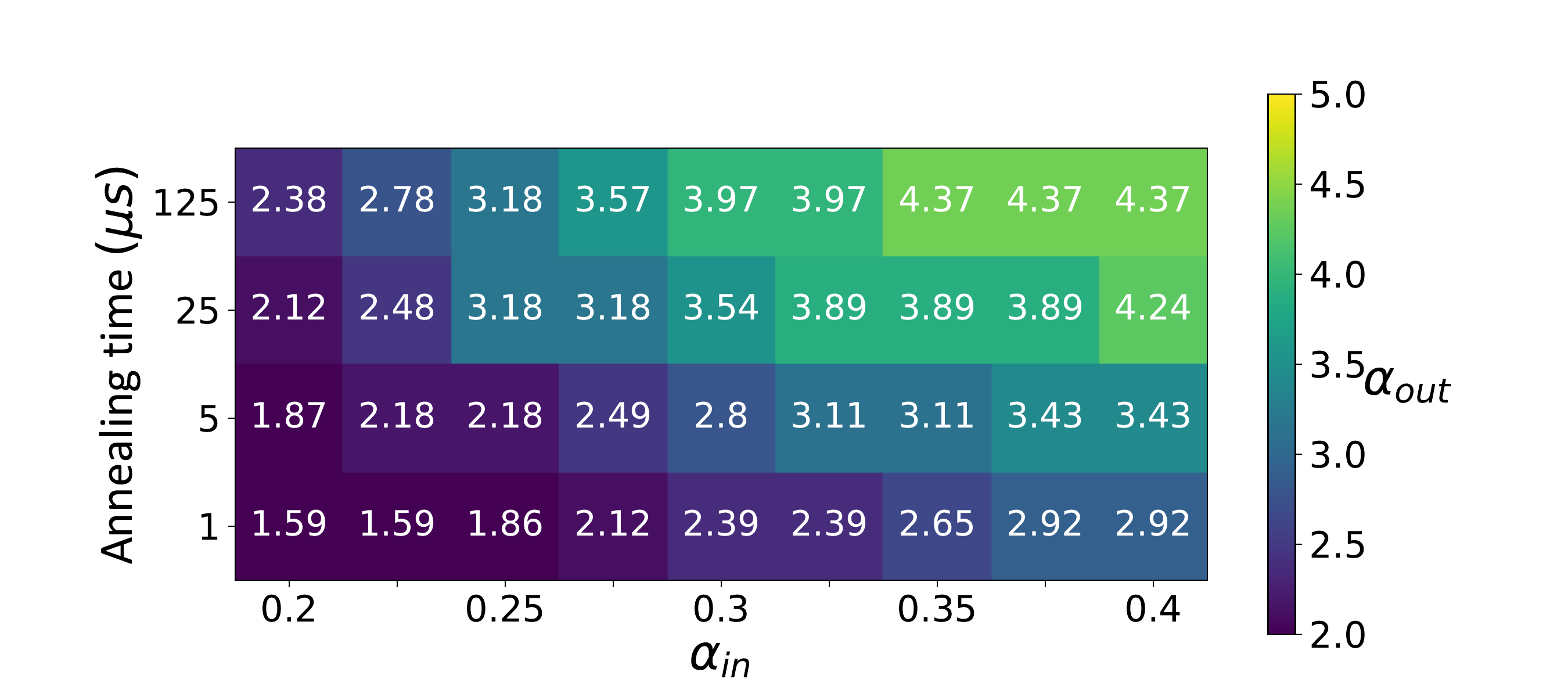}
        \caption{GSD-F-6}
    \end{subfigure}
    \caption{The dependence of effective temperature $\alpha_{out}$ on $\alpha_{in}$ and annealing time for all GSD-F models.}
\end{figure*}

\end{document}